%% file: main.tex
\pdfoutput=1
\documentclass[12pt,a4paper]{article}

\usepackage{ifthen} 
\usepackage{placeins}
\usepackage{siunitx}
\usepackage{xfrac}
\usepackage{array}
\usepackage[percent]{overpic}
\usepackage{multirow}
\newboolean{pdflatex}
\setboolean{pdflatex}{true} 
\usepackage{tikz}
\usetikzlibrary{external}
\newboolean{articletitles}
\setboolean{articletitles}{true} 

\newboolean{uprightparticles}
\setboolean{uprightparticles}{false} 

\def\paperauthors{LHCb collaboration} 
\def\paperasciititle{Search for prompt production of pentaquarks in charm hadron final states} 
\def\papertitle{Search for prompt production of pentaquarks in charm hadron final states} 
\def\paperkeywords{{High Energy Physics}, {LHCb}} 
\def\papercopyright{\the\year\ CERN for the benefit of the LHCb collaboration} 
\def\paperlicence{CC BY 4.0 licence}
\def\paperlicenceurl{https://creativecommons.org/licenses/by/4.0/}

\input{preamble}
\input{my-symbols.tex}
\DeclareCaptionFormat{cont}{#1 (cont.)#2#3\par}
\begin{document}

\renewcommand{\thefootnote}{\fnsymbol{footnote}}
\setcounter{footnote}{1}

\input{title-LHCb-PAPER}


\renewcommand{\thefootnote}{\arabic{footnote}}
\setcounter{footnote}{0}

\cleardoublepage


\pagestyle{plain} 
\setcounter{page}{1}
\pagenumbering{arabic}


\input{body}

\input{acknowledgements}

\clearpage

\input{appendix}


\clearpage

\addcontentsline{toc}{section}{References}
\bibliographystyle{LHCb}
\bibliography{main,standard,LHCb-PAPER,LHCb-CONF,LHCb-DP,LHCb-TDR, my-bib}

\clearpage

\input{Authorship_LHCb-PAPER-2023-018}
\end{document}

%% file: preamble.tex
\usepackage[top=1in, bottom=1.25in, left=1in, right=1in]{geometry}

\usepackage{microtype}
\usepackage{lineno}  
\usepackage{xspace} 
\usepackage{caption} 

\usepackage{graphicx}  
\usepackage{color}
\usepackage{colortbl}
\graphicspath{{./figs/}} 

\usepackage{amsmath} 
\usepackage{amssymb}
\usepackage{amsfonts}
\usepackage{upgreek} 

\newcommand*\patchAmsMathEnvironmentForLineno[1]{%
\expandafter\let\csname old#1\expandafter\endcsname\csname #1\endcsname
\expandafter\let\csname oldend#1\expandafter\endcsname\csname
end#1\endcsname
 \renewenvironment{#1}%
   {\linenomath\csname old#1\endcsname}%
   {\csname oldend#1\endcsname\endlinenomath}%
}
\newcommand*\patchBothAmsMathEnvironmentsForLineno[1]{%
  \patchAmsMathEnvironmentForLineno{#1}%
  \patchAmsMathEnvironmentForLineno{#1*}%
}
\AtBeginDocument{%
\patchBothAmsMathEnvironmentsForLineno{equation}%
\patchBothAmsMathEnvironmentsForLineno{align}%
\patchBothAmsMathEnvironmentsForLineno{flalign}%
\patchBothAmsMathEnvironmentsForLineno{alignat}%
\patchBothAmsMathEnvironmentsForLineno{gather}%
\patchBothAmsMathEnvironmentsForLineno{multline}%
\patchBothAmsMathEnvironmentsForLineno{eqnarray}%
}

\usepackage{hyperxmp}

\usepackage[pdftex,
            pdfauthor={\paperauthors},
            pdftitle={\paperasciititle},
            pdfkeywords={\paperkeywords},
            pdfcopyright={Copyright (C) \papercopyright},
            pdflicenseurl={\paperlicenceurl}]{hyperref}

\usepackage[colorinlistoftodos,textsize=scriptsize]{todonotes}

\usepackage[bottom,flushmargin,hang,multiple]{footmisc}

\usepackage[all]{hypcap} 

\input{lhcb-symbols-def} 

\usepackage{cite} 
\usepackage{mciteplus}

%% file: lhcb-symbols-def.tex
\usepackage{xspace} 
\usepackage{upgreek}

\def\lhcb   {\mbox{LHCb}\xspace}

\def\MagUp {\mbox{\em Mag\kern -0.05em Up}\xspace}

\ifthenelse{\boolean{uprightparticles}}%
{

 \def\Ppi         {\ensuremath{\uppi}\xspace}

 \def\Ppsi        {\ensuremath{\uppsi}\xspace}

 \def\PDelta      {\ensuremath{\Delta}\xspace}                 
 \def\PXi         {\ensuremath{\Xi}\xspace}                 
 \def\PLambda     {\ensuremath{\Lambda}\xspace}                 
 \def\PSigma      {\ensuremath{\Sigma}\xspace}                 
 \def\POmega      {\ensuremath{\Omega}\xspace}                 
 \def\PUpsilon    {\ensuremath{\Upsilon}\xspace}
 \let\oldPi\Pi
 \def\PPi         {\ensuremath{\oldPi}\xspace}

 \def\PB      {\ensuremath{\mathrm{B}}\xspace}                 
                  
 \def\PD      {\ensuremath{\mathrm{D}}\xspace}

 \def\PJ      {\ensuremath{\mathrm{J}}\xspace}                 
 \def\PK      {\ensuremath{\mathrm{K}}\xspace}

 \def\Pb      {\ensuremath{\mathrm{b}}\xspace}                 
 \def\Pc      {\ensuremath{\mathrm{c}}\xspace}                 
 \def\Pd      {\ensuremath{\mathrm{d}}\xspace}

 \def\Pi      {\ensuremath{\mathrm{i}}\xspace}

 \def\Pp      {\ensuremath{\mathrm{p}}\xspace}

 \def\Ps      {\ensuremath{\mathrm{s}}\xspace}                 
                  
 \def\Pu      {\ensuremath{\mathrm{u}}\xspace}

 \def\thebaroffset{0.0em}
}
{

 \def\Ppi         {\ensuremath{\pi}\xspace}

 \def\Ppsi        {\ensuremath{\psi}\xspace}                 
                  
 \mathchardef\PDelta="7101
 \mathchardef\PXi="7104
 \mathchardef\PLambda="7103
 \mathchardef\PSigma="7106
 \mathchardef\POmega="710A
 \mathchardef\PUpsilon="7107
 \mathchardef\PPi="7105
                  
 \def\PB      {\ensuremath{B}\xspace}                 
                  
 \def\PD      {\ensuremath{D}\xspace}

 \def\PJ      {\ensuremath{J}\xspace}                 
 \def\PK      {\ensuremath{K}\xspace}

 \def\Pb      {\ensuremath{b}\xspace}                 
 \def\Pc      {\ensuremath{c}\xspace}                 
 \def\Pd      {\ensuremath{d}\xspace}

 \def\Pi      {\ensuremath{i}\xspace}

 \def\Pp      {\ensuremath{p}\xspace}

 \def\Ps      {\ensuremath{s}\xspace}                 
                  
 \def\Pu      {\ensuremath{u}\xspace}

 \def\thebaroffset{0.18em}
}
\newcommand{\offsetoverline}[2][\thebaroffset]{\kern #1\overline{\kern -#1 #2}}%

\makeatletter
\ifcase \@ptsize \relax
  \newcommand{\miniscule}{\@setfontsize\miniscule{4}{5}}
\or
  \newcommand{\miniscule}{\@setfontsize\miniscule{5}{6}}
\or
  \newcommand{\miniscule}{\@setfontsize\miniscule{5}{6}}
\fi
\makeatother

\DeclareRobustCommand{\optbar}[1]{\shortstack{{\miniscule (\rule[.5ex]{1.25em}{.18mm})}
  \\ [-.7ex] $#1$}}

\def\uquark    {{\ensuremath{\Pu}}\xspace}

\def\dquark    {{\ensuremath{\Pd}}\xspace}

\def\squark    {{\ensuremath{\Ps}}\xspace}

\def\cquark    {{\ensuremath{\Pc}}\xspace}
\def\cquarkbar {{\ensuremath{\overline \cquark}}\xspace}

\def\bquark    {{\ensuremath{\Pb}}\xspace}

\def\pion   {{\ensuremath{\Ppi}}\xspace}

\def\pip    {{\ensuremath{\pion^+}}\xspace}
\def\pim    {{\ensuremath{\pion^-}}\xspace}
\def\pipm   {{\ensuremath{\pion^\pm}}\xspace}

\def\kaon    {{\ensuremath{\PK}}\xspace}

\def\KorKbar {\kern \thebaroffset\optbar{\kern -\thebaroffset \PK}{}\xspace}

\def\Km      {{\ensuremath{\kaon^-}}\xspace}

\def\Dbar    {{\ensuremath{\offsetoverline{\PD}}}\xspace}
\def\D       {{\ensuremath{\PD}}\xspace}
\def\Db      {{\ensuremath{\Dbar}}\xspace}
\def\DorDbar {\kern \thebaroffset\optbar{\kern -\thebaroffset \PD}\xspace}
\def\Dz      {{\ensuremath{\D^0}}\xspace}
\def\Dzb     {{\ensuremath{\Dbar{}^0}}\xspace}
\def\Dp      {{\ensuremath{\D^+}}\xspace}
\def\Dm      {{\ensuremath{\D^-}}\xspace}

\def\DpDm    {\ensuremath{\Dp {\kern -0.16em \Dm}}\xspace}
\def\Dstar   {{\ensuremath{\D^*}}\xspace}

\def\Dstarzb {{\ensuremath{\Dbar{}^{*0}}}\xspace}

\def\Dstarp  {{\ensuremath{\D^{*+}}}\xspace}
\def\Dstarm  {{\ensuremath{\D^{*-}}}\xspace}

\def\B       {{\ensuremath{\PB}}\xspace}

\def\BorBbar {\kern \thebaroffset\optbar{\kern -\thebaroffset \PB}\xspace}

\def\Bd      {{\ensuremath{\B^0}}\xspace}

\def\BdorBdbar {\kern \thebaroffset\optbar{\kern -\thebaroffset \Bd}\xspace}

\def\Bs      {{\ensuremath{\B^0_\squark}}\xspace}

\def\BsorBsbar {\kern \thebaroffset\optbar{\kern -\thebaroffset \Bs}\xspace}

\def\jpsi     {{\ensuremath{{\PJ\mskip -3mu/\mskip -2mu\Ppsi}}}\xspace}

\def\Y#1S{\ensuremath{\PUpsilon{(#1S)}}\xspace}

\def\proton      {{\ensuremath{\Pp}}\xspace}

\def\Lz          {{\ensuremath{\PLambda}}\xspace}

\def\LorLbar     {\kern \thebaroffset\optbar{\kern -\thebaroffset \PLambda}\xspace}

\def\Sigmares    {{\ensuremath{\PSigma}}\xspace}

\def\Xires       {{\ensuremath{\PXi}}\xspace}

\def\Lc          {{\ensuremath{\Lz^+_\cquark}}\xspace}

\def\Xicz        {{\ensuremath{\Xires^0_\cquark}}\xspace}

\def\Xicc        {{\ensuremath{\Xires_{\cquark\cquark}}}\xspace}

\def\Lb           {{\ensuremath{\Lz^0_\bquark}}\xspace}

\newcommand{\decay}[2]{\ensuremath{#1\!\to #2}\xspace} 

\def\to                 {\ensuremath{\rightarrow}\xspace}

\def\AT#1     {\ensuremath{A_{\mathrm{T}}^{#1}}\xspace}           

\def\C#1      {\ensuremath{\mathcal{C}_{#1}}\xspace}                       
\def\Cp#1     {\ensuremath{\mathcal{C}_{#1}^{'}}\xspace}                    
\def\Ceff#1   {\ensuremath{\mathcal{C}_{#1}^{\mathrm{(eff)}}}\xspace}        
\def\Cpeff#1  {\ensuremath{\mathcal{C}_{#1}^{'\mathrm{(eff)}}}\xspace}       
\def\Ope#1    {\ensuremath{\mathcal{O}_{#1}}\xspace}                       
\def\Opep#1   {\ensuremath{\mathcal{O}_{#1}^{'}}\xspace}                    

\newcommand{\nospaceunit}[1]{\ensuremath{\text{#1}}}       
\newcommand{\aunit}[1]{\ensuremath{\text{\,#1}}}       

\newcommand{\tev}{\aunit{Te\kern -0.1em V}\xspace}
\newcommand{\gev}{\aunit{Ge\kern -0.1em V}\xspace}
\newcommand{\mev}{\aunit{Me\kern -0.1em V}\xspace}
\newcommand{\kev}{\aunit{ke\kern -0.1em V}\xspace}
\newcommand{\ev}{\aunit{e\kern -0.1em V}\xspace}
 
\newcommand{\mevc}{\ensuremath{\aunit{Me\kern -0.1em V\!/}c}\xspace}
\newcommand{\gevc}{\ensuremath{\aunit{Ge\kern -0.1em V\!/}c}\xspace}
\newcommand{\mevcc}{\ensuremath{\aunit{Me\kern -0.1em V\!/}c^2}\xspace}
\newcommand{\gevcc}{\ensuremath{\aunit{Ge\kern -0.1em V\!/}c^2}\xspace}

\def\mum  {\ensuremath{\,\upmu\nospaceunit{m}}\xspace}

\def\fb   {\ensuremath{\aunit{fb}}\xspace}
\def\invfb   {\ensuremath{\fb^{-1}}\xspace}

\def\gsim{{~\raise.15em\hbox{$>$}\kern-.85em
          \lower.35em\hbox{$\sim$}~}\xspace}
\def\lsim{{~\raise.15em\hbox{$<$}\kern-.85em
          \lower.35em\hbox{$\sim$}~}\xspace}

\def\pt         {\ensuremath{p_{\mathrm{T}}}\xspace}

\def\ptot       {\ensuremath{p}\xspace}

\def\evtgen     {\mbox{\textsc{EvtGen}}\xspace}

\def\geant      {\mbox{\textsc{Geant4}}\xspace}

\def\photos     {\mbox{\textsc{Photos}}\xspace}

\def\pythia     {\mbox{\textsc{Pythia}}\xspace}

\def\tell1  {TELL1\xspace}
\def\ukl1   {UKL1\xspace}

\newcommand{\eg}{\mbox{\itshape e.g.}\xspace}

\newcommand{\phz}{\phantom{0}}

\newcommand{\lhcborcid}[1]{\href{https://orcid.org/#1}{\hspace*{0.1em}\raisebox{-0.45ex}{\includegraphics[width=1em]{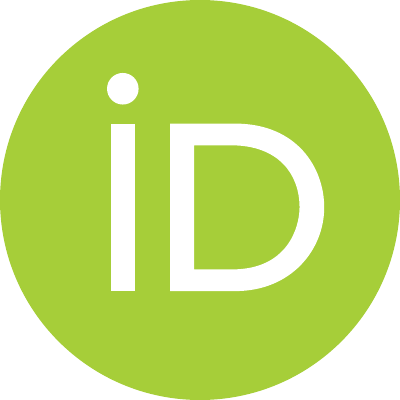}}}}


%% file: my-symbols.tex
\usepackage{xspace} 
\usepackage{upgreek}

\def\Sigmares    {{\ensuremath{\PSigma}}\xspace}
\def\cquark      {{\ensuremath{\Pc}}\xspace}
\def\Sigmac      {{\ensuremath{\Sigmares_\cquark}}\xspace}

\def\Sigmacp     {{\ensuremath{\Sigmares_\cquark^{+}}}\xspace}
\def\Sigmacpp    {{\ensuremath{\Sigmares_\cquark^{++}}}\xspace}

\def\Sigmacz     {{\ensuremath{\Sigmares_\cquark^{0}}}\xspace}
\def\Sigmacstpp    {{\ensuremath{\Sigmares_\cquark^{*++}}}\xspace}

\def\Sigmacbstbpp    {{\ensuremath{\Sigmares_\cquark^{(*)++}}}\xspace}
\def\Sigmacbstbz     {{\ensuremath{\Sigmares_\cquark^{(*)0}}}\xspace}

\def\Sigmacst    {{\ensuremath{\Sigmares_\cquark^{*}}}\xspace}
\def\Sigmacstz     {{\ensuremath{\Sigmares_\cquark^{*0}}}\xspace}

\def\Isospin     {{\ensuremath{I_3}}\xspace}
\def\Hypercharge {{\ensuremath{Y}}\xspace}

\def\pval {{$p$\hspace{0pt}-\hspace{0pt}value}\xspace}
\def\pvals {{$p$\hspace{0pt}-\hspace{0pt}values}\xspace}
\def\qval {{$Q$\hspace{0pt}-\hspace{0pt}value}\xspace}

%% file: title-LHCb-PAPER.tex

\begin{titlepage}
\pagenumbering{roman}

\vspace*{-1.5cm}
\centerline{\large EUROPEAN ORGANIZATION FOR NUCLEAR RESEARCH (CERN)}
\vspace*{1.5cm}
\noindent
\begin{tabular*}{\linewidth}{lc@{\extracolsep{\fill}}r@{\extracolsep{0pt}}}
\ifthenelse{\boolean{pdflatex}}
{\vspace*{-1.5cm}\mbox{\!\!\!\includegraphics[width=.14\textwidth]{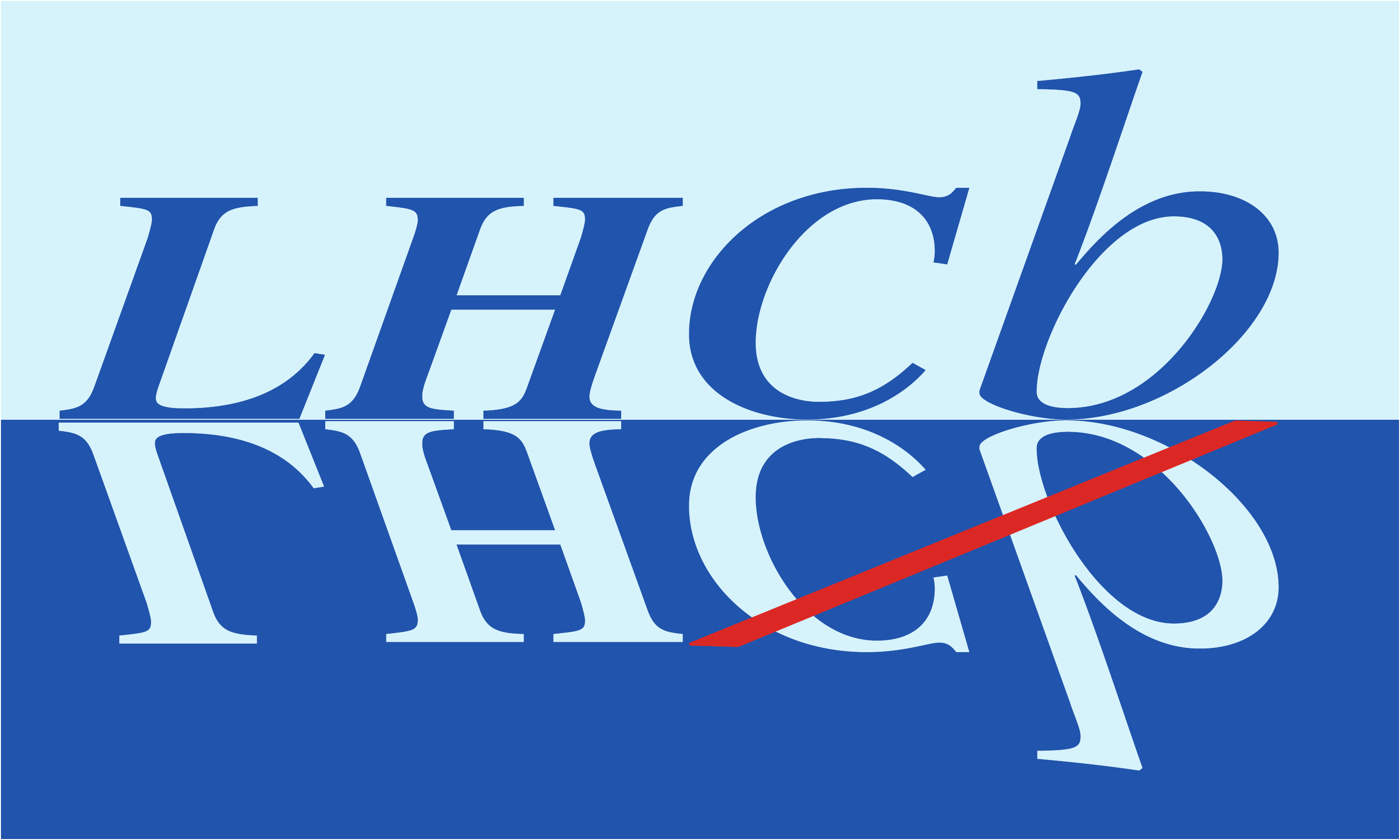}} & &}%
{\vspace*{-1.2cm}\mbox{\!\!\!\includegraphics[width=.12\textwidth]{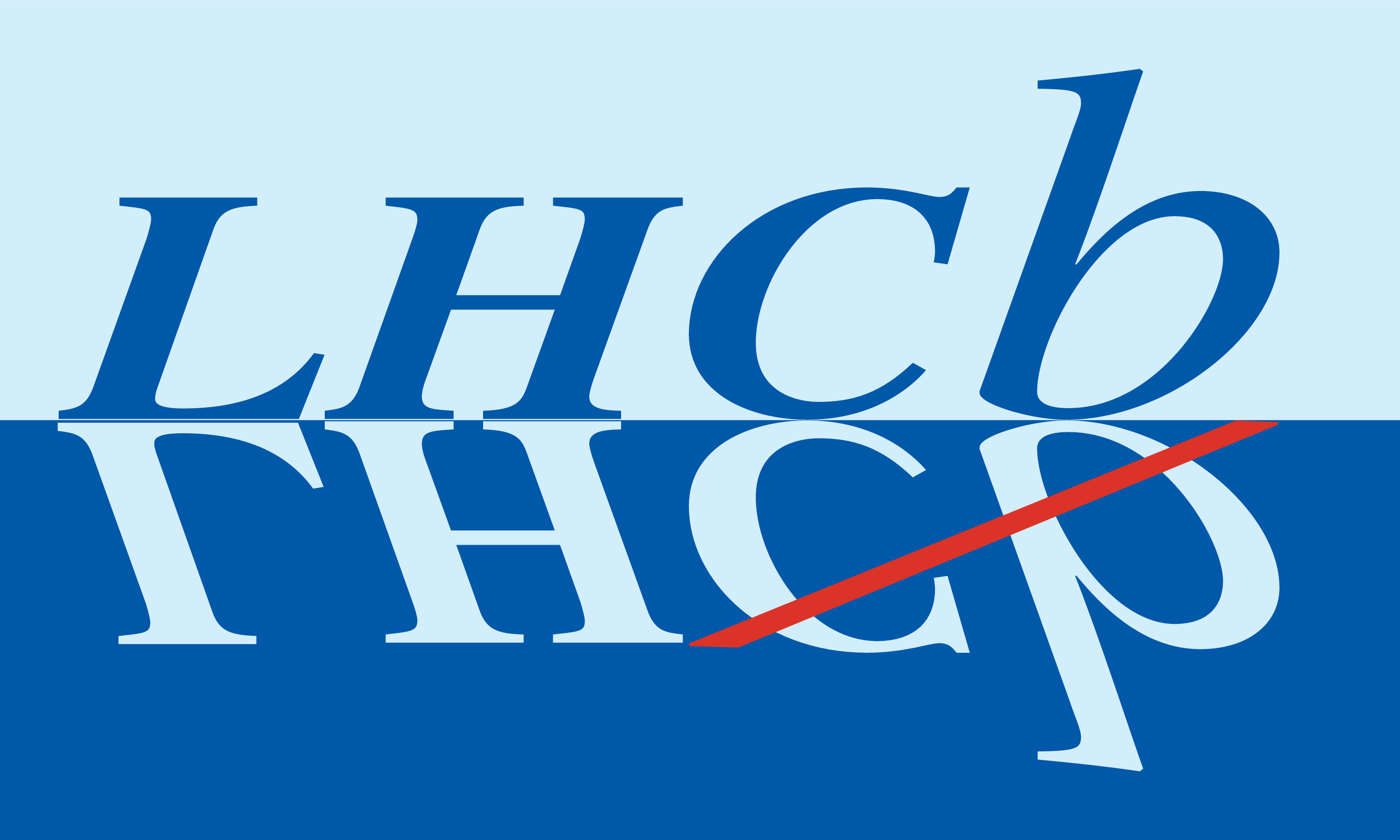}} & &}%
\\
 & & CERN-EP-2024-071 \\  
 & & LHCb-PAPER-2023-018 \\  
 & & August 2, 2024\\ 
 & & \\
\end{tabular*}

\vspace*{4.0cm}

{\normalfont\bfseries\boldmath\huge
\begin{center}
  \papertitle 
\end{center}
}

\vspace*{2.0cm}

\begin{center}
\paperauthors\footnote{Authors are listed at the end of this paper.}
\end{center}

\vspace{\fill}

\begin{abstract}
  \noindent
  A search for hidden-charm pentaquark states decaying to a range of $\Sigmares_c\Dbar$ and $\Lc \Dbar$ final states,
  as well as doubly-charmed pentaquark states to $\Sigmares_c\D$ and $\Lc \D$, is made using samples of proton-proton collision data corresponding to an integrated luminosity of $5.7\invfb$ recorded by the LHCb detector at $\sqrt{s} = 13\tev$. Since no significant signals are found, upper limits are set on the pentaquark yields relative to that of the \Lc baryon in the \decay{\Lc}{\proton\Km\pip} decay mode. The known pentaquark states are also investigated, and their signal yields are found to be consistent with zero in all cases.
  
\end{abstract}

\vspace*{2.0cm}

\begin{center}
  Published in Phys. Rev. D 110 (2024) 032001
\end{center}

\vspace{\fill}

{\footnotesize 
\centerline{\copyright~\papercopyright. \href{\paperlicenceurl}{\paperlicence}.}}
\vspace*{2mm}

\end{titlepage}


\newpage
\setcounter{page}{2}
\mbox{~}
%
%
%
%

%% file: body.tex
\section{Introduction}
\label{sec:Introduction}

Since the formulation of the quark model~\cite{Gell-Mann:1964ewy, Zweig:1964jf}, hadronic states beyond the conventional $q\bar q$ mesons and $qqq$ baryons have been proposed. Hadrons with different combinations of quarks $q$ and gluons $g$, such as pentaquarks ($q \bar q q q q$), tetraquarks ($q \bar q q \bar q$)~\cite{Gell-Mann:1964ewy, Zweig:1964jf}, six-quark H-dibaryons ($q \bar q q \bar q q \bar q$)~\cite{Jaffe:1976yi}, hybrids ($q\bar q g$)~\cite{Horn:1977rq} and glueballs ($ggg$)~\cite{Fritzsch:1973pi} have been predicted by QCD-based models. The existence of such ``exotic'' hadrons had been debated for several years without a consensus being reached. In the early 2000s, new hadrons with unexpected features were observed, such as the $D^*_{s0}(2317)^+$ \cite{BaBar:2003oey} and $\chi_{c1}(3872)$ \cite{Belle:2003nnu} mesons,  followed shortly after by the discovery of many other charmonium-like and bottomo\-nium-like states. While it is still not possible to rule out firmly a conventional nature for the majority of such states, the observation of manifestly exotic hadrons such as the $Z_c(4430)^+$ meson~\cite{Belle:2007hrb}, an electrically charged charmonium-like state, three $P_c^+$ baryons~\cite{LHCb-PAPER-2015-029, LHCB-PAPER-2016-009, LHCb-PAPER-2019-014}, with a minimal quark content $c \bar c uud$, and the $T_{cc}^+$ state~\cite{LHCb-PAPER-2021-031,LHCb-PAPER-2021-032}, a meson containing two charm quarks, established the existence of QCD exotics. Many models have been proposed to explain the exotic nature of such states: {\it hadronic molecules} \cite{Close:2003sg, Karliner:2015ina}, whose constituents are color-singlet hadrons bound by residual nuclear forces; {\it tetraquarks and pentaquarks} \cite{Maiani:2004vq, Maiani:2015vwa}, bound states where diquarks and diantiquarks are building blocks; {\it hadro-quarkonium} \cite{Dubynskiy:2008mq}, a cloud of light quarks and gluons bound to a heavy $Q \offsetoverline Q$ core state via van-der-Waals forces ($Q$ represents a heavy quark, such as the \bquark quark), and {\it threshold effects}, enhancements caused by threshold cusps or rescattering processes \cite{Bugg:2011jr, Pakhlov:2014qva}. An intriguing feature of many exotic hadrons is the proximity to hadron-hadron thresholds. For example, the three pentaquark states $P_c(4312)^+$ , $P_c(4440)^+$ and $P_c(4457)^+$, observed in the $\jpsi \proton$ projection of $\decay{\Lb}{\jpsi \proton \Km}$ decays~\cite{LHCb-PAPER-2019-014}, have masses that are just below the $\Sigmacp\Dzb$ and $\Sigmacp\Dstarzb$ thresholds.\footnote{The $\Sigmares_c(2455)$ and $\Sigmares_c(2520)$ baryons are referred to as $\Sigmares_c$ and $\Sigmacst$, respectively.} More experimental and theoretical scrutiny is required to understand if this is just coincidental or related to the internal structure of the states as bound systems of a baryon and a meson. As for the conventional hadrons, the observation of new decay modes can shed light on the binding mechanism of exotic hadrons.  In addition, many models predict doubly-charmed pentaquark states, where doubly-charmed refers to a state with two units of charm quantum number, decaying to a range of \Sigmac\D and \Lc\D combinations~\cite{Yang_2020}. Excited \Xicc baryons could be observed in such final states as well~\cite{Padmanath:2013laa}.

This paper presents a search for the $P_c(4312)^+$, $P_c(4440)^+$ and $P_c(4457)^+$ baryons and other pentaquarks with hidden charm, meaning charm quantum number equal to zero, in the prompt $\Sigmac \Db^{(*)}$, $\Sigmacst \Db^{(*)}$, $\Lc \Db^{(*)}$,  and $\Lc \pi \Db^{(*)}$ mass spectra.\footnote{The inclusion of charge-conjugate processes is implied throughout.} A search for pentaquarks containing two charm quarks has also been carried out in the $\Sigmac \D^{(*)}$,  $\Sigmacst \D^{(*)}$, $\Lc \D^{(*)}$ and $\Lc \pi \D^{(*)}$ mass spectra. Upper limits on the yields in these spectra relative to the \decay{\Lc}{\proton\Km\pip} normalisation channel are presented. The measurements are based on samples of proton-proton ($pp$) collision data corresponding to an integrated luminosity of $5.7\invfb$ at centre-of-mass energies of $13\tev$ recorded by the LHCb experiment between 2016 and 2018.

\section{The \texorpdfstring{\lhcb}{LHCb} detector}
The \lhcb detector~\cite{LHCb-DP-2008-001,LHCb-DP-2014-002} is a single-arm forward
spectrometer covering the \mbox{pseudorapidity} range $2<\eta <5$,
designed for the study of particles containing \bquark or \cquark
quarks. The detector includes a high-precision tracking system
consisting of a silicon-strip vertex detector surrounding the $pp$
interaction region, a large-area silicon-strip detector located
upstream of a dipole magnet with a bending power of about
$4{\mathrm{\,Tm}}$ and three stations of silicon-strip detectors and straw
drift tubes placed downstream of the magnet. The tracking system provides a measurement of the momentum, \ptot, of charged particles with a relative uncertainty that varies from $0.5\%$ at low momentum to $1.0\%$ at $200\gevc$. The minimum distance of a track to a primary $pp$ collision vertex (PV), the impact parameter (IP), is measured with a resolution of $(15+29/\pt)\mum$, where \pt is the component of the momentum transverse to the beam, in$\gevc$. Different types of charged hadrons are distinguished using information from two ring-imaging Cherenkov detectors. Photons, electrons and hadrons are identified by a calorimeter system consisting of scintillating-pad and preshower detectors, an electromagnetic and a hadronic calorimeter. Muons are identified by a system composed of alternating layers of iron and multiwire proportional chambers. 

The online event selection is performed by a trigger, which consists of a hardware stage followed by a two-level software stage \cite{LHCb-DP-2019-001}. At the hardware trigger stage, events are required to have a muon with high \pt or hadron, photon or electron with high transverse energy in the calorimeters. In between the two software trigger stages, an alignment and calibration of the detector is performed in near real-time and their results are used in the trigger~\cite{Borghi:2017hfp}. The same alignment and calibration information is propagated to the offline reconstruction, ensuring consistent and high-quality particle identification (PID) information between the trigger and offline software. The identical performance of the online and offline reconstruction offers the opportunity to perform physics analyses directly using candidates reconstructed in the trigger \cite{LHCb-DP-2012-004,LHCb-DP-2016-001}. This analysis exploits this by using $\Dz$, $\Dp$ and $\Lc$ candidates fully reconstructed in the trigger, as well as single pions for certain final states.

Simulated $pp$ collisions are generated using \pythia~\cite{Sjostrand:2006za,*Sjostrand:2007gs}  with a specific \lhcb configuration~\cite{LHCb-PROC-2010-056}.  Decays of hadronic particles are described by \evtgen~\cite{Lange:2001uf}, in which
final-state radiation is generated using \photos~\cite{Golonka:2005pn}. The interaction of the generated
particles with the detector, and its response, are implemented using the \geant toolkit~\cite{Allison:2006ve,*Agostinelli:2002hh} as described in Ref.~\cite{LHCb-PROC-2011-006}. 

  \section{Selection}
\begin{figure}[tb]
    \centering
    \begin{overpic}[width=0.48\linewidth]{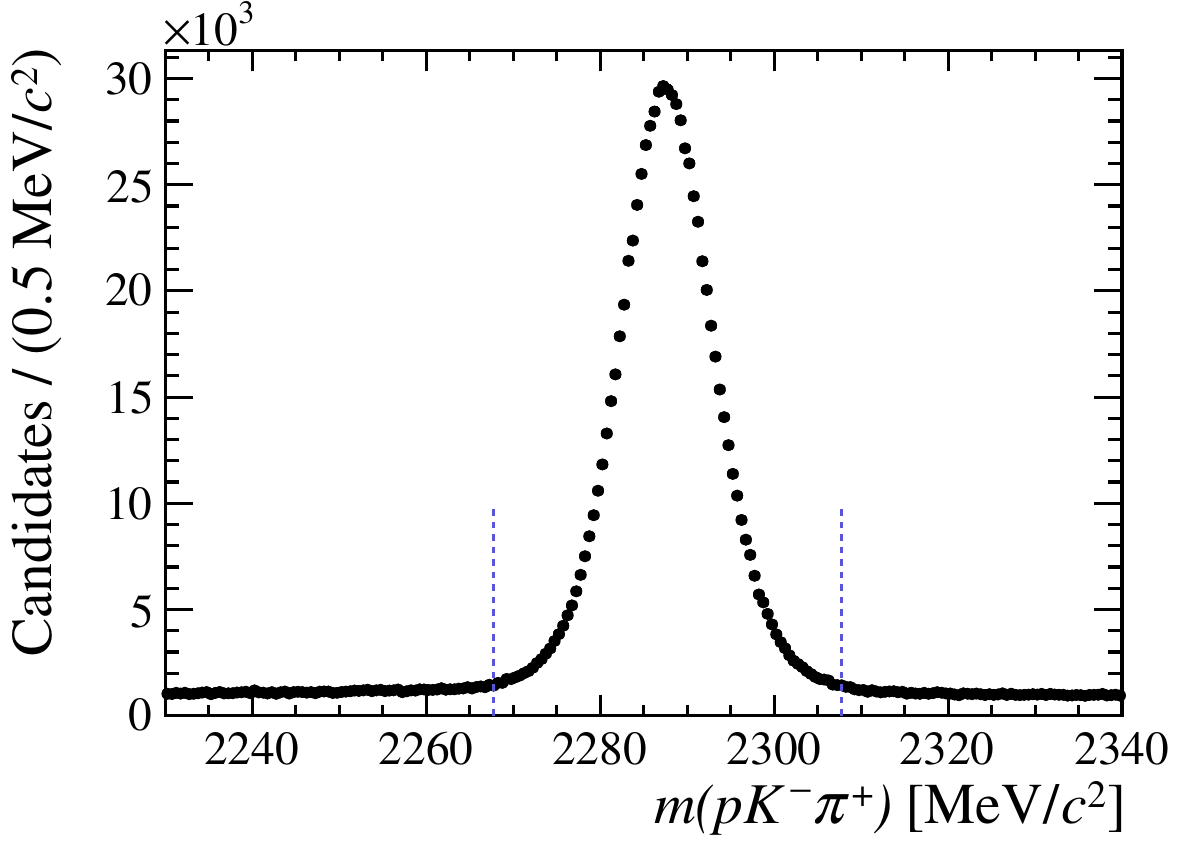}
    \put(70,50) {\begin{tabular}{@{}l@{}}\lhcb \\ 5.7\invfb\end{tabular}}
    \put(30,50){(a)}
    \end{overpic}
    \begin{overpic}[width=0.48\linewidth]{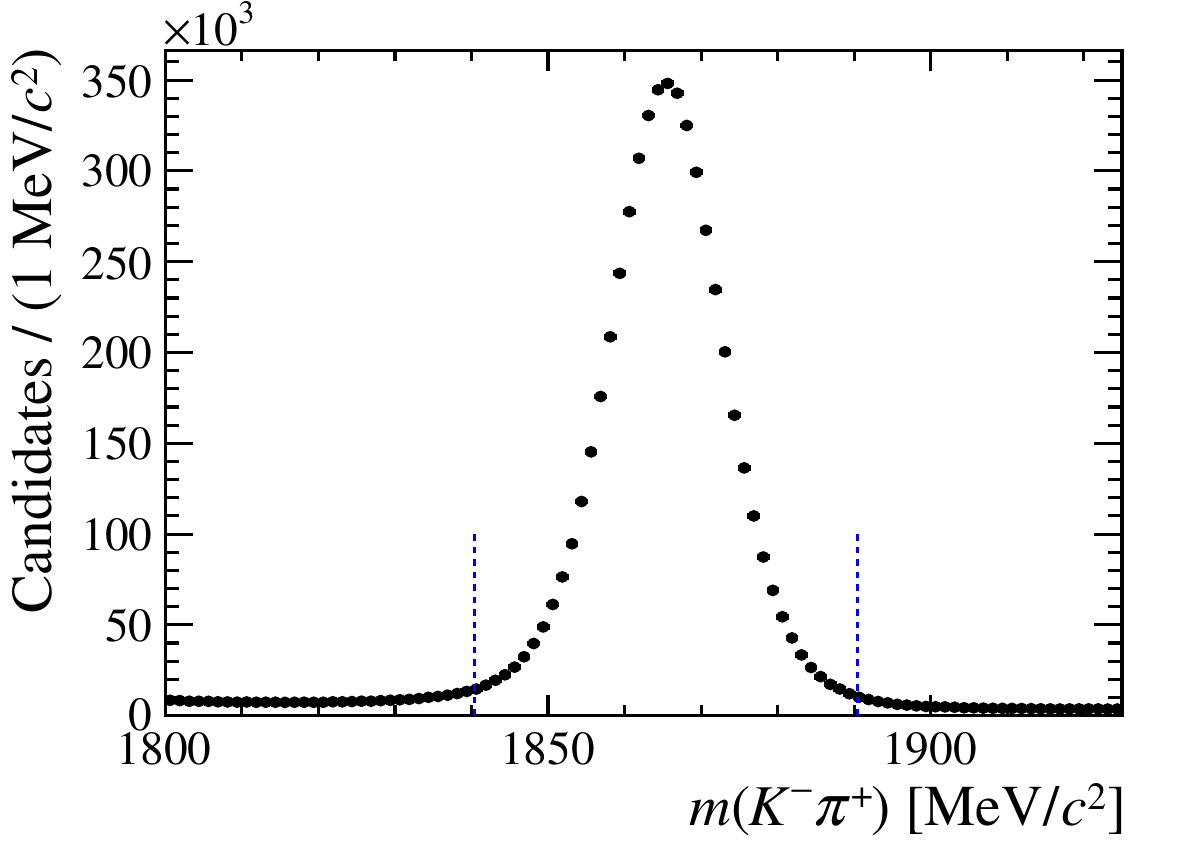}
    \put(70,50) {\begin{tabular}{@{}l@{}}\lhcb \\ 5.7\invfb\end{tabular}}
    \put(30,50){(b)}
    \end{overpic}
    \begin{overpic}[width=0.48\linewidth]{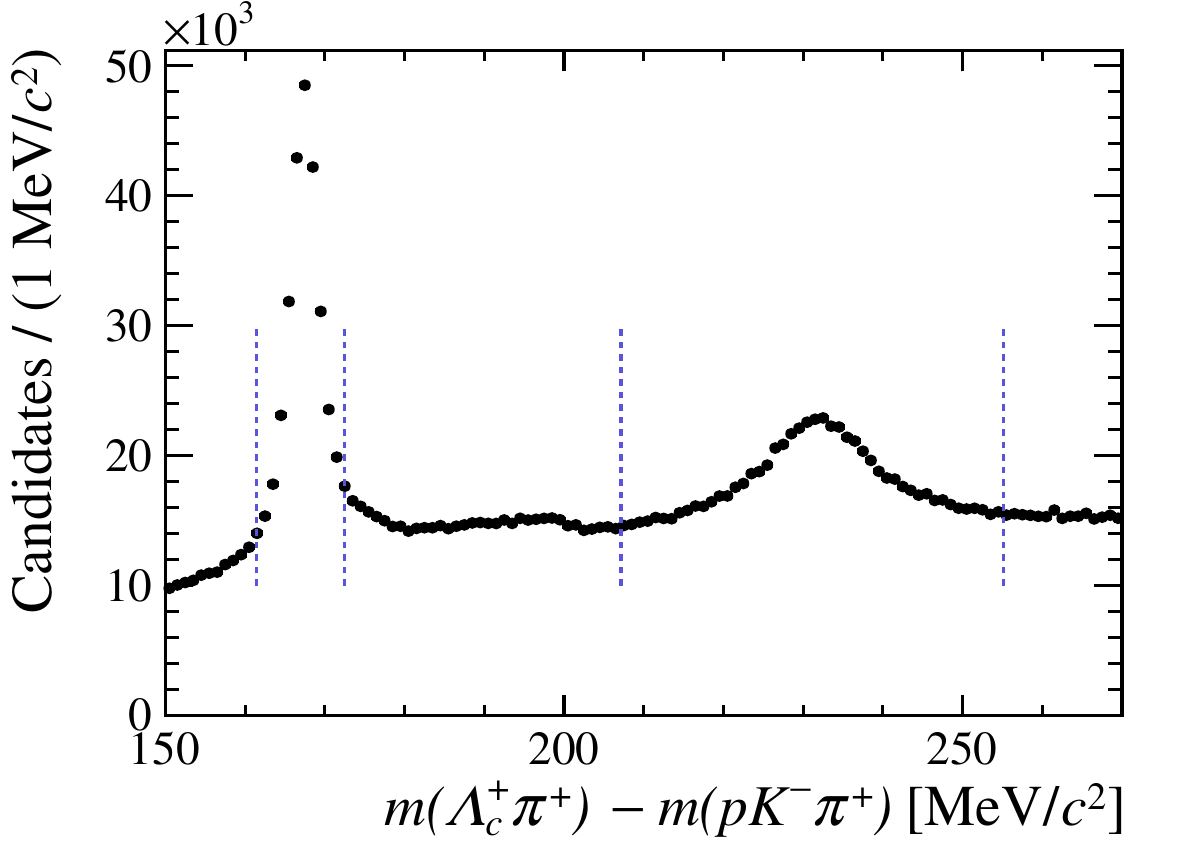}
    \put(70,50) {\begin{tabular}{@{}l@{}}\lhcb \\ 5.7\invfb\end{tabular}}
    \put(30,50){(c)}
    \end{overpic}
    \begin{overpic}[width=0.48\linewidth]{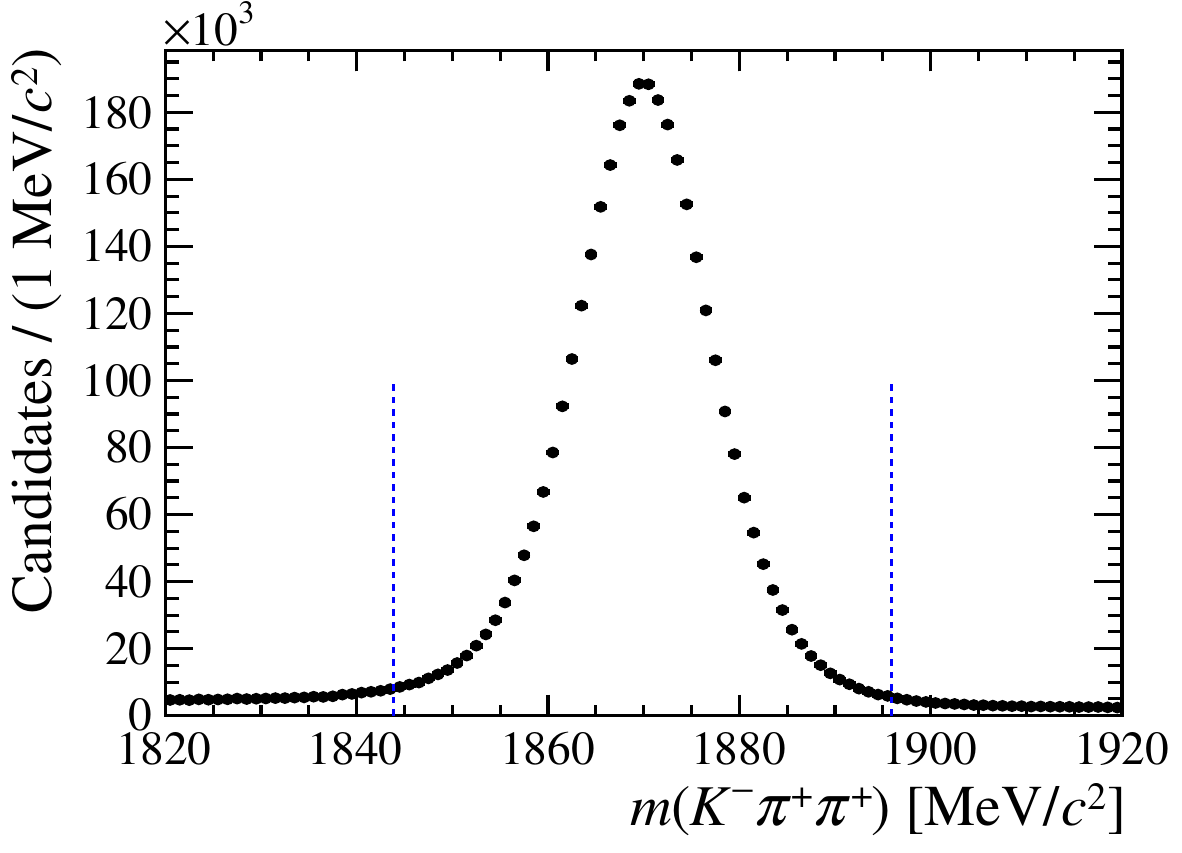}
    \put(70,50) {\begin{tabular}{@{}l@{}}\lhcb \\ 5.7\invfb\end{tabular}}
    \put(30,50){(d)}
    \end{overpic}
    \begin{overpic}[width=0.48\linewidth]{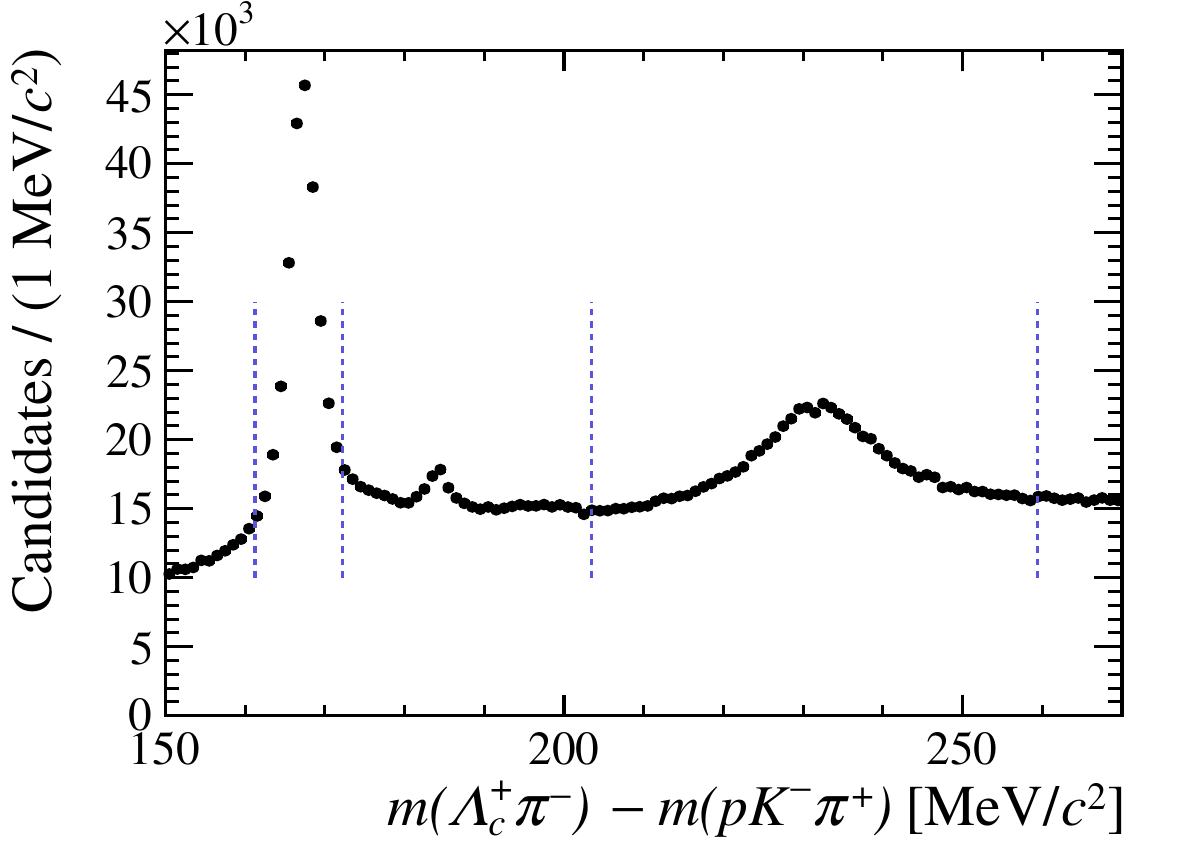}
    \put(70,50) {\begin{tabular}{@{}l@{}}\lhcb \\ 5.7\invfb\end{tabular}}
    \put(30,50){(e)}
    \end{overpic}
    \begin{overpic}[width=0.48\linewidth]{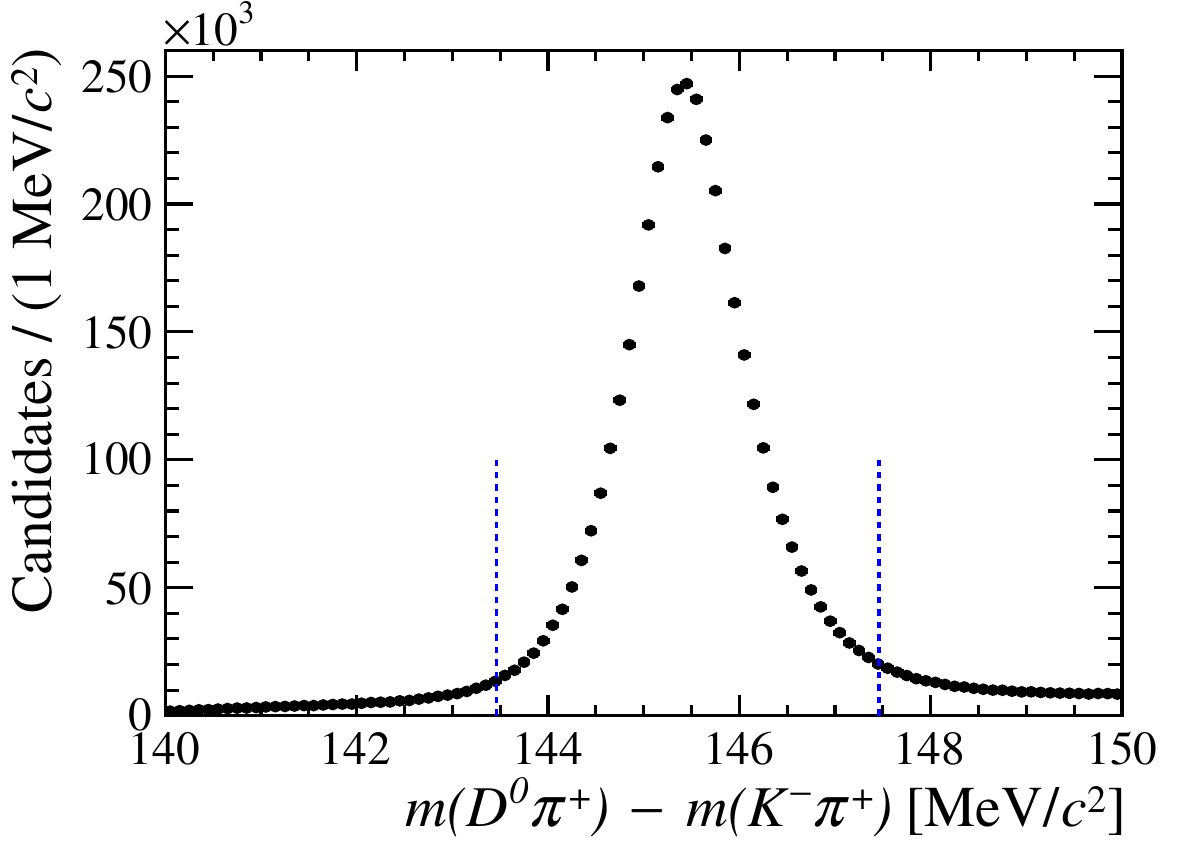}
    \put(70,50) {\begin{tabular}{@{}l@{}}\lhcb \\ 5.7\invfb\end{tabular}}
    \put(30,50){(f)}
    \end{overpic}
    
    \caption{Invariant mass distributions of the (a)  \decay{\Lc}{\proton\Km\pip}, (b) \decay{\Dz}{\Km\pip}, \mbox{(c) \decay{\Sigmacbstbpp}{\Lc\pip}}, (d) \decay{\Dp}{\Km\pip\pip}, (e) \decay{\Sigmacbstbz}{\Lc\pim}  and (f) \decay{\Dstarp}{\Dz\pip} decays. Note that in (c), (e) and (f) the mass of the charm hadron is subtracted from the mass distribution to minimise detector resolution effects. In (e) a contribution can also be seen at around $185\mevcc$ from the fully reconstructed \decay{\Xicz}{\Lc\pim} decay. The blue dashed lines show the chosen signal windows around the peaks.}
    \label{fig:massplots}
\end{figure}

All signal mode combinations are listed in Table~\ref{tab:PentaquarkMultiplet}. It is interesting to note that the quark content of each of these states is comparable to the quark content of the \jpsi\proton combination (\cquark\cquarkbar\uquark\uquark\dquark), since all combinations consist of only the up, down and charm quarks. 
\begin{table}[tb]
\caption{All possible combinations of \Sigmac or \Lc baryons with $\D^{(*)}$ mesons to produce the isospin multiplet. Combinations of \Lc baryons, pions and \D mesons are also considered. From each \Sigmac\D combination the corresponding \Lc\pion\D combination can be derived. The charge of the corresponding pentaquark state is given, along with the isospin, hypercharge and charm quantum numbers. The last column indicates whether a mode has an upper limit set in this paper. The combinations are split by the charm value.}
    \centering
    \resizebox{0.65\columnwidth}{!}{
    \begin{tabular}{llccccc}
    \multirow{2}{*}{Hadron 1} & \multirow{2}{*}{Hadron 2} & \multirow{2}{*}{Charge} & \multirow{2}{*}{\Isospin} & \multirow{2}{*}{\Hypercharge} & \multirow{2}{*}{C} & Limit \\
                             & & & & & & Set \\
    \hline
        \Lc  & \Dzb & $+1$ & $\sfrac{1}{2}$          & 1 & 0 & \checkmark \rule{0pt}{2.6ex}\\
    \Lc & \Dm                          & $\phantom{+}0$ & $\sfrac{-1}{2}\phz\phz$ & 1 & 0 & \checkmark \\
    \Lc & \Dstarm                      & $\phantom{+}0$ & $\sfrac{-1}{2}\phz\phz$ & 1 & 0 & \checkmark \\
    \Sigmacpp  &  \Dzb  & +2 & $\sfrac{3}{2}$ & 1 & 0 & \checkmark\\
    \Sigmacpp  & \Dm  & +1 &  $\sfrac{1}{2}$ & 1 & 0 & \checkmark \\
    \Sigmacpp   &  \Dstarm  & $+1$ & $\sfrac{1}{2}$          & 1 & 0 & \texttimes\\
    \Sigmacz   & \Dzb & \phantom{+}0 & $\sfrac{-1}{2}\phz\phz$ & 1 & 0 & \checkmark\\
    \Sigmacz   & \Dm & $-1$ & $\sfrac{-3}{2}\phz\phz$ & 1 & 0 & \checkmark \\
    \Sigmacz    &  \Dstarm  & $-1$ & $\sfrac{-3}{2}$\phz\phz & 1 & 0 & \texttimes\\
     \Sigmacstpp &  \Dzb & $+2$ & $\sfrac{3}{2}$ & 1 & 0 & \checkmark\\
    \Sigmacstpp & \Dm & $+1$ & $\sfrac{1}{2}$ & 1 & 0 & \checkmark \\
    \Sigmacstpp &  \Dstarm  & $+1$ & $\sfrac{1}{2}$ & 1 & 0 & \checkmark\\
    \Sigmacstz &  \Dzb & $\phantom{+}0$ & $\sfrac{-1}{2}\phz\phz$ & 1 & 0 & \checkmark\\
    \Sigmacstz & \Dm & $-1$ & $\sfrac{-3}{2}\phz\phz$ & 1 & 0 & \checkmark \\
    \Sigmacstz  &  \Dstarm & $-1$ & $\sfrac{-3}{2}\phz\phz$ & 1 & 0 & \checkmark \\
    \hline
    \Lc & \Dz                          & $+1$ & $\sfrac{-1}{2}\phz\phz$ & 3 & 2 & \checkmark \rule{0pt}{2.6ex} \\
    \Lc & \Dp                          & $+2$ & $\sfrac{1}{2}$          & 3 & 2 & \checkmark \\
    \Lc & \Dstarp                      & $+2$ & $\sfrac{1}{2}$          & 3 & 2 & \checkmark \\
    \Sigmacpp   &  \Dz      & $+2$ & $\sfrac{1}{2}$          & 3 & 2 & \texttimes\\
    \Sigmacpp   &  \Dp      & $+3$ & $\sfrac{3}{2}$          & 3 & 2 & \texttimes\\
    \Sigmacpp   &  \Dstarp  & $+3$ & $\sfrac{3}{2}$          & 3 & 2 & \texttimes\\
    \Sigmacz    &  \Dz      & $\phantom{+}0$ & $\sfrac{-3}{2}\phz\phz$ & 3 & 2 & \texttimes\\
    \Sigmacz    &  \Dp      & $+1$ & $\sfrac{-1}{2}\phz\phz$ & 3 & 2 & \texttimes\\
    \Sigmacz    &  \Dstarp  & $+1$ & $\sfrac{-1}{2}\phz\phz$ & 3 & 2 & \texttimes\\
    \Sigmacstpp &  \Dz  & $+2$ & $\sfrac{1}{2}$ & 3 & 2 & \checkmark\\
    \Sigmacstpp & \Dp & $+3$ & $\sfrac{3}{2}$ & 3 & 2 & \checkmark \\
    \Sigmacstpp &  \Dstarp  & $+3$ & $\sfrac{3}{2}$          & 3 & 2 & \texttimes\\
    \Sigmacstz &  \Dz  & $\phantom{+}0$ & $\sfrac{-3}{2}\phz\phz$ & 3 & 2 & \checkmark\\
    \Sigmacstz & \Dp & $+1$ & $\sfrac{-1}{2}\phz\phz$ & 3 & 2 & \checkmark \\
    \Sigmacstz  &  \Dstarp  & $+1$ & $\sfrac{-1}{2}\phz\phz$ & 3 & 2 & \texttimes\\
    \end{tabular}}
    \label{tab:PentaquarkMultiplet}
\end{table}
In the selection process, open-charm hadrons are selected by the trigger. The \Lc baryon is reconstructed in the \decay{\Lc}{\proton\Km\pip} decay mode for both signal and normalisation channels, the $\Dz$ meson in the decay \decay{\Dz}{\Km\pip} and the $\Dp$ meson in the decay \decay{\Dp}{\Km\pip\pip}. To improve the signal purity, stringent PID and vertex quality requirements are imposed. To further suppress the background in the $\Lc$ decay mode, the gradient-boosted decision tree classifier trained on data for the analysis described in Ref.~\cite{LHCb-PAPER-2020-004} is used. This identifies $\Lc$ candidates with high efficiency by combining information on the vertex quality and the PID. The classifier is applied directly in the case of the modes involving a direct $\Lc$ decay and is used as an input to the multivariate algorithm that is used to select \Sigmac candidates in the \decay{\Sigmac}{\Lc\pipm} decay mode. In Fig.~\ref{fig:massplots} the invariant mass distribution for each of these open-charm hadrons is shown.
Candidate $\Dstar$ ($\Sigmac$) hadrons are built by combining the $\Dz$ ($\Lc$) candidates with a well-identified charged pion with \pt above $200\mevc$. 

To improve the purity of the $\Sigmac$ selection, a multi-layer perceptron algorithm provided by the TMVA package~\cite{TMVA4} is trained using the \mbox{$\decay{\Sigmacpp}{\Lc\pip}$}
simulated signal sample and a data sample at high mass, well above the expected signal, in order to describe the shape of the combinatoric background events. As input, the network uses the $\pt$ and rapidity of the pion and of the \Sigmacpp and $\Lc$ baryons, together with information on the pion PID and the $\Lc$ classifier response described above.    

Once the corresponding baryon and meson combinations have been built, the signal regions are selected. These regions are based on fits to the baryon and meson spectra and require the mass is within a $3\,\sigma$ window around the mean mass value found from the fits, where $\sigma$ is the resolution of the signal peak. These regions are dependent on the baryon and meson combination the signal mode consists of. The chosen windows are highlighted in Fig.~\ref{fig:massplots}. The background regions are selected as a wider window around the fitted mean value to accurately describe the background as close to the signal region as possible. This corresponds to a $3-4.5\,\sigma$ window around the fitted mean value of the mass. Selection criteria are applied on the opening angle between each pair of charged particles to ensure there are no candidates where the same track is used multiple times. If two or more candidates are selected, all but one candidate is discarded at random. Any signal combinations with fewer than 30 candidates within the mass range of interest are not analysed further. This corresponds to 10 modes in total, as summarised in Table~\ref{tab:PentaquarkMultiplet}. Note that all \Lc\D\pion combinations exceed this threshold.

  \section{Limit setting procedure}
\begin{figure}[!tb]
\centering
\begin{overpic}[width=0.48\linewidth]{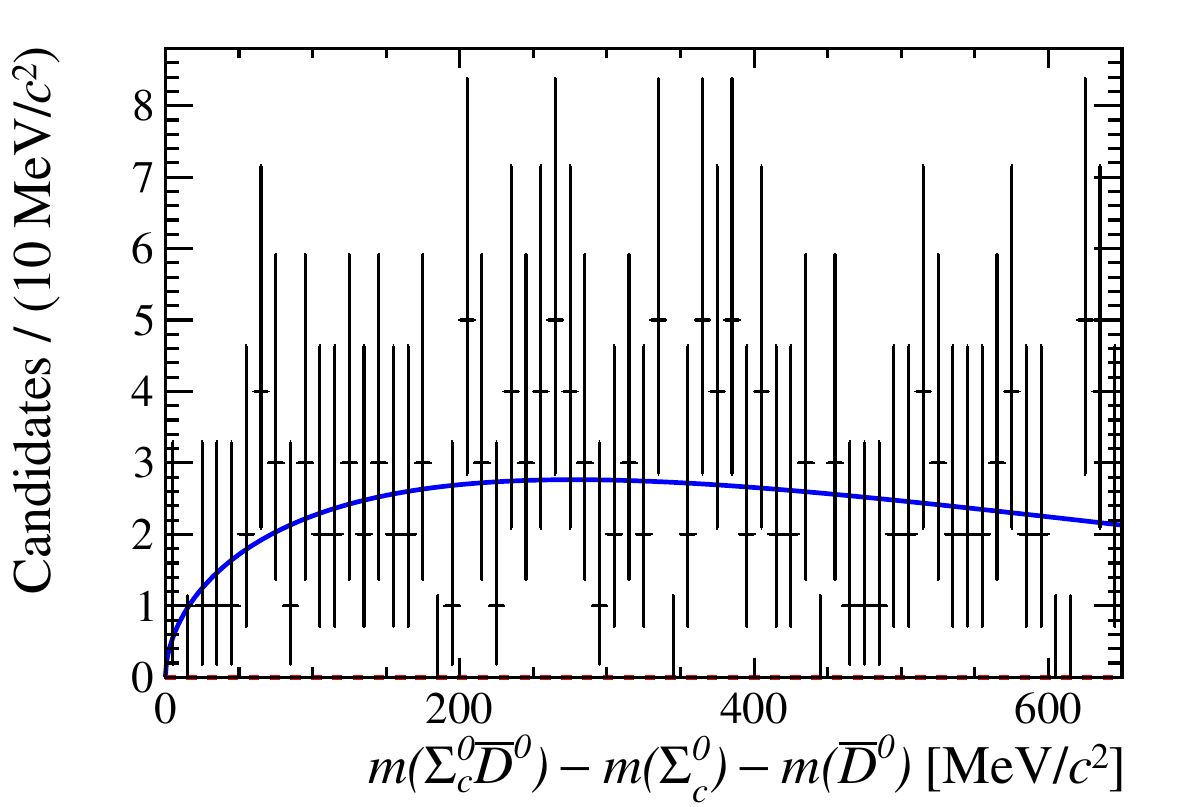}
\put(18, 55) {(a)}
\put(73,56) {\scriptsize \begin{tabular}{@{}l@{}}\lhcb \\ 5.7\invfb\end{tabular}}
\end{overpic}
\begin{overpic}[width=0.48\linewidth]{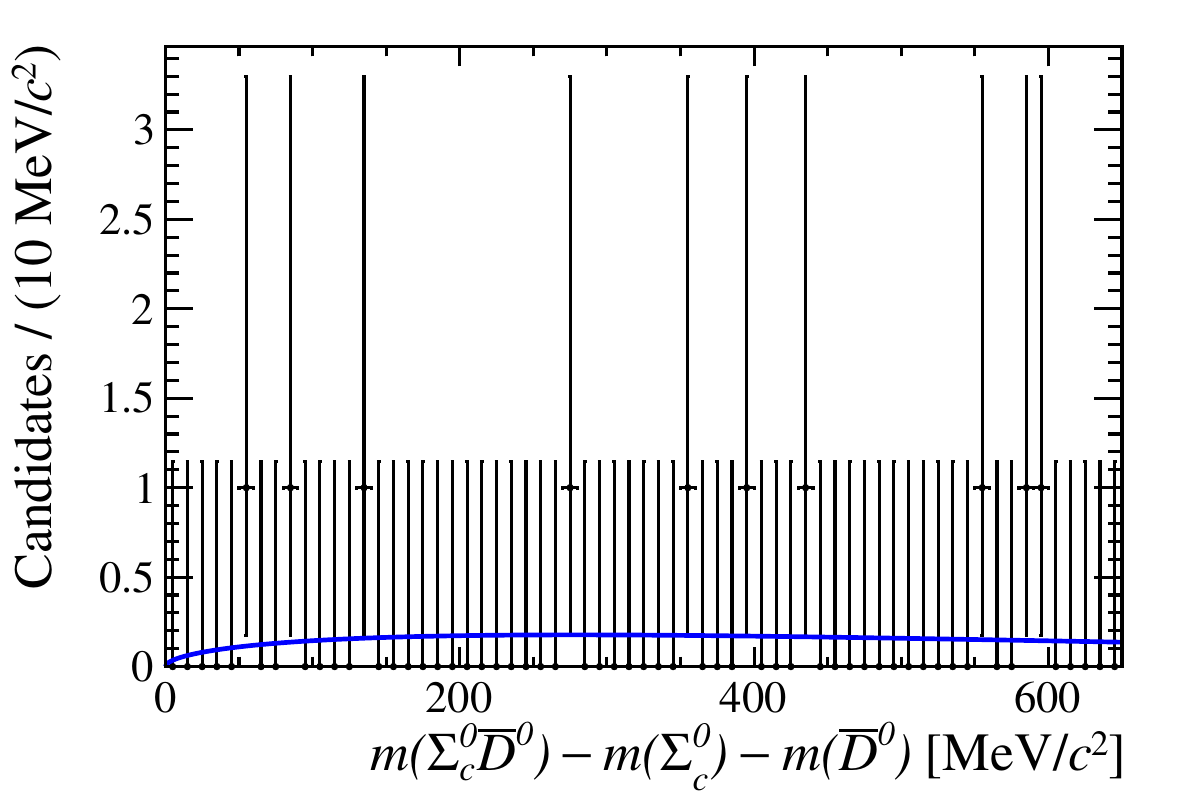}
\put(33, 55) {(b)}
\put(69,56) {\scriptsize \begin{tabular}{@{}l@{}}\lhcb \\ 5.7\invfb\end{tabular}}
\end{overpic}
\begin{overpic}[width=0.48\linewidth]{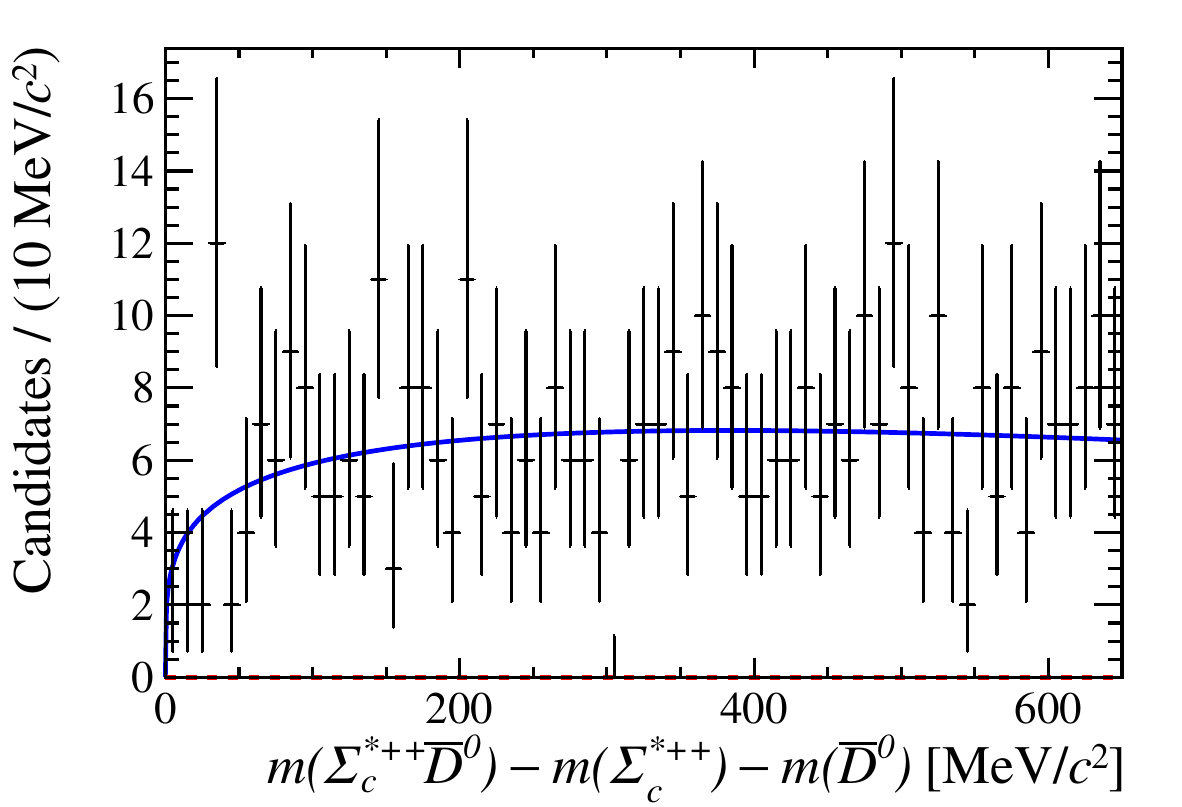}
\put(18, 55) {(c)}
\put(43,53) {\scriptsize \begin{tabular}{@{}l@{}}\lhcb \\ 5.7\invfb\end{tabular}}
\end{overpic}
\begin{overpic}[width=0.48\linewidth]{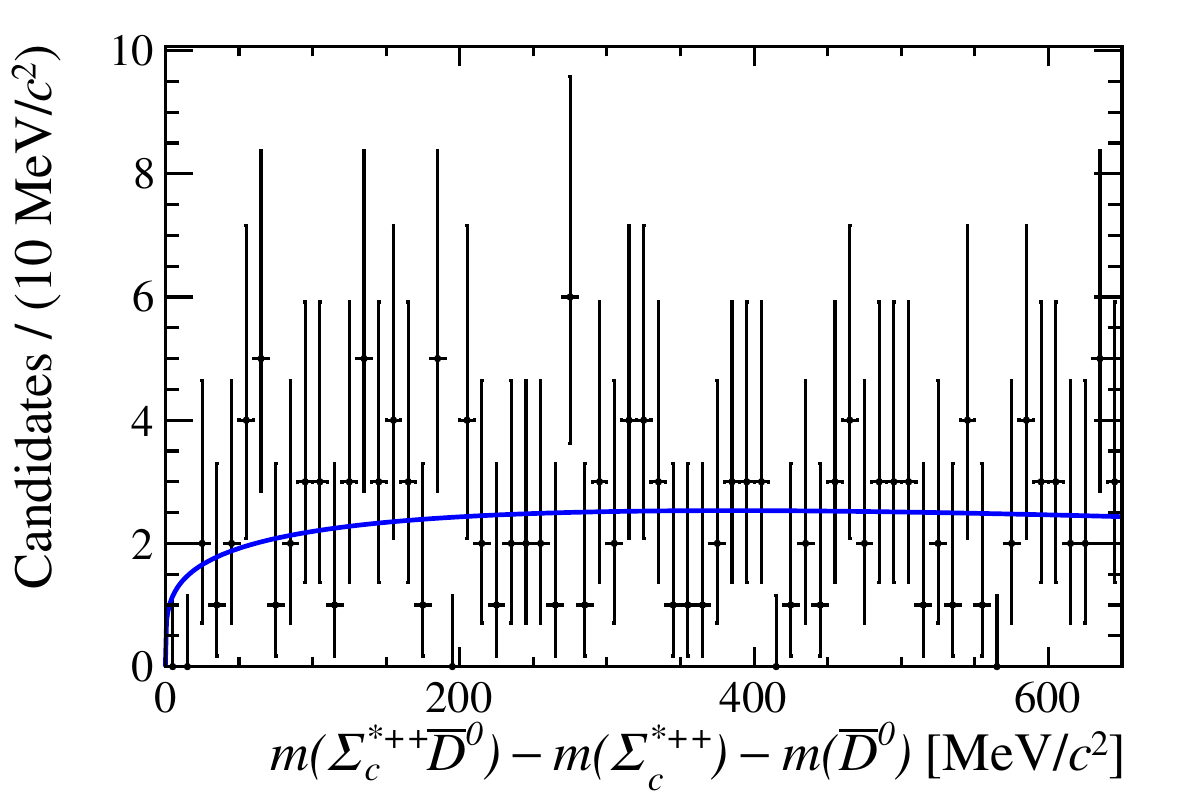}
\put(18, 55) {(d)}
\put(75,55) {\scriptsize \begin{tabular}{@{}l@{}}\lhcb \\ 5.7\invfb\end{tabular}}
\end{overpic}
\begin{overpic}[width=0.48\linewidth]{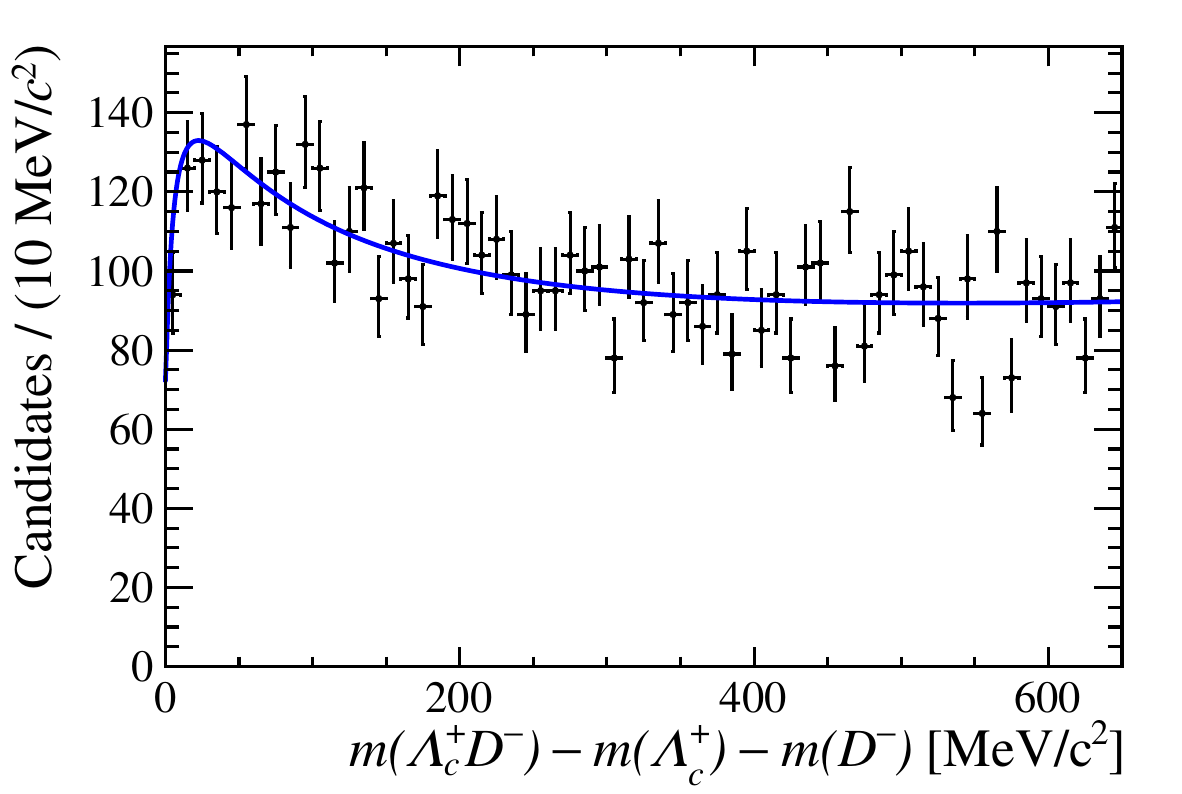}
\put(65, 53) {(e)}
\put(75,53) {\scriptsize \begin{tabular}{@{}l@{}}\lhcb \\ 5.7\invfb\end{tabular}}
\end{overpic}
\begin{overpic}[width=0.48\linewidth]{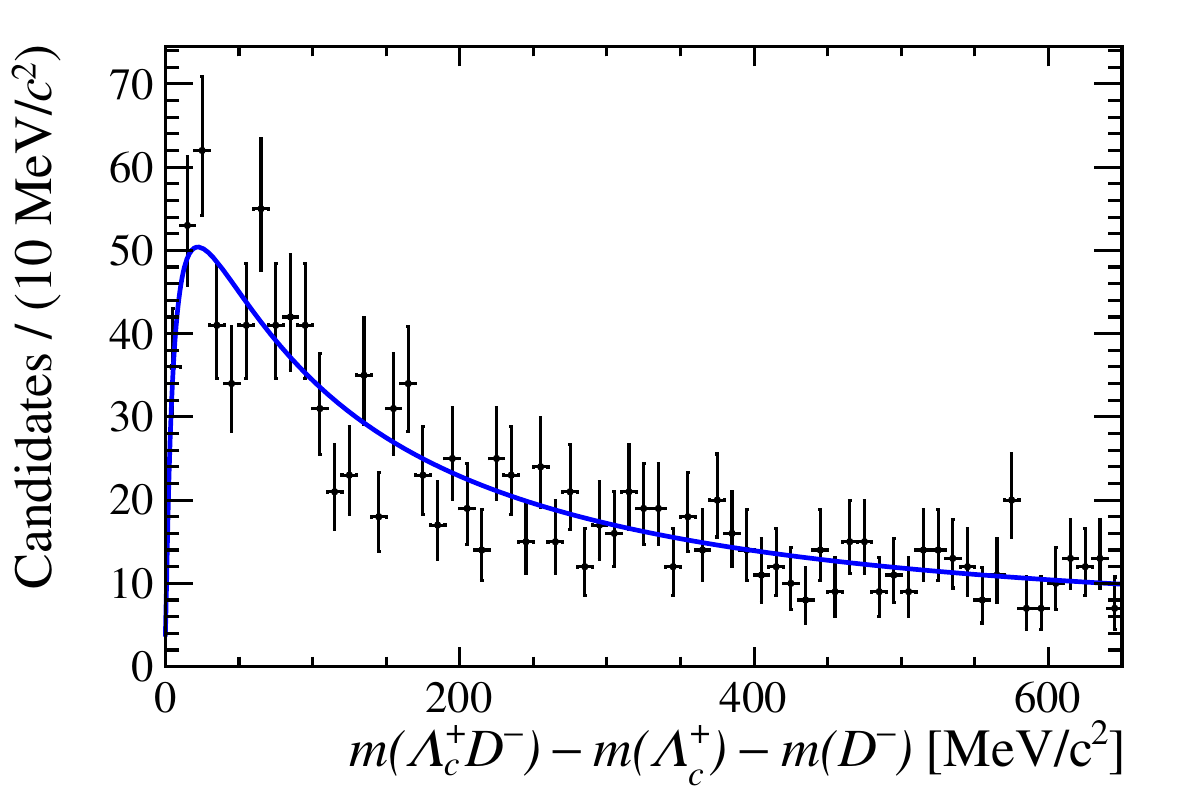}
\put(65, 53) {(f)}
\put(75,53) {\scriptsize \begin{tabular}{@{}l@{}}\lhcb \\ 5.7\invfb\end{tabular}}
\end{overpic}
\begin{overpic}[width=0.48\linewidth] {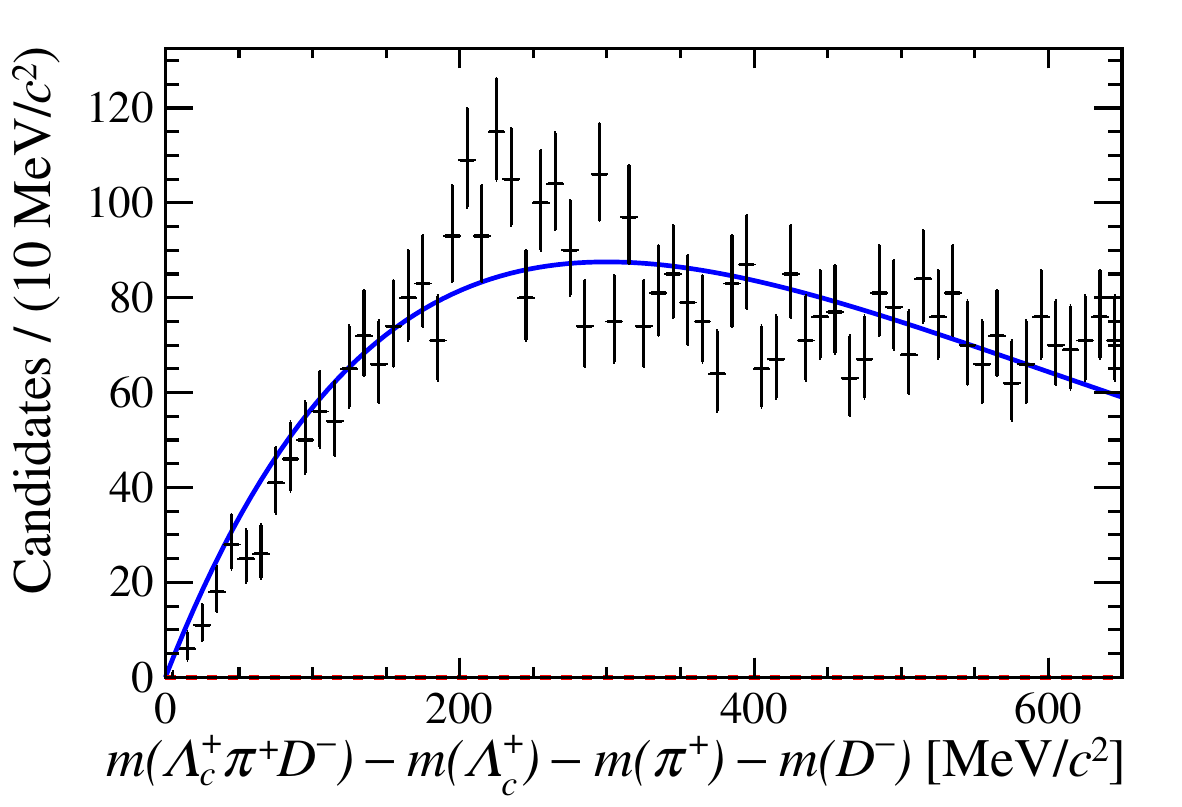}
\put(25, 53) {(g)}
\put(75,53) {\scriptsize \begin{tabular}{@{}l@{}}\lhcb \\ 5.7\invfb\end{tabular}}
\end{overpic}
\begin{overpic}[width=0.48\linewidth]{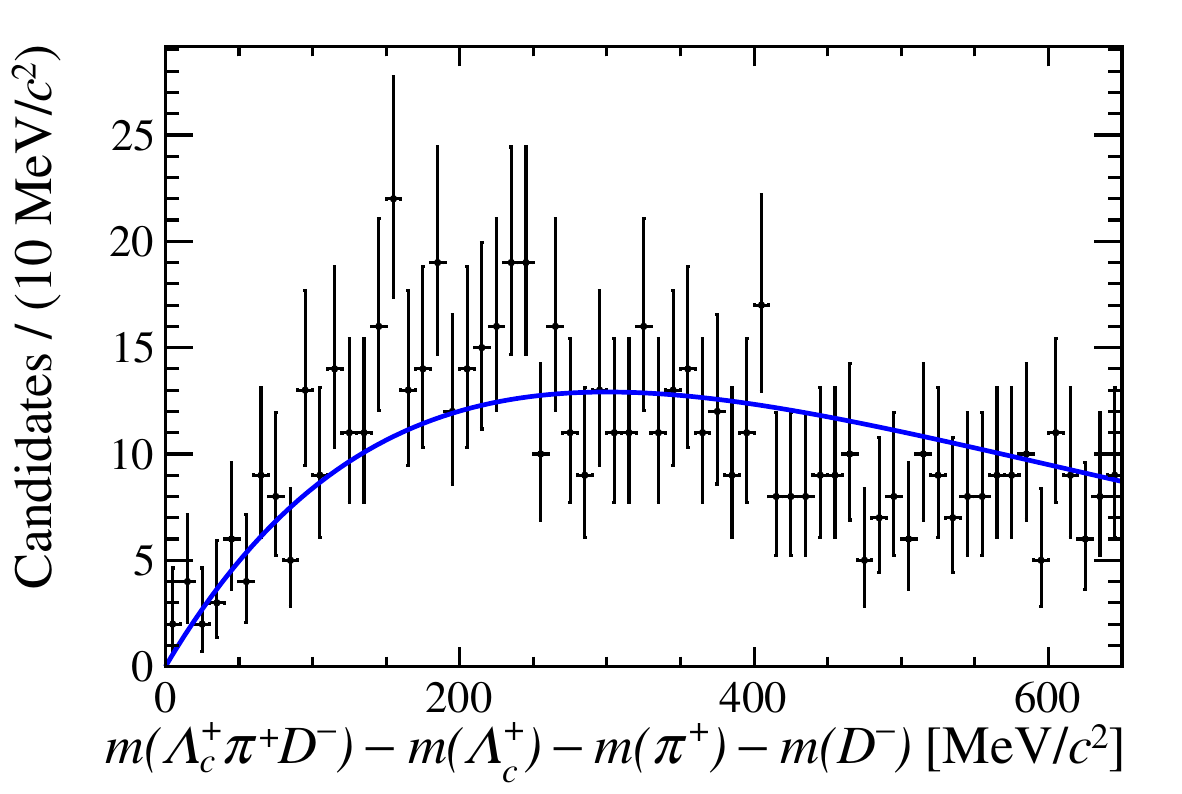}
\put(25, 53) {(h)}
\put(75,53) {\scriptsize \begin{tabular}{@{}l@{}}\lhcb \\ 5.7\invfb\end{tabular}}
\end{overpic}
\caption{\small Distributions of the \qval spectra for the \Sigmacz\Dzb (a) signal and (b) background regions, for the \Sigmacstpp\Dzb (c) signal and (d) background regions, the \Lc\Dm (e) signal and (f) background regions and the \Lc\pip\Dm (g) signal and (h) background regions. The fits for the background-only hypotheses are overlaid.}
\label{fig:bkgfits}
\end{figure}

For each mode, a kinematic fit is done to constrain the mass of intermediate charm hadrons to their known values, and to constrain them to originate at the same PV. The \qval spectrum, where the mass of each charmed hadron from the kinematic fit is subtracted from the decaying particle, is fitted using a simultaneous extended unbinned maximum-likelihood fit to the background and signal regions, where the background shape is shared between the two. The background normalisation region is a combination of the upper and lower sideband regions and for the \Sigmac\D and \Sigmacst\D modes is selected from the \Sigmac sideband region. For the \Lc\pion\D and \Lc\D modes the background region used is from the \Lc sideband region. For the \Sigmac\D, \Sigmacst\D and \Lc\pion\D modes, the background model used is a threshold function with all parameters shared between signal and sideband regions, while for the \Lc\D modes, the background model is the sum of a first-order Chebyshev polynomial and a log-normal distribution with all parameters shared between signal and sideband regions apart from the fraction between the two functions, which is independent. An example of the fit in the background-only hypothesis for each signal mode category (\Sigmac\D, \Sigmacst\D, \Lc\D and \Lc\pion\D) and the corresponding background is shown in Fig.~\ref{fig:bkgfits}. Four different signal models are investigated: one using a Gaussian function where the resolution is fixed to the detector resolution found using simulation, and three using Voigtian functions, built from the convolution of the same Gaussian function with a Breit--Wigner distribution, with fixed widths of 5, 10 and $15\mevcc$, in order to provide greater sensitivity to pentaquark states with broader width. Larger widths are not considered, since the pentaquark states are predicted to be narrow~\cite{Karliner:2015ina}.

\begin{figure}[!tb]
    \centering
    \begin{overpic}[width=0.68\linewidth]{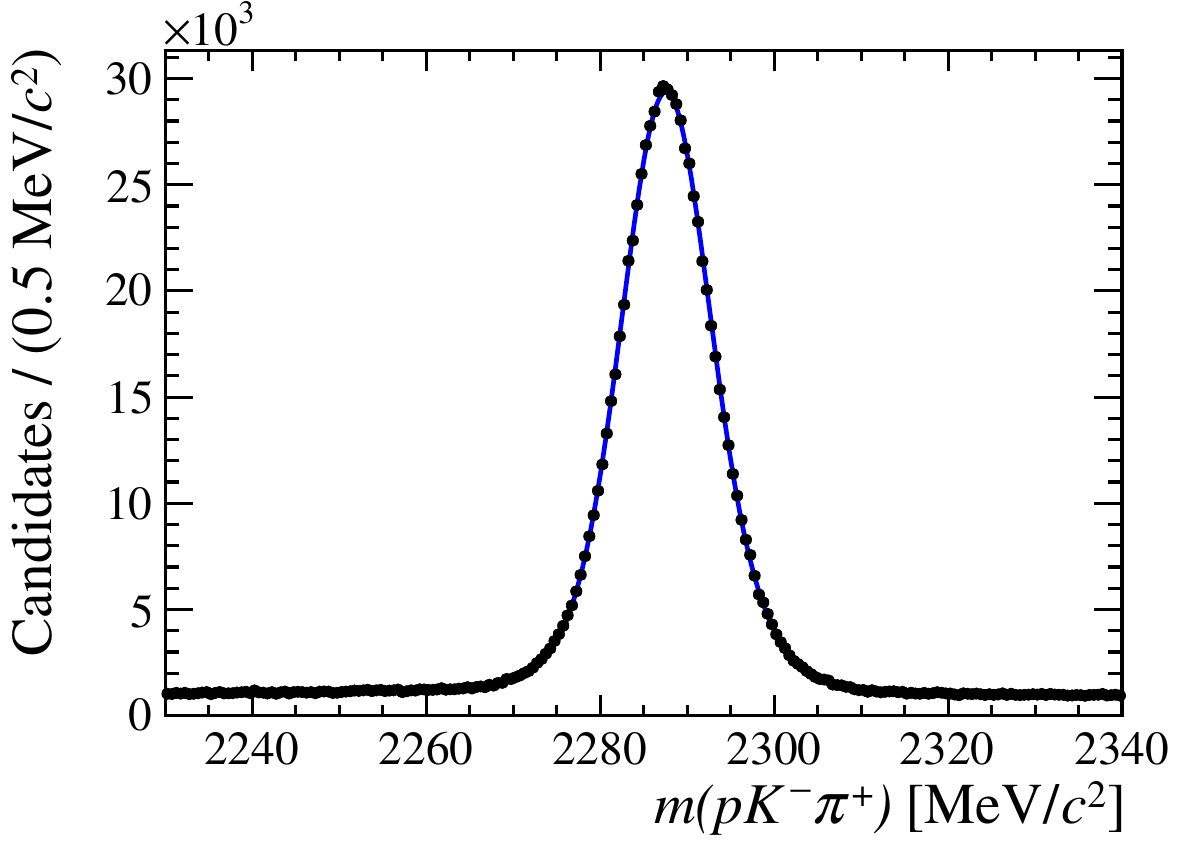}
    \put(70,55) {\begin{tabular}{@{}l@{}}\lhcb \\ 5.7\invfb\end{tabular}}
    \end{overpic}
    \caption{Invariant mass distribution of the \decay{\Lc}{\proton\Km\pip} decay. The fit results are overlaid.}
    \label{fig:Lcfit}
\end{figure}

The invariant mass distribution of \decay{\Lc}{\proton\Km\pip} candidates (the normalisation channel) is shown in Fig.~\ref{fig:Lcfit}. The distribution is fitted using the sum of a Gaussian function with a Crystal Ball function~\cite{Skwarnicki:1986xj} for the signal model, and a first-order Chebyshev polynomial for the background. The obtained signal yield is $789\,200\pm1\,300$.

\begin{figure}[tb]
    \centering
    \begin{overpic}[width=0.48\linewidth] {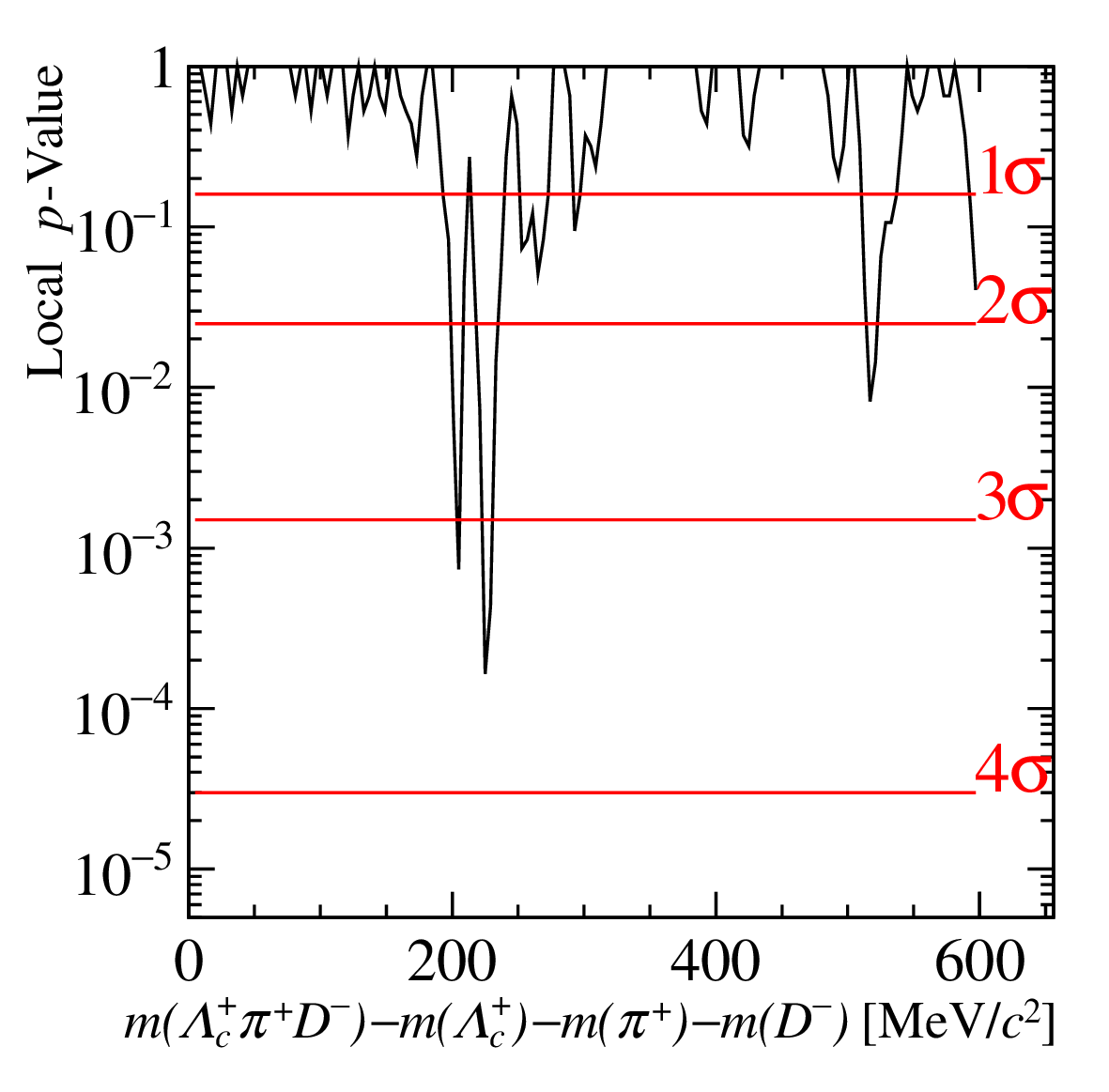}
    \put(25, 35) {(a)}
\put(75,35) {\begin{tabular}{@{}l@{}}\lhcb \\ 5.7\invfb\end{tabular}}
    \end{overpic}
    \begin{overpic}[width=0.48\linewidth]{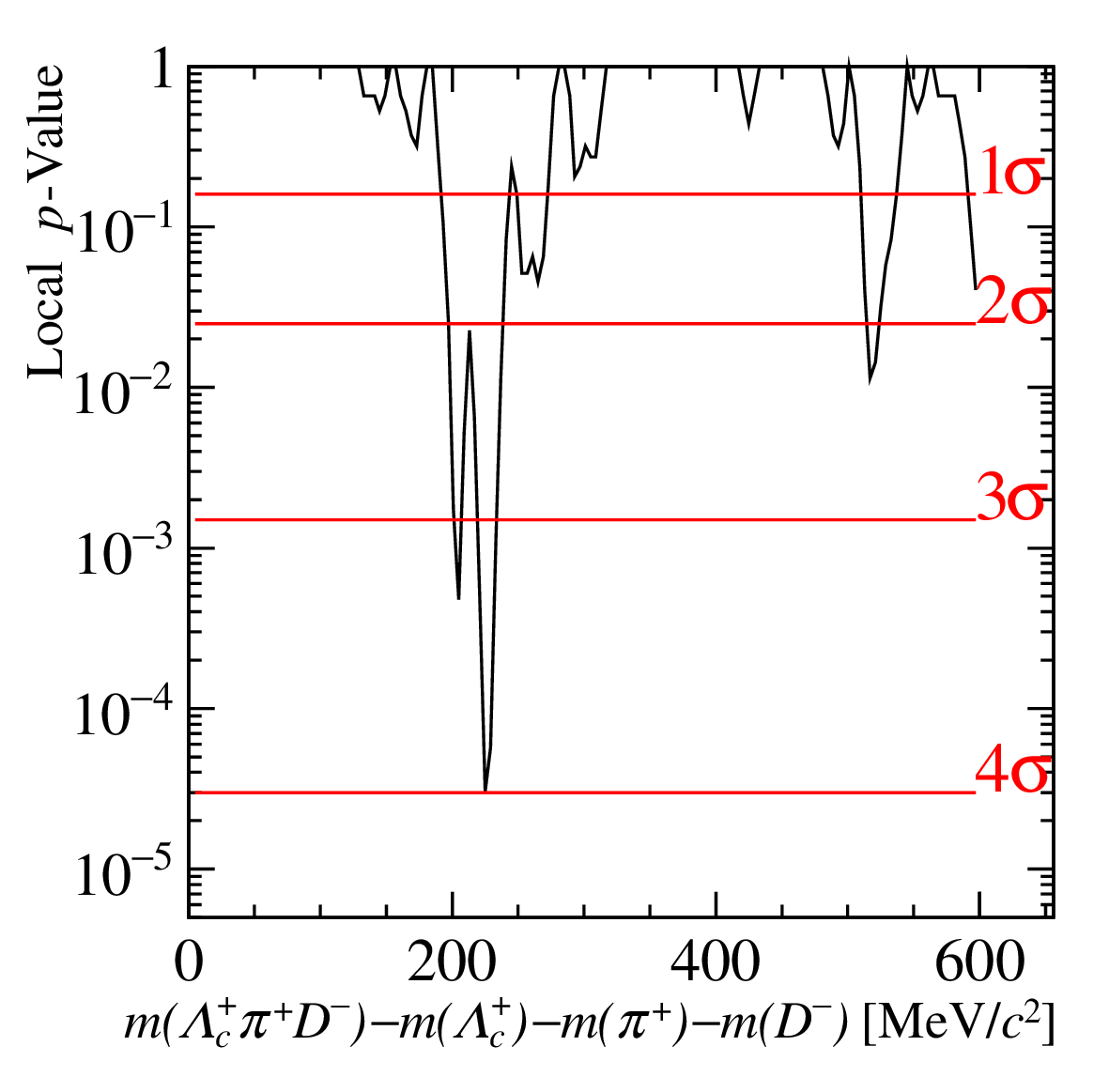}
    \put(25, 35) {(b)}
\put(75,35) {\begin{tabular}{@{}l@{}}\lhcb \\ 5.7\invfb\end{tabular}}
    \end{overpic}
    \begin{overpic}[width=0.48\linewidth]{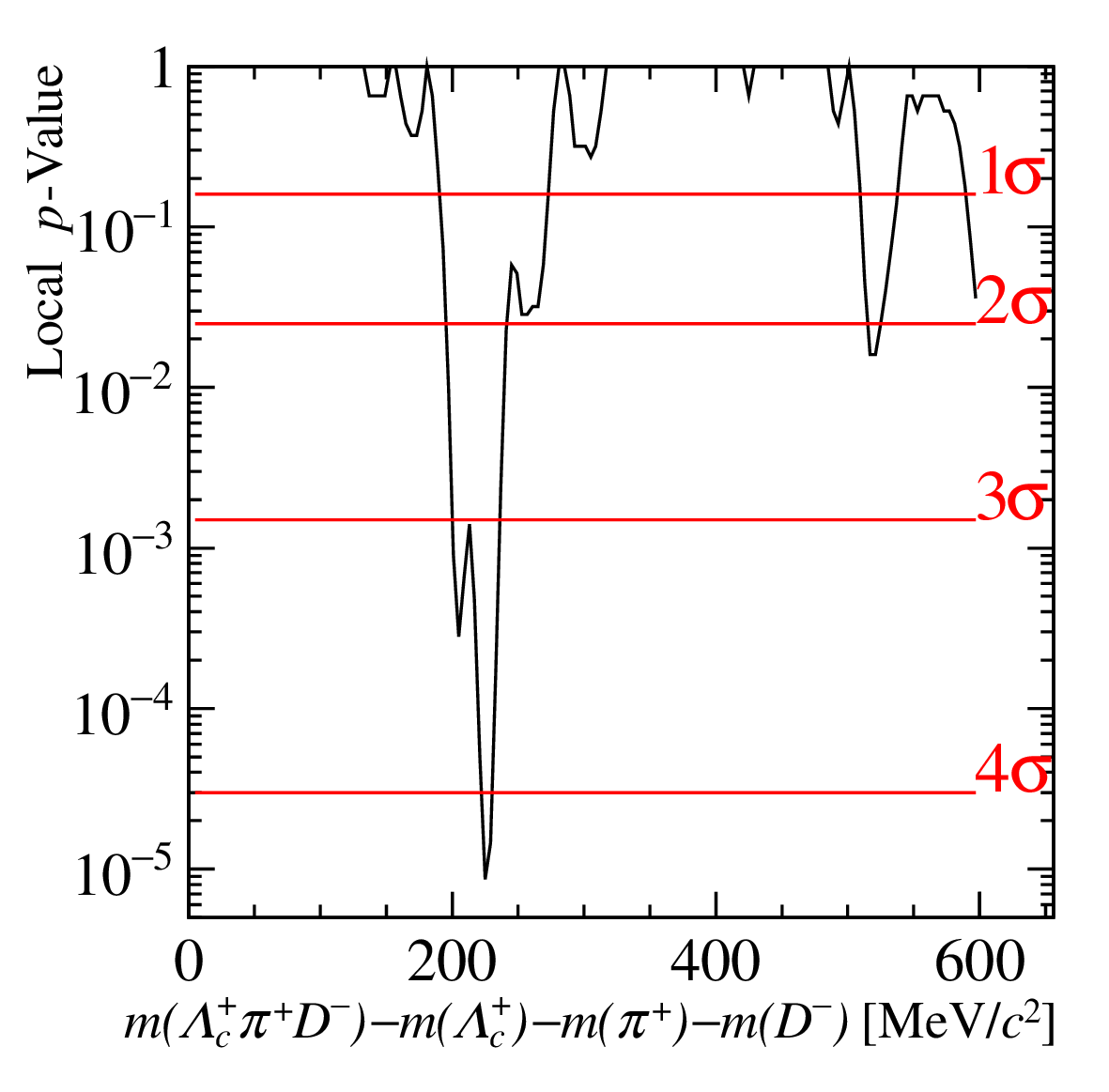}
    \put(25, 35) {(c)}
\put(75,35) {\begin{tabular}{@{}l@{}}\lhcb \\ 5.7\invfb\end{tabular}}
    \end{overpic}
    \begin{overpic}[width=0.48\linewidth]{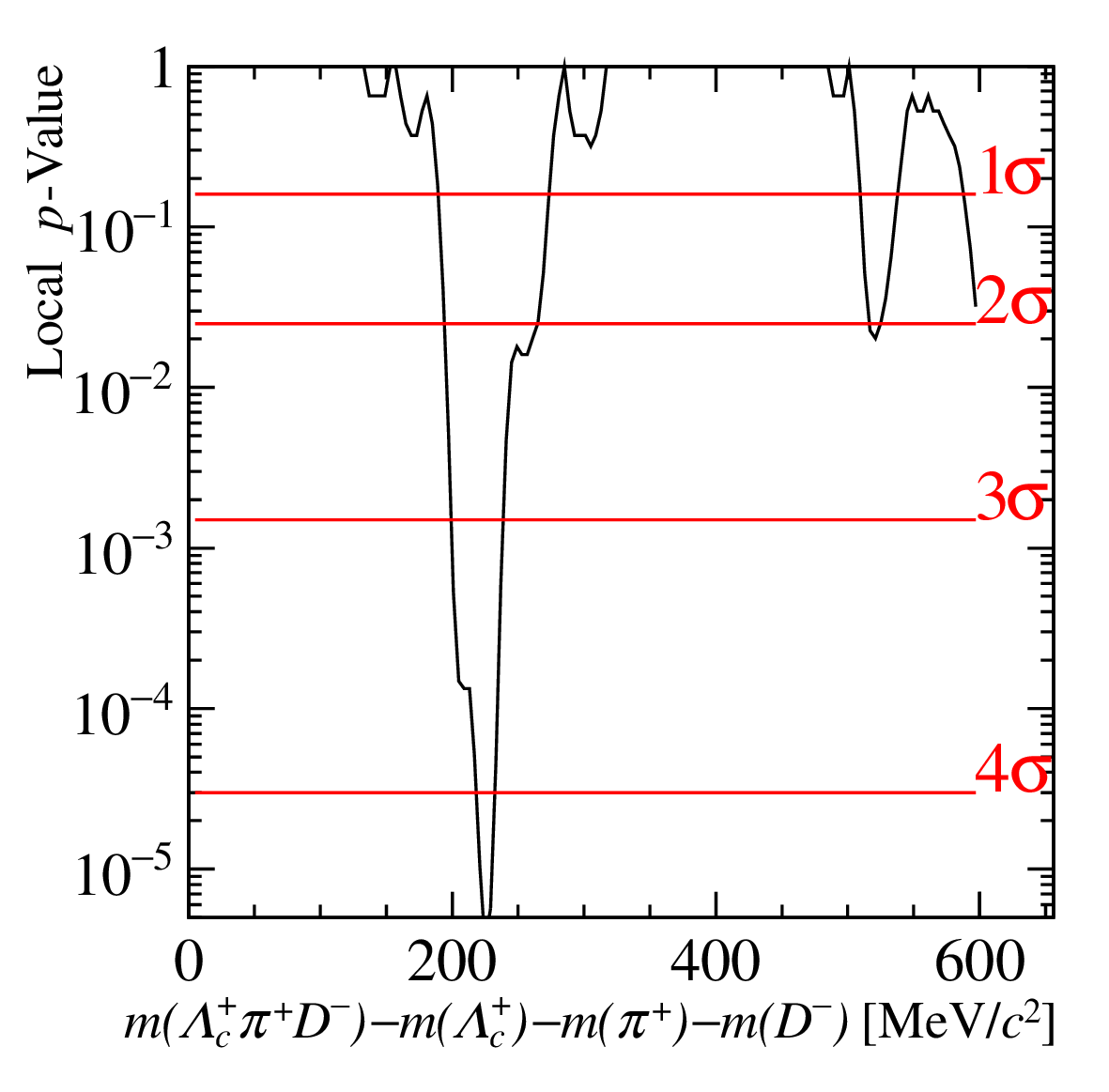}
    \put(25, 35) {(d)}
\put(75,35) {\begin{tabular}{@{}l@{}}\lhcb \\ 5.7\invfb\end{tabular}}
    \end{overpic}
    \caption{\small Local \pval distributions for the \Lc\pip\Dm mode with different signal models: (a) Gaussian function, (b) Voigtian function with 5\mevcc width, (c) Voigtian function with 10\mevcc width, and (d) Voigtian function with 15\mevcc width. The red lines correspond to the levels of local significance.}
    \label{fig:pvals}
\end{figure}

To search for possible pentaquark contributions, a scan of the \qval distribution in each signal mode is made from the kinematic threshold up to 600\mevcc in steps of $4\mevcc$, which corresponds roughly to the signal resolution. At each point, an extended unbinned maximum-likelihood fit is performed. The local \pval is determined from the difference between the negative log-likelihood of each fit and a fit with the background-only hypothesis and is assumed to be a one-tailed distribution. An example of the \pval distribution across the scan range is shown for each signal category for the \Lc\pim\Dzb mode in Fig.~\ref{fig:pvals}. The minimum local \pval in each channel varies from 0.041 in the \Sigmacstpp\Dstarm channel with the Voigtian signal model with a $15\mevcc$ width, to \num{3.36e-6} in the \Lc\pip\Dm channel with the Voigtian signal model with a $15\mevcc$ width. The latter corresponds to a local significance of $4.50\,\sigma$.

The local \pval is corrected in order to account for the look-elsewhere effect~\cite{Gross:2010qma} (LEE) using 
\begin{equation}
    p_{corr} = p_{loc} + \langle N(c_0)\rangle\exp\left({-\frac{c-c_0}{2}}\right),
\end{equation}
where $p_{corr}$ and $p_{loc}$ are the corrected and local \pval, respectively, $c$ is the profile likelihood ratio $2\Delta\ln \mathcal{L}$, and $c_0$ is a reference level set to 0.5 where the average number of `upcrossings' $\langle N(c_0)\rangle$ is found. An upcrossing is defined as when the profile likelihood ratio crosses the chosen threshold with a positive slope. The corrected \pval varies from 1 (corresponding to no observation of any significant signal) in many channels, to \num{1.45e-4} in the \Lc\pip\Dm channel with the Voigtian signal model where the width is set to 15\mevcc. To evaluate the interpretation of these \pvals the scanning procedure is repeated using 1000 background-only pseudo-experiments for each channel and the Voigtian signal model with a width of 15\mevcc. The average number of fluctuations above $3\,\sigma$ significance across all channels is 7.0 with a standard deviation of 5.0. In the data, five channels, namely the \Lc\pip\Dzb, \Lc\pip\Dm, \Lc\pim\Dzb, \Lc\pim\Dm and \Lc\pim\Dp channels, are observed with local significances greater than $3\,\sigma$, obtained with the Voigtian signal model with a width of 15\mevcc. Thus, we conclude that these significances are consistent with background fluctuations, as determined in the study with pseudo-experiments.

The previously observed pentaquark states are also investigated, namely the $P_c(4312)^+$, $P_c(4440)^+$ and $P_c(4457)^+$ states. This is only carried out for states with the same total charge and hidden charm quark content. By setting the mass and width in the Voigtian function to the known values of these states~\cite{LHCb-PAPER-2019-014} and carrying out the same fitting procedure previously described, the significance of the states is found, and the limits are set using the procedure described below. 

In this study, several sources of systematic uncertainties are considered. Since the number of tracks is different for signal and normalization, an important source of systematic uncertainty is due to the knowledge of the tracking efficiency, which is affected by hadronic interactions in the detector and overlaps between tracks or occupancy. Further uncertainties arise and are quantified from possible differences in the reconstruction between data and simulation, such as differences in selection and PID efficiencies. The performance of the classifier for the \Sigmac baryon, and the behaviour of the trigger in data and simulation are also considered as sources of uncertainty. The knowledge of the branching fractions of the decay modes used leads to a further uncertainty~\cite{PDG2022}. The effect of varying the background model on the signal yield is also investigated, \eg by using either a threshold function or Chebyshev polynomial summed with a log-normal distribution or by varying which background parameters are shared between the signal and background regions. This is done for the combination with the highest signal yield, and the uncertainty is applied for all signal combinations. The range of values found for each contribution is summarised in Table~\ref{tab:systematics}.

\begin{table}[tb]
    \centering
    \caption{\small Range of values of each systematic uncertainty contribution and the total combination for the different signal modes.}
    \begin{tabular}{lc}
      Source & Uncertainty (\%)         \\
    \hline
    Tracking                  & $3.00-6.20$  \\
    Reconstruction efficiency & $0.70-3.70$  \\
    Generation efficiency     & $0.10-0.20$  \\
    \Sigmac classifier performance           & $0.54$  \\
    L0 trigger efficiency     & $1.40-7.90$ \\
     Branching fraction       & $0.01-0.03$ \\
     Background model         & $11.6$ \\
    \hline
    Total                     & $12.10 - 15.79$  \\
    \end{tabular}
    \label{tab:systematics}
\end{table}

\begin{figure}[tb]
    \centering
    \begin{overpic}[width=0.48\linewidth]{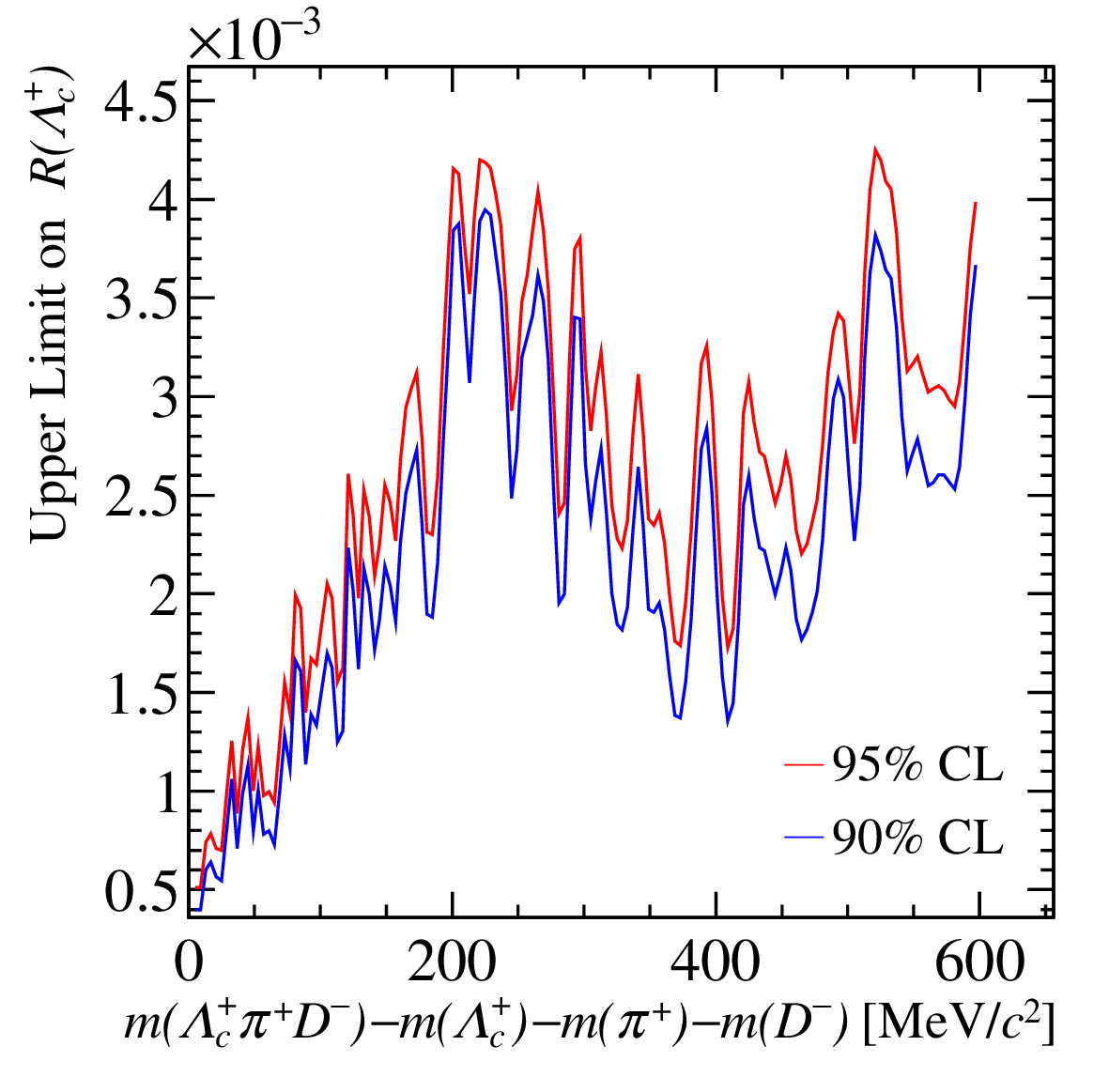}
    \put(27, 85) {(a)}
\put(55,80) {\begin{tabular}{@{}l@{}}\lhcb \\ 5.7\invfb\end{tabular}}
    \end{overpic}
    \begin{overpic}[width=0.48\linewidth]{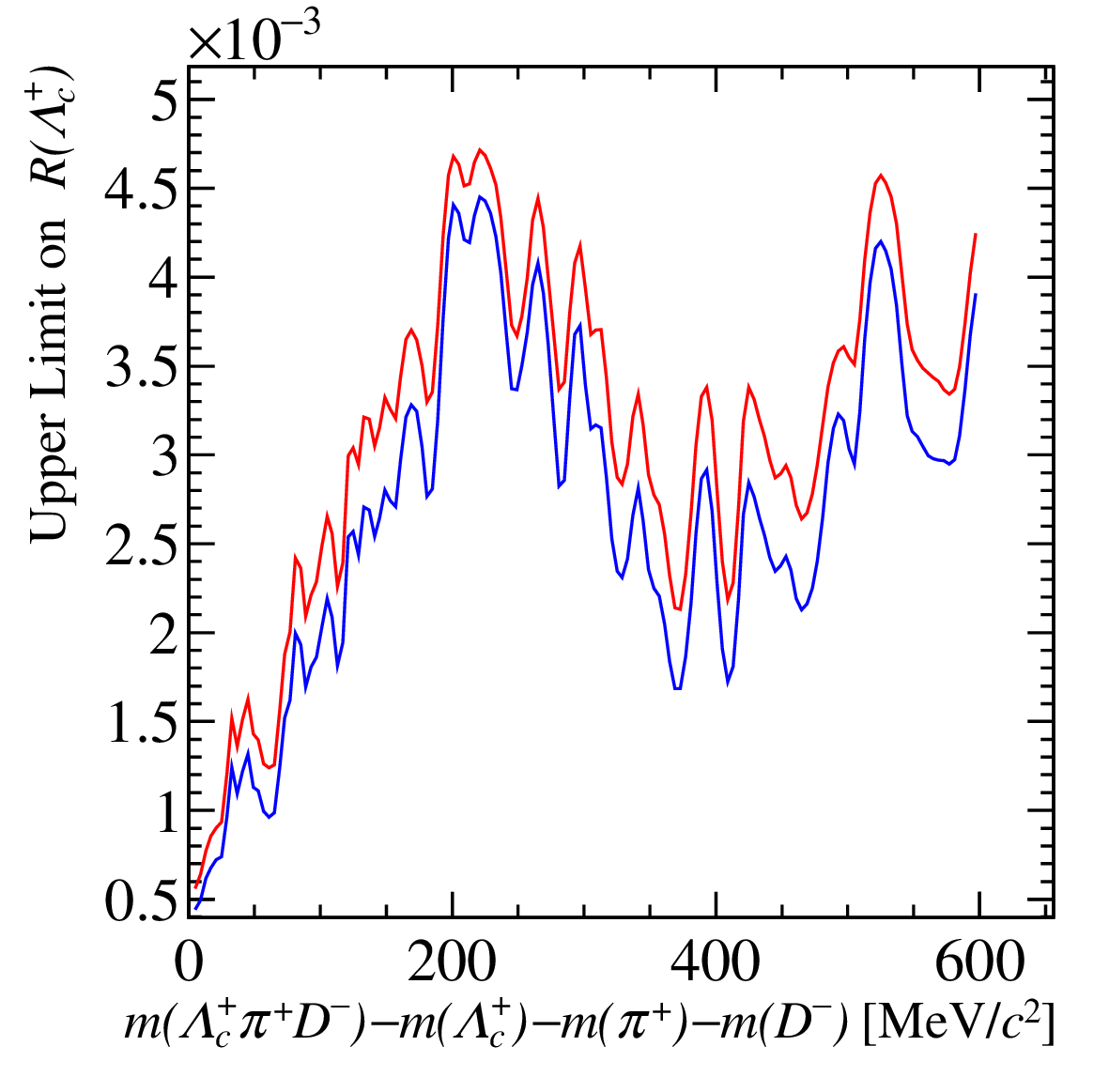}
    \put(27, 85) {(b)}
\put(75,30) {\begin{tabular}{@{}l@{}}\lhcb \\ 5.7\invfb\end{tabular}}
    \end{overpic}
    \begin{overpic}[width=0.48\linewidth]{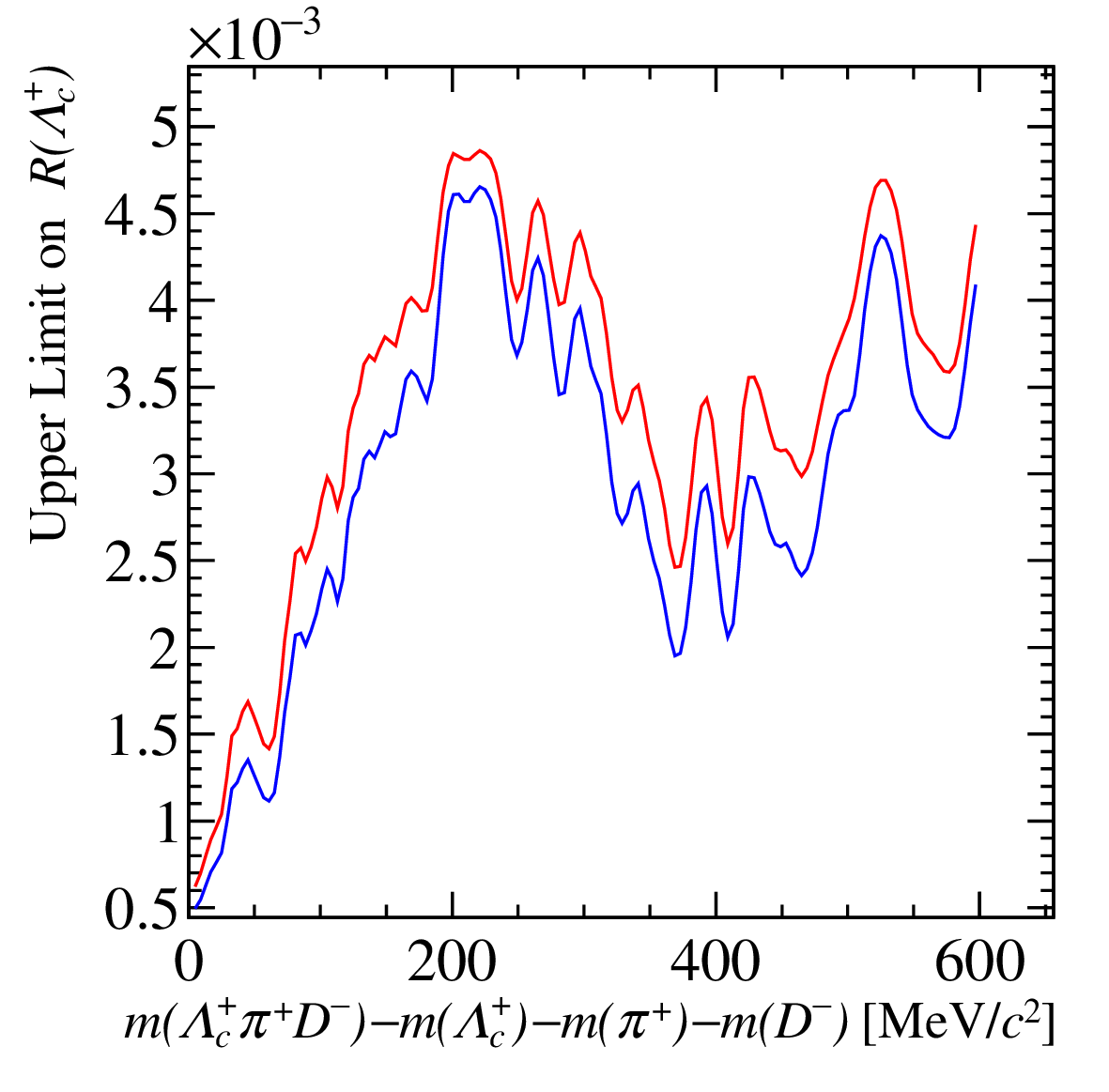}
    \put(27, 85) {(c)}
\put(75,30) {\begin{tabular}{@{}l@{}}\lhcb \\ 5.7\invfb\end{tabular}}
    \end{overpic}
    \begin{overpic}[width=0.48\linewidth]{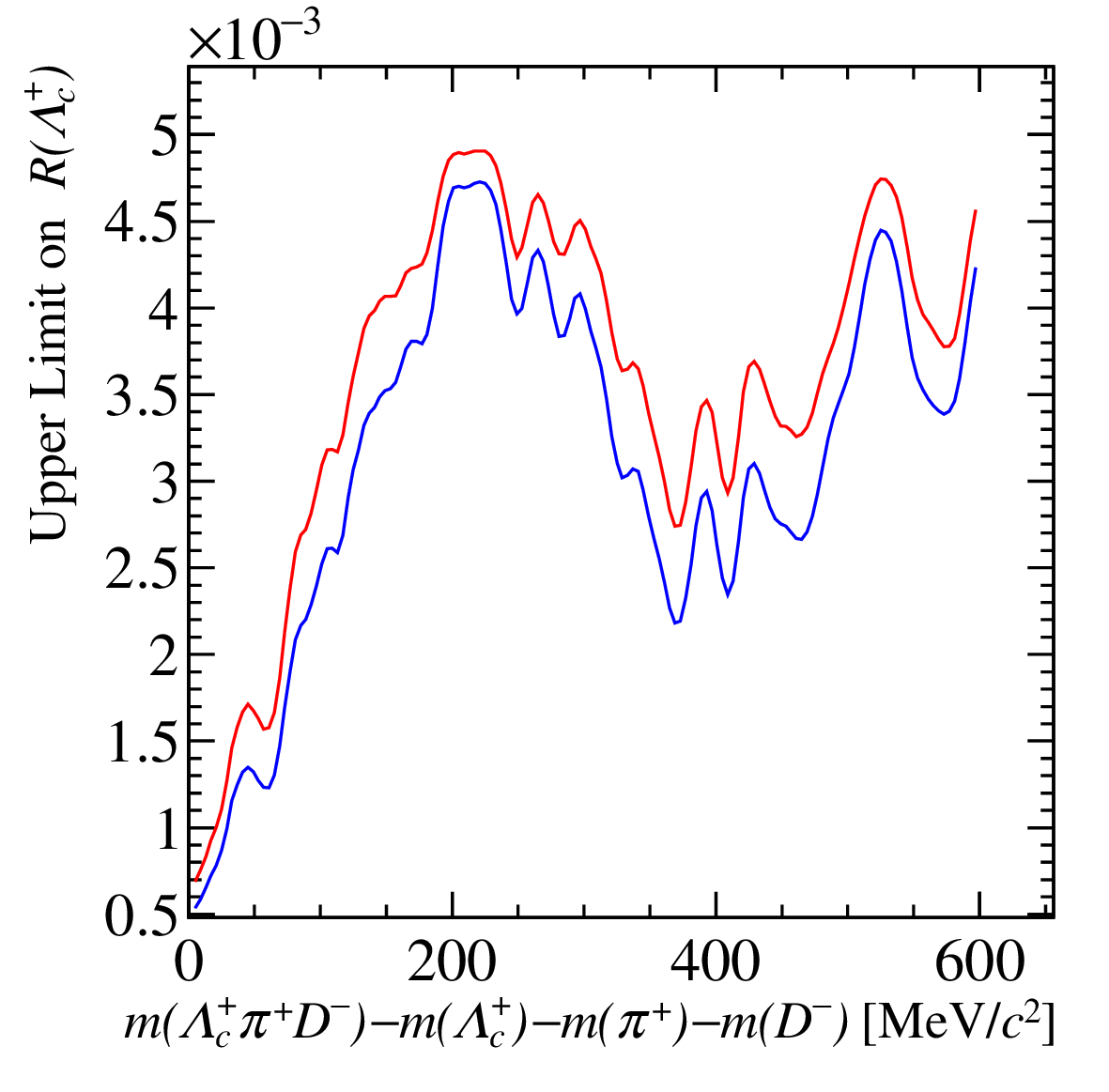}
    \put(27, 85) {(d)}
\put(75,30) {\begin{tabular}{@{}l@{}}\lhcb \\ 5.7\invfb\end{tabular}}
    \end{overpic}
    \caption{\small Upper limits on $R({\Lc})$ distribution, at 90\%  and 95\% CL, for the \Lc\pip\Dm mode with different signal models: (a) Gaussian function, (b) Voigtian function with $5\mevcc$ width, (c) Voigtian function with $10\mevcc$ width, and (d) Voigtian function with $15\mevcc$ width.}
    \label{fig:ULs}
\end{figure}

An upper limit (UL) is set on $R$(\Lc), which is defined as
\begin{equation}
    R(\Lc) = \frac{N_{P}}{N_{\Lc}}\times\frac{\epsilon_{\Lc}}{\epsilon_{P}},
    \label{eq:uleq}
\end{equation}
where $N$ represents the \Lc or pentaquark ($P$) yield and $\epsilon$ is the combined trigger and selection efficiency as found using simulated events. The limit is then determined at 90\% and 95\% confidence levels (CL). The likelihood profile is assumed to be parabolic, and is determined for five equal-sized steps in the signal yield around the value found from the fit. It is then convolved with a Gaussian function with a mean of zero and a width set to $\sigma = \sigma_{syst.}\cdot\mu$, where $\sigma_{syst.}$ is the systematic uncertainty on the signal yield and $\mu$ is the most probable signal yield. The convolved function takes the form
\begin{equation}
    L(N') = \int_{0}^{\infty}L(N)\frac{1}{\sqrt{2\pi}\sigma}\exp\left(\frac{-(N - N')^{2}}{2\sigma^{2}}\right)dN,
\end{equation}
where $L(N)$ is the normalised likelihood function. This likelihood profile is numerically integrated to find the value of the integral at 90\% or 95\% of the physical region, corresponding to the 90\% and 95\% CL upper limits on the signal yield and is then converted to an upper limit on $R(\Lc)$ using Eq.~\ref{eq:uleq}. An example of the upper limit for each signal category is shown for the \Lc\pim\Dzb mode in Fig.~\ref{fig:ULs}.

\begin{figure}[tb]
    \centering
    \begin{overpic}[width=0.68\linewidth]{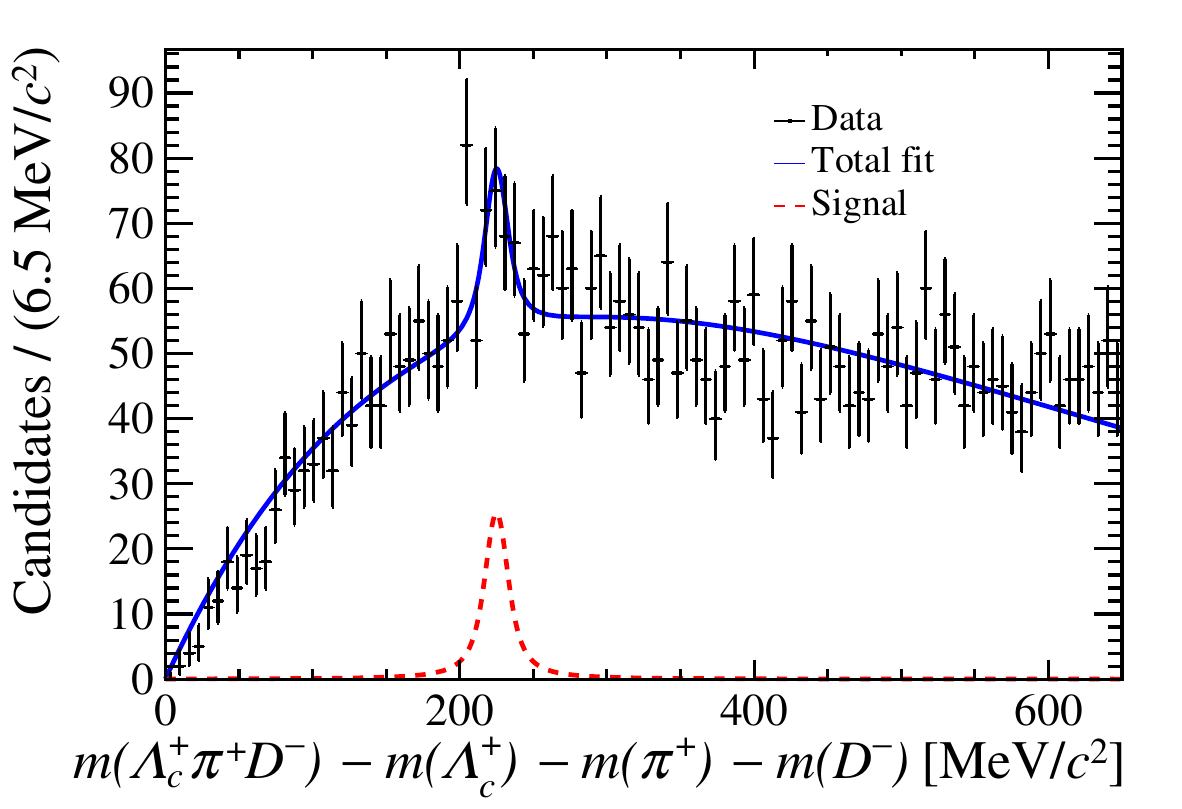}
    \put(18,53) {\begin{tabular}{@{}l@{}}\lhcb \\ 5.7\invfb\end{tabular}}
    \end{overpic}
    \caption{Distribution of the Q-value in the \Lc\pip\Dm channel, where the most significant signal is seen. The fit result is overlaid.}
    \label{fig:sigPlot}
\end{figure}

The results for the scan across the \qval spectrum in each signal combination are summarised in Appendix~\ref{sec:restofTabs}. All channels show a signal yield consistent with the background-only hypothesis. The most significant deviation is seen in the \Lc\pip\Dm channel. The fit result for this channel is shown in Fig.~\ref{fig:sigPlot}. When fitting with the mass and width of the known pentaquark states, the local significance in these spectra is found to be close to (or equal to) zero in all cases and is summarised in Table~\ref{tab:knownStates}.

\begin{table}[tb] 
\centering 
\caption{Upper limits on $R({\Lc})$ at $90\%$ and $95\%$ CL for the modes with the same total charge as the known pentaquark states, and with hidden-charm quark content. The local \pval and significance are listed as well as the signal yield, where the error on the signal yield is statistical only.}
\label{tab:knownStates} 
\sisetup{exponent-mode = scientific}
\resizebox{\columnwidth}{!}{ 
\begin{tabular}{lcccrrr}  
\multirow{2}{*}{Decay Mode} & Pentaquark &\multirow{2}{*}{\pval} & \multirow{2}{*}{Significance ($\sigma$)} &  \multirow{2}{*}{Signal Yield} & \multicolumn{2}{c}{Upper Limit ($\times10^{-3}$)} \\ 
 & Hypothesis & & & & (90\% CL) & (95\% CL) \\ 
\hline 
\multirow{3}{*}{\Lc\Dzb} & $P_c(4312)^+$ & 0.32 & 0.48 &  $19.78 \pm 22.27$ & 1.17 & 1.29 \\
                         & $P_c(4440)^+$ & 0.44 & 0.15 &  $26.91 \pm 28.17$ & 1.41 & 1.53 \\
                         & $P_c(4457)^+$ & 0.53 & 0.00 &  $6.20 \pm 13.60$ & 1.27 & 1.43 \\
\hline
\multirow{2}{*}{\Lc\pip\Dstarm} & $P_c(4440)^+$ & 1.00 & 0.00 &  $0.00 \pm 0.96$ & 0.72 & 0.91 \\
                                & $P_c(4457)^+$ & 1.00 & 0.00 &  $0.00 \pm 1.73$ & 0.77 & 0.97 \\
\hline
\multirow{2}{*}{\Lc\pim\Dstarm} & $P_c(4440)^+$ & 1.00 & 0.00 &  $0.00 \pm 0.80$ & 0.63 & 0.80 \\
                                & $P_c(4457)^+$ & 1.00 & 0.00 &  $0.00 \pm 0.74$ & 0.59 & 0.74 \\
\hline
\multirow{3}{*}{\Lc\pip\Dm} & $P_c(4312)^+$ & 1.00 & 0.00 &  $0.00 \pm \phz1.56$ & 0.69 & 0.88 \\
                            & $P_c(4440)^+$ & 0.65 & 0.00 &  $4.43 \pm 11.67$ & 3.71 & 4.24 \\
                            & $P_c(4457)^+$ & 0.65 & 0.00 &  $5.94 \pm 12.68$  & 3.13 & 3.61 \\
\hline
\multirow{3}{*}{\Lc\pim\Dm} & $P_c(4312)^+$ & 1.00 & 0.00 &  $0.00 \pm \phz1.42$ & 0.67 & 0.86 \\
                            & $P_c(4440)^+$ & 0.53 & 0.00 &  $12.52 \pm 15.89$ & 3.91 & 4.37 \\
                            & $P_c(4457)^+$ & 0.53 & 0.00 &  $8.60 \pm 12.22$ & 3.10 & 3.51 \\
\hline
\multirow{2}{*}{\Sigmacz\Dm} & $P_c(4440)^+$ & 1.00 & 0.00 &  $0.00 \pm 2.47$ & 0.82 & 1.03 \\
                             & $P_c(4457)^+$ & 1.00 & 0.00 &  $0.00 \pm 1.05$ & 0.63 & 0.81 \\
\hline
\multirow{2}{*}{\Sigmacpp\Dm} & $P_c(4440)^+$ & 0.80 & 0.00 &  $0.61 \pm 4.52$ & 1.13 & 1.37 \\
                              & $P_c(4457)^+$ & 0.59 & 0.00 &  $0.66 \pm1.79$ & 0.80 & 0.99 \\
\hline
\multirow{2}{*}{\Sigmacstz\Dm} & $P_c(4440)^+$ & 0.31 & 0.49 & $3.23 \pm 3.53$ & 1.89 & 2.24 \\
                               & $P_c(4457)^+$ & 1.00 & 0.00 &  $0.00 \pm 3.09$  & 0.91 & 1.13 \\
\hline
\multirow{2}{*}{\Sigmacstpp\Dm} & $P_c(4440)^+$ & 0.75 & 0.00 &  $1.20 \pm 3.81$ & 1.38 & 1.67 \\
                                & $P_c(4457)^+$ & 1.00 & 0.00 &  $0.00 \pm 5.74$ & 0.87 & 1.08 \\
 \end{tabular}} 
\end{table}
\FloatBarrier
\section{Conclusion}
Using an integrated luminosity of $5.7\invfb$ of $pp$ collision data collected by the \lhcb detector, a large range of combinations of open-charm and hidden-charm hadronic states are investigated for possible pentaquark decay channels. A signal model based on the known pentaquark states is also fitted, and the signal yield is found to be consistent with zero in all cases. When scanning the \qval distribution from threshold to $600\mevcc$, the upper limit on the yield relative to the normalisation channel is set at $90\%$ and $95\%$ confidence levels. The highest significance is seen in the \Lc\pip\Dm final state, however using studies of background-only pseudo-experiments this is found to be consistent with the background-only hypothesis.

%% file: acknowledgements.tex
\section*{Acknowledgements}
%
%
\noindent We express our gratitude to our colleagues in the CERN
accelerator departments for the excellent performance of the LHC. We
thank the technical and administrative staff at the LHCb
institutes.
We acknowledge support from CERN and from the national agencies:
CAPES, CNPq, FAPERJ and FINEP (Brazil); 
MOST and NSFC (China); 
CNRS/IN2P3 (France); 
BMBF, DFG and MPG (Germany); 
INFN (Italy); 
NWO (Netherlands); 
MNiSW and NCN (Poland); 
MCID/IFA (Romania); 
MICINN (Spain); 
SNSF and SER (Switzerland); 
NASU (Ukraine); 
STFC (United Kingdom); 
DOE NP and NSF (USA).
We acknowledge the computing resources that are provided by CERN, IN2P3
(France), KIT and DESY (Germany), INFN (Italy), SURF (Netherlands),
PIC (Spain), GridPP (United Kingdom), 
CSCS (Switzerland), IFIN-HH (Romania), CBPF (Brazil),
and Polish WLCG (Poland).
We are indebted to the communities behind the multiple open-source
software packages on which we depend.
Individual groups or members have received support from
ARC and ARDC (Australia);
Key Research Program of Frontier Sciences of CAS, CAS PIFI, CAS CCEPP, 
Fundamental Research Funds for the Central Universities, 
and Sci. \& Tech. Program of Guangzhou (China);
Minciencias (Colombia);
EPLANET, Marie Sk\l{}odowska-Curie Actions, ERC and NextGenerationEU (European Union);
A*MIDEX, ANR, IPhU and Labex P2IO, and R\'{e}gion Auvergne-Rh\^{o}ne-Alpes (France);
AvH Foundation (Germany);
ICSC (Italy); 
GVA, XuntaGal, GENCAT, Inditex, InTalent and Prog.~Atracci\'on Talento, CM (Spain);
SRC (Sweden);
the Leverhulme Trust, the Royal Society
 and UKRI (United Kingdom).

%% file: appendix.tex

\section*{Appendices}

\appendix
\section{Summary tables}
Tables~\ref{tab:ULtable5},~\ref{tab:ULtable6} and~\ref{tab:ULtable7} summarise the upper limits set for each signal channel, along with the signal yields with corresponding significances, as well as the \qval that these values occur at.
\label{sec:restofTabs}
\input{tabs/tab1.txt}

%% file: tabs/tab1.txt
\begin{table}[!htb] 
\centering 
\caption{Upper limits on $R({\Lc})$ at $90\%$ and $95\%$ CL are shown for eleven of the hidden-charm modes, for their lowest \pval and highest significance (both local and corrected for LEE) found in the mass scan for all signal models. The \qval and signal yield for these points are also summarised, where the error on the signal yield is statistical only. The Q-value can be converted to the absolute mass by adding the masses of each decay product in the respective combination.} 
\label{tab:ULtable5} 
\sisetup{exponent-mode = scientific}
\resizebox{\columnwidth}{!}{
\begin{tabular}{lccccccrrr} 
\multirow{2}{*}{Decay Mode} & Width & \multicolumn{2}{c}{Lowest \pval} & \multicolumn{2}{c}{Significance ($\sigma$)} & \qval & \multirow{2}{*}{Signal Yield} & \multicolumn{2}{c}{UL ($\times10^{-3}$)} \\ 
 & (\mevcc) & Local & Corrected & Local & Corrected & (\mevcc) & & $90\%$ CL & $95\%$ CL \\ 
\hline 
\multirow{4}{*}{\Lc\Dzb} & 0 & \num[tight-spacing=true]{0.0047} & 0.36 & 2.60 & 0.36 & 353 & $22.7 \pm 30.6$ & \num[drop-exponent]{0.00141} & \num[drop-exponent]{0.00154} \\
  & 5 & \num[tight-spacing=true]{0.0034} & 0.20 & 2.71 & 0.86 & 353 & $31.4 \pm 12.8$ & \num[drop-exponent]{0.00153} & \num[drop-exponent]{0.00164} \\
  & 10 & \num[tight-spacing=true]{0.0027} & 0.13 & 2.78 & 1.15 & 349 & $39.4 \pm 34.8$ & \num[drop-exponent]{0.00154} & \num[drop-exponent]{0.00165} \\
  & 15 & \num[tight-spacing=true]{0.0022} & 0.09 & 2.85 & 1.36 & 349 & $46.8 \pm 91.0$ & \num[drop-exponent]{0.00157} & \num[drop-exponent]{0.00167} \\
\hline 
\multirow{4}{*}{\Lc\Dm} & 0 & \num[tight-spacing=true]{0.0047} & 0.38 & 2.60 & 0.31 & 501 & $24.0 \pm 12.9$ & \num[drop-exponent]{0.00205} & \num[drop-exponent]{0.00225} \\
  & 5 & \num[tight-spacing=true]{0.0037} & 0.22 & 2.67 & 0.77 & 497 & $26.7 \pm 22.2$ & \num[drop-exponent]{0.00209} & \num[drop-exponent]{0.00227} \\
  & 10 & \num[tight-spacing=true]{0.0027} & 0.13 & 2.78 & 1.15 & 497 & $33.1 \pm 23.0$ & \num[drop-exponent]{0.00217} & \num[drop-exponent]{0.00235} \\
  & 15 & \num[tight-spacing=true]{0.003} & 0.11 & 2.75 & 1.21 & 497 & $38.0 \pm 25.5$ & \num[drop-exponent]{0.00223} & \num[drop-exponent]{0.00239} \\ 
\hline 
\multirow{4}{*}{\Lc\Dstarm} & 0 & \num[tight-spacing=true]{0.0014} & 0.12 & 2.99 & 1.17 & 417 & $17.5 \pm \phz6.9$ & \num[drop-exponent]{0.00248} & \num[drop-exponent]{0.00272} \\
  & 5 & \num[tight-spacing=true]{0.0022} & 0.13 & 2.85 & 1.13 & 417 & $20.8 \pm \phz9.3$ & \num[drop-exponent]{0.00274} & \num[drop-exponent]{0.00294} \\
  & 10 & \num[tight-spacing=true]{0.003} & 0.13 & 2.75 & 1.11 & 421 & $23.8 \pm \phz9.6$ & \num[drop-exponent]{0.00280} & \num[drop-exponent]{0.00296} \\
  & 15 & \num[tight-spacing=true]{0.0047} & 0.16 & 2.60 & 0.99 & 421 & $26.3 \pm 10.8$ & \num[drop-exponent]{0.00285} & \num[drop-exponent]{0.00300} \\ 
\hline 
\multirow{4}{*}{\Sigmacpp\Dzb} & 0 & \num[tight-spacing=true]{0.0012} & 0.10 & 3.04 & 1.27 & 301 & $\phz7.0 \pm \phz3.2$ & \num[drop-exponent]{0.00110} & \num[drop-exponent]{0.00122} \\
  & 5 & \num[tight-spacing=true]{0.0043} & 0.23 & 2.63 & 0.73 & 301 & $\phz7.7 \pm \phz3.6$ & \num[drop-exponent]{0.00123} & \num[drop-exponent]{0.00137} \\
  & 10 & \num[tight-spacing=true]{0.014} & 0.48 & 2.21 & 0.04 & 301 & $\phz7.9 \pm \phz4.1$ & \num[drop-exponent]{0.00132} & \num[drop-exponent]{0.00149} \\
  & 15 & \num[tight-spacing=true]{0.032} & 0.81 & 1.85 & 0.00 & 301 & $\phz7.9 \pm \phz4.5$ & \num[drop-exponent]{0.00140} & \num[drop-exponent]{0.00159} \\ 
\hline 
\multirow{4}{*}{\Lc\pip\Dzb} & 0 & \num[tight-spacing=true]{0.00074} & 0.06 & 3.18 & 1.58 & 245 & $41.9 \pm 13.7$ & \num[drop-exponent]{0.00287} & \num[drop-exponent]{0.00306} \\
  & 5 & \num[tight-spacing=true,round-mode=figures,round-precision=3]{9.69438e-05} &  \num[tight-spacing=true,round-mode=figures,round-precision=3]{0.00576309} &  \num[tight-spacing=true,round-mode=figures,round-precision=3]{3.72685027} & \num[tight-spacing=true,round-mode=figures,round-precision=3]{2.52632785} & 245 & $67.6 \pm 19.2$ & \num[drop-exponent]{0.00322} & \num[drop-exponent]{0.00335} \\
  & 10 & \num[tight-spacing=true,round-mode=figures,round-precision=3]{2.45787e-05} &  \num[tight-spacing=true,round-mode=figures,round-precision=3]{0.00112118} &  \num[tight-spacing=true,round-mode=figures,round-precision=3]{4.05959797} & \num[tight-spacing=true,round-mode=figures,round-precision=3]{3.05609965} & 245 & $91.6 \pm 24.1$ & \num[drop-exponent]{0.00329} & \num[drop-exponent]{0.00339} \\
  & 15 & \num[tight-spacing=true,round-mode=figures,round-precision=3]{8.61234e-06} &  \num[tight-spacing=true,round-mode=figures,round-precision=3]{0.00031072} &  \num[tight-spacing=true,round-mode=figures,round-precision=3]{4.29812813} & \num[tight-spacing=true,round-mode=figures,round-precision=3]{3.42207623} & 245 & $115.0 \pm 28.5$ & \num[drop-exponent]{0.00330} & \num[drop-exponent]{0.00340} \\ 
\hline 
\multirow{4}{*}{\Sigmacpp\Dm} & 0 & \num[tight-spacing=true]{0.0052} & 0.41 & 2.56 & 0.24 & 181 & $\phz3.9 \pm \phz2.3$ & \num[drop-exponent]{0.00107} & \num[drop-exponent]{0.00121} \\
  & 5 & \num[tight-spacing=true]{0.0041} & 0.22 & 2.65 & 0.76 & 177 & $\phz6.8 \pm \phz3.3$ & \num[drop-exponent]{0.00146} & \num[drop-exponent]{0.00163} \\
  & 10 & \num[tight-spacing=true]{0.0037} & 0.15 & 2.68 & 1.02 & 177 & $\phz7.9 \pm \phz3.7$ & \num[drop-exponent]{0.00164} & \num[drop-exponent]{0.00184} \\
  & 15 & \num[tight-spacing=true]{0.0047} & 0.15 & 2.60 & 1.03 & 177 & $\phz8.4 \pm \phz4.0$ & \num[drop-exponent]{0.00176} & \num[drop-exponent]{0.00198} \\ 
\hline 
\multirow{4}{*}{\Lc\pip\Dm} & 0 & \num[tight-spacing=true]{0.00016} & 0.01 & 3.59 & 2.21 & 225 & $41.6 \pm 12.6$ & \num[drop-exponent]{0.00395} & \num[drop-exponent]{0.00419} \\
  & 5 & \num[tight-spacing=true,round-mode=figures,round-precision=3]{3.03075e-05} &  \num[tight-spacing=true,round-mode=figures,round-precision=3]{0.00195585} &  \num[tight-spacing=true,round-mode=figures,round-precision=3]{4.01040363} & \num[tight-spacing=true,round-mode=figures,round-precision=3]{2.88519669} & 225 & $64.7 \pm 17.4$ & \num[drop-exponent]{0.00443} & \num[drop-exponent]{0.00469} \\
  & 10 & \num[tight-spacing=true,round-mode=figures,round-precision=3]{8.61234e-06} &  \num[tight-spacing=true,round-mode=figures,round-precision=3]{0.00044365} &  \num[tight-spacing=true,round-mode=figures,round-precision=3]{4.29812813} & \num[tight-spacing=true,round-mode=figures,round-precision=3]{3.32402039} & 225 & $87.1 \pm 21.6$ & \num[drop-exponent]{0.00464} & \num[drop-exponent]{0.00485} \\
  & 15 & \num[tight-spacing=true,round-mode=figures,round-precision=3]{3.35578e-06} &  \num[tight-spacing=true,round-mode=figures,round-precision=3]{0.00014459} &  \num[tight-spacing=true,round-mode=figures,round-precision=3]{4.50263643} & \num[tight-spacing=true,round-mode=figures,round-precision=3]{3.62481284} & 225 & $108.2 \pm 25.3$ & \num[drop-exponent]{0.00472} & \num[drop-exponent]{0.00490} \\ 
\hline 
\multirow{4}{*}{\Lc\pip\Dstarm} & 0 & \num[tight-spacing=true]{0.0027} & 0.20 & 2.78 & 0.86 & 213 & $12.8 \pm \phz5.3$ & \num[drop-exponent]{0.00342} & \num[drop-exponent]{0.00375} \\
  & 5 & \num[tight-spacing=true]{0.00045} & 0.02 & 3.32 & 1.96 & 213 & $22.3 \pm \phz7.8$ & \num[drop-exponent]{0.00452} & \num[drop-exponent]{0.00476} \\
  & 10 & \num[tight-spacing=true,round-mode=figures,round-precision=3]{0.000144738} &  \num[tight-spacing=true,round-mode=figures,round-precision=3]{0.00628987} &  \num[tight-spacing=true,round-mode=figures,round-precision=3]{3.62454009} & \num[tight-spacing=true,round-mode=figures,round-precision=3]{2.49545026} & 213 & $30.4 \pm \phz9.7$ & \num[drop-exponent]{0.00475} & \num[drop-exponent]{0.00493} \\
  & 15 & \num[tight-spacing=true,round-mode=figures,round-precision=3]{7.42165e-05} &  \num[tight-spacing=true,round-mode=figures,round-precision=3]{0.00259775} &  \num[tight-spacing=true,round-mode=figures,round-precision=3]{3.79367661} & \num[tight-spacing=true,round-mode=figures,round-precision=3]{2.79465628} & 209 & $37.7 \pm 11.4$ & \num[drop-exponent]{0.00485} & \num[drop-exponent]{0.00502} \\ 
\hline 
\multirow{4}{*}{\Sigmacz\Dzb} & 0 & \num[tight-spacing=true]{0.013} & 0.95 & 2.22 & 0.00 & \phz65 & $\phz2.9 \pm \phz2.0$ & 0.68 & 0.78 \\
  & 5 & \num[tight-spacing=true]{0.017} & 0.79 & 2.13 & 0.00 & \phz65 & $\phz3.7 \pm 68.8$ & 0.90 & \num[drop-exponent]{0.00103} \\
  & 10 & \num[tight-spacing=true]{0.022} & 0.74 & 2.02 & 0.00 & \phz65 & $\phz4.4 \pm \phz4.3$ & \num[drop-exponent]{0.00102} & \num[drop-exponent]{0.00118} \\
  & 15 & \num[tight-spacing=true]{0.028} & 0.74 & 1.90 & 0.00 & \phz65 & $\phz4.8 \pm \phz4.2$ & \num[drop-exponent]{0.00113} & \num[drop-exponent]{0.00129} \\ 
\hline 
\multirow{4}{*}{\Lc\pim\Dzb} & 0 & \num[tight-spacing=true]{0.00048} & 0.04 & 3.30 & 1.72 & 597 & $54.0 \pm 17.2$ & \num[drop-exponent]{0.00279} & \num[drop-exponent]{0.00298} \\
  & 5 & \num[tight-spacing=true,round-mode=figures,round-precision=3]{9.65435e-05} &  \num[tight-spacing=true,round-mode=figures,round-precision=3]{0.00710847} &  \num[tight-spacing=true,round-mode=figures,round-precision=3]{3.72789335} & \num[tight-spacing=true,round-mode=figures,round-precision=3]{2.45173478} & 597 & $78.8 \pm 21.9$ & \num[drop-exponent]{0.00302} & \num[drop-exponent]{0.00320} \\
  & 10 & \num[tight-spacing=true,round-mode=figures,round-precision=3]{2.71248e-05} &  \num[tight-spacing=true,round-mode=figures,round-precision=3]{0.00163471} &  \num[tight-spacing=true,round-mode=figures,round-precision=3]{4.03652143} & \num[tight-spacing=true,round-mode=figures,round-precision=3]{2.94120216} & 597 & $104.0 \pm 26.3$ & \num[drop-exponent]{0.00315} & \num[drop-exponent]{0.00330} \\
  & 15 & \num[tight-spacing=true,round-mode=figures,round-precision=3]{9.50019e-06} &  \num[tight-spacing=true,round-mode=figures,round-precision=3]{0.00048262} &  \num[tight-spacing=true,round-mode=figures,round-precision=3]{4.27632523} & \num[tight-spacing=true,round-mode=figures,round-precision=3]{3.30046892} & 597 & $128.5 \pm 30.4$ & \num[drop-exponent]{0.00320} & \num[drop-exponent]{0.00333} \\ 
\hline 
\multirow{4}{*}{\Sigmacz\Dm} & 0 & \num[tight-spacing=true]{0.0023} & 0.19 & 2.84 & 0.88 & 261 & $\phz4.4 \pm \phz2.7$ & \num[drop-exponent]{0.00124} & \num[drop-exponent]{0.00139} \\
  & 5 & \num[tight-spacing=true]{0.0034} & 0.18 & 2.71 & 0.90 & 261 & $\phz5.7 \pm \phz3.0$ & \num[drop-exponent]{0.00150} & \num[drop-exponent]{0.00169} \\
  & 10 & \num[tight-spacing=true]{0.006} & 0.23 & 2.51 & 0.74 & 261 & $\phz6.4 \pm \phz3.4$ & \num[drop-exponent]{0.00166} & \num[drop-exponent]{0.00187} \\
  & 15 & \num[tight-spacing=true]{0.01} & 0.30 & 2.32 & 0.53 & 261 & $\phz7.0 \pm \phz3.8$ & \num[drop-exponent]{0.00178} & \num[drop-exponent]{0.00203} \\
\end{tabular}} 
\end{table}

\begin{table}[!htb]
\centering 
\caption{Upper limits on $R({\Lc})$ at $90\%$ and $95\%$ CL are shown for eight of the hidden-charm modes, for their lowest \pval and highest significance (both local and corrected for LEE) found in the mass scan for all signal models. The \qval and signal yield for these points are also summarised, where the error on the signal yield is statistical only. The Q-value can be converted to the absolute mass by adding the masses of each decay product in the respective combination.} 
\label{tab:ULtable6} 
\sisetup{exponent-mode = scientific}
\resizebox{\columnwidth}{!}{
\begin{tabular}{lccccccrrr} 
\multirow{2}{*}{Decay Mode} & Width & \multicolumn{2}{c}{Lowest \pval} & \multicolumn{2}{c}{Significance ($\sigma$)} & \qval & \multirow{2}{*}{Signal Yield} & \multicolumn{2}{c}{UL ($\times10^{-3}$)} \\ 
 & (\mev) & Local & Corrected & Local & Corrected & (\mevcc) & & $90\%$ CL & $95\%$ CL \\
 \hline
\multirow{4}{*}{\Lc\pim\Dm} & 0 & \num[tight-spacing=true]{0.00039} & 0.03 & 3.36 & 1.90 & 257 & $38.1 \pm 12.4$ & \num[drop-exponent]{0.00428} & \num[drop-exponent]{0.00456} \\
  & 5 & \num[tight-spacing=true,round-mode=figures,round-precision=3]{5.71353e-05} &  \num[tight-spacing=true,round-mode=figures,round-precision=3]{0.00333092} &  \num[tight-spacing=true,round-mode=figures,round-precision=3]{3.85810280} & \num[tight-spacing=true,round-mode=figures,round-precision=3]{2.71329188} & 253 & $62.1 \pm 17.1$ & \num[drop-exponent]{0.00462} & \num[drop-exponent]{0.00483} \\
  & 10 & \num[tight-spacing=true,round-mode=figures,round-precision=3]{1.45405e-05} &  \num[tight-spacing=true,round-mode=figures,round-precision=3]{0.00069169} &  \num[tight-spacing=true,round-mode=figures,round-precision=3]{4.18054581} & \num[tight-spacing=true,round-mode=figures,round-precision=3]{3.19809771} & 249 & $83.7 \pm 21.2$ & \num[drop-exponent]{0.00472} & \num[drop-exponent]{0.00488} \\
  & 15 & \num[tight-spacing=true,round-mode=figures,round-precision=3]{4.59283e-06} &  \num[tight-spacing=true,round-mode=figures,round-precision=3]{0.00018269} &  \num[tight-spacing=true,round-mode=figures,round-precision=3]{4.43550491} & \num[tight-spacing=true,round-mode=figures,round-precision=3]{3.56390572} & 249 & $103.5 \pm 24.6$ & \num[drop-exponent]{0.00477} & \num[drop-exponent]{0.00492} \\ 
\hline 
\multirow{4}{*}{\Lc\pim\Dstarm} & 0 & \num[tight-spacing=true]{0.0044} & 0.31 & 2.62 & 0.48 & 197 & $12.0 \pm \phz5.3$ & \num[drop-exponent]{0.00311} & \num[drop-exponent]{0.00345} \\
  & 5 & \num[tight-spacing=true]{0.0071} & 0.31 & 2.45 & 0.51 & 197 & $16.8 \pm \phz7.3$ & \num[drop-exponent]{0.00408} & \num[drop-exponent]{0.00453} \\
  & 10 & \num[tight-spacing=true]{0.0086} & 0.27 & 2.38 & 0.61 & 197 & $21.2 \pm \phz9.1$ & \num[drop-exponent]{0.00469} & \num[drop-exponent]{0.00515} \\
  & 15 & \num[tight-spacing=true]{0.0089} & 0.22 & 2.37 & 0.78 & 197 & $25.5 \pm 10.8$ & \num[drop-exponent]{0.00511} & \num[drop-exponent]{0.00556} \\ 
\hline 
\multirow{4}{*}{\Sigmacstpp\Dzb} & 0 & \num[tight-spacing=true]{0.01} & 0.75 & 2.32 & 0.00 & \phz37 & $\phz5.0 \pm \phz2.8$ & 0.96 & \num[drop-exponent]{0.00109} \\
  & 5 & \num[tight-spacing=true]{0.012} & 0.62 & 2.24 & 0.00 & \phz37 & $\phz7.8 \pm \phz4.0$ & \num[drop-exponent]{0.00132} & \num[drop-exponent]{0.00149} \\
  & 10 & \num[tight-spacing=true]{0.027} & 0.92 & 1.92 & 0.00 & 205 & $\phz7.0 \pm 20.6$ & \num[drop-exponent]{0.00157} & \num[drop-exponent]{0.00178} \\
  & 15 & \num[tight-spacing=true]{0.027} & 0.73 & 1.92 & 0.00 & 485 & $12.5 \pm \phz6.7$ & \num[drop-exponent]{0.00223} & \num[drop-exponent]{0.00249} \\ 
\hline 
\multirow{4}{*}{\Sigmacstpp\Dm} & 0 & \num[tight-spacing=true]{0.0012} & 0.11 & 3.03 & 1.21 & 537 & $\phz6.5 \pm \phz3.3$ & \num[drop-exponent]{0.00163} & \num[drop-exponent]{0.00182} \\
  & 5 & \num[tight-spacing=true]{0.0016} & 0.10 & 2.95 & 1.30 & 497 & $11.8 \pm \phz5.0$ & \num[drop-exponent]{0.00252} & \num[drop-exponent]{0.00279} \\
  & 10 & \num[tight-spacing=true]{0.0025} & 0.11 & 2.81 & 1.24 & 497 & $13.0 \pm \phz5.7$ & \num[drop-exponent]{0.00282} & \num[drop-exponent]{0.00312} \\
  & 15 & \num[tight-spacing=true]{0.0043} & 0.14 & 2.63 & 1.07 & 497 & $13.9 \pm \phz6.3$ & \num[drop-exponent]{0.00302} & \num[drop-exponent]{0.00337} \\ 
\hline 
\multirow{4}{*}{\Sigmacstpp\Dstarm} & 0 & \num[tight-spacing=true]{0.023} & 1.40 & 2.00 & 0.00 & 193 & $\phz2.5 \pm \phz1.8$ & \num[drop-exponent]{0.00108} & \num[drop-exponent]{0.00123} \\
  & 5 & \num[tight-spacing=true]{0.035} & 1.44 & 1.81 & 0.00 & 449 & $\phz2.9 \pm \phz2.1$ & \num[drop-exponent]{0.00126} & \num[drop-exponent]{0.00145} \\
  & 10 & \num[tight-spacing=true]{0.035} & 1.08 & 1.81 & 0.00 & 453 & $\phz3.2 \pm \phz2.3$ & \num[drop-exponent]{0.00136} & \num[drop-exponent]{0.00157} \\
  & 15 & \num[tight-spacing=true]{0.041} & 0.99 & 1.74 & 0.00 & 453 & $\phz3.3 \pm \phz2.4$ & \num[drop-exponent]{0.00145} & \num[drop-exponent]{0.00166} \\ 
\hline 
\multirow{4}{*}{\Sigmacstz\Dzb} & 0 & \num[tight-spacing=true]{0.0034} & 0.27 & 2.71 & 0.63 & 341 & $11.4 \pm \phz5.0$ & \num[drop-exponent]{0.00164} & \num[drop-exponent]{0.00183} \\
  & 5 & \num[tight-spacing=true]{0.0026} & 0.14 & 2.80 & 1.07 & 341 & $16.8 \pm \phz6.8$ & \num[drop-exponent]{0.00229} & \num[drop-exponent]{0.00256} \\
  & 10 & \num[tight-spacing=true]{0.0022} & 0.09 & 2.84 & 1.31 & 341 & $21.3 \pm \phz8.3$ & \num[drop-exponent]{0.00273} & \num[drop-exponent]{0.00300} \\
  & 15 & \num[tight-spacing=true]{0.0018} & 0.06 & 2.90 & 1.52 & 337 & $26.0 \pm \phz9.6$ & \num[drop-exponent]{0.00302} & \num[drop-exponent]{0.00327} \\ 
\hline 
\multirow{4}{*}{\Sigmacstz\Dm} & 0 & \num[tight-spacing=true]{0.0072} & 0.53 & 2.45 & 0.00 & 113 & $\phz6.2 \pm \phz3.2$ & \num[drop-exponent]{0.00144} & \num[drop-exponent]{0.00162} \\
  & 5 & \num[tight-spacing=true]{0.0014} & 0.08 & 2.99 & 1.40 & 537 & $11.5 \pm \phz4.8$ & \num[drop-exponent]{0.00260} & \num[drop-exponent]{0.00285} \\
  & 10 & \num[tight-spacing=true]{0.0011} & 0.05 & 3.06 & 1.66 & 537 & $13.6 \pm \phz5.4$ & \num[drop-exponent]{0.00299} & \num[drop-exponent]{0.00327} \\
  & 15 & \num[tight-spacing=true]{0.0012} & 0.04 & 3.02 & 1.70 & 537 & $15.1 \pm \phz6.0$ & \num[drop-exponent]{0.00323} & \num[drop-exponent]{0.00354} \\ 
\hline 
\multirow{4}{*}{\Sigmacstz\Dstarm} & 0 & \num[tight-spacing=true]{0.0099} & 0.51 & 2.33 & 0.00 & 109 & $\phz2.6 \pm \phz1.7$ & 0.96 & \num[drop-exponent]{0.00111} \\
  & 5 & \num[tight-spacing=true]{0.0032} & 0.16 & 2.73 & 0.99 & \phz17 & $\phz2.8 \pm \phz1.9$ & \num[drop-exponent]{0.00107} & \num[drop-exponent]{0.00122} \\
  & 10 & \num[tight-spacing=true]{0.0041} & 0.16 & 2.64 & 1.01 & \phz17 & $\phz2.9 \pm \phz2.1$ & \num[drop-exponent]{0.00113} & \num[drop-exponent]{0.00130} \\
  & 15 & \num[tight-spacing=true]{0.0061} & 0.17 & 2.51 & 0.94 & \phz17 & $\phz3.0 \pm \phz2.4$ & \num[drop-exponent]{0.00118} & \num[drop-exponent]{0.00134} \\
\end{tabular}} 
\end{table}

\begin{table}[!htb]
\centering 
\caption{Upper limits on $R({\Lc})$ at $90\%$ and $95\%$ CL are shown for each doubly-charmed mode, for their lowest \pval and highest significance (both local and corrected for LEE) found in the mass scan for all signal models. The \qval and signal yield for these points are also summarised, where the error on the signal yield is statistical only. The Q-value can be converted to the absolute mass by adding the masses of each decay product in the respective combination.} 
\label{tab:ULtable7} 
\sisetup{exponent-mode = scientific}
\resizebox{\columnwidth}{!}{
\begin{tabular}{lccccccrrr} 
\multirow{2}{*}{Decay Mode} & Width & \multicolumn{2}{c}{Lowest \pval} & \multicolumn{2}{c}{Significance ($\sigma$)} & \qval & \multirow{2}{*}{Signal Yield} & \multicolumn{2}{c}{UL ($\times10^{-3}$)} \\ 
 & (\mev) & Local & Corrected & Local & Corrected & (\mevcc) & & $90\%$ CL & $95\%$ CL \\
 \hline
\multirow{4}{*}{\Lc\Dz} & 0 & \num[tight-spacing=true]{0.0082} & 0.59 & 2.40 & 0.00 & \phz37 & $15.0 \pm \phz7.3$ & 0.97 & \num[drop-exponent]{0.00106} \\
  & 5 & \num[tight-spacing=true]{0.0047} & 0.26 & 2.60 & 0.64 & 153 & $26.8 \pm 11.0$ & \num[drop-exponent]{0.00119} & \num[drop-exponent]{0.00126} \\
  & 10 & \num[tight-spacing=true]{0.0027} & 0.12 & 2.78 & 1.17 & 153 & $36.5 \pm 14.0$ & \num[drop-exponent]{0.00127} & \num[drop-exponent]{0.00135} \\
  & 15 & \num[tight-spacing=true]{0.0024} & 0.09 & 2.82 & 1.34 & 153 & $45.0 \pm 16.5$ & \num[drop-exponent]{0.00133} & \num[drop-exponent]{0.00143} \\ 
\hline 
\multirow{4}{*}{\Lc\Dp} & 0 & \num[tight-spacing=true]{0.02} & 1.43 & 2.05 & 0.00 & 133 & $\phz9.4 \pm \phz5.0$ & \num[drop-exponent]{0.00101} & \num[drop-exponent]{0.00113} \\
  & 5 & \num[tight-spacing=true]{0.0047} & 0.27 & 2.60 & 0.61 & 169 & $11.0 \pm \phz9.9$ & \num[drop-exponent]{0.00125} & \num[drop-exponent]{0.00141} \\
  & 10 & \num[tight-spacing=true]{0.0042} & 0.18 & 2.64 & 0.90 & 169 & $13.5 \pm 24.5$ & \num[drop-exponent]{0.00152} & \num[drop-exponent]{0.00172} \\
  & 15 & \num[tight-spacing=true]{0.0058} & 0.20 & 2.52 & 0.85 & 169 & $14.9 \pm \phz7.6$ & \num[drop-exponent]{0.00171} & \num[drop-exponent]{0.00192} \\ 
\hline 
\multirow{4}{*}{\Lc\Dstarp} & 0 & \num[tight-spacing=true]{0.021} & 1.25 & 2.04 & 0.00 & \phz29 & $\phz3.5 \pm \phz2.3$ & 0.69 & 0.80 \\
  & 5 & \num[tight-spacing=true]{0.01} & 0.49 & 2.31 & 0.02 & \phz33 & $\phz6.2 \pm \phz4.2$ & \num[drop-exponent]{0.00115} & \num[drop-exponent]{0.00132} \\
  & 10 & \num[tight-spacing=true]{0.0081} & 0.31 & 2.41 & 0.51 & \phz33 & $\phz8.8 \pm \phz5.3$ & \num[drop-exponent]{0.00145} & \num[drop-exponent]{0.00164} \\
  & 15 & \num[tight-spacing=true]{0.0071} & 0.22 & 2.45 & 0.76 & \phz33 & $10.5 \pm \phz6.2$ & \num[drop-exponent]{0.00166} & \num[drop-exponent]{0.00188} \\ 
\hline 
\multirow{4}{*}{\Lc\pip\Dz} & 0 & \num[tight-spacing=true]{0.0027} & 0.22 & 2.78 & 0.77 & 193 & $18.0 \pm \phz7.2$ & \num[drop-exponent]{0.00202} & \num[drop-exponent]{0.00219} \\
  & 5 & \num[tight-spacing=true]{0.0016} & 0.09 & 2.95 & 1.32 & 193 & $28.0 \pm 10.4$ & \num[drop-exponent]{0.00232} & \num[drop-exponent]{0.00246} \\
  & 10 & \num[tight-spacing=true]{0.0014} & 0.07 & 2.99 & 1.51 & 193 & $36.7 \pm 12.9$ & \num[drop-exponent]{0.00246} & \num[drop-exponent]{0.00262} \\
  & 15 & \num[tight-spacing=true]{0.0014} & 0.05 & 2.99 & 1.61 & 197 & $44.8 \pm 15.3$ & \num[drop-exponent]{0.00270} & \num[drop-exponent]{0.00287} \\ 
\hline 
\multirow{4}{*}{\Lc\pip\Dp} & 0 & \num[tight-spacing=true]{0.0087} & 0.49 & 2.38 & 0.03 & 225 & $12.2 \pm \phz5.9$ & \num[drop-exponent]{0.00204} & \num[drop-exponent]{0.00229} \\
  & 5 & \num[tight-spacing=true]{0.0025} & 0.11 & 2.81 & 1.23 & 229 & $21.8 \pm \phz8.5$ & \num[drop-exponent]{0.00305} & \num[drop-exponent]{0.00332} \\
  & 10 & \num[tight-spacing=true]{0.00099} & 0.03 & 3.09 & 1.81 & 229 & $29.1 \pm 10.3$ & \num[drop-exponent]{0.00344} & \num[drop-exponent]{0.00370} \\
  & 15 & \num[tight-spacing=true]{0.00059} & 0.02 & 3.25 & 2.12 & 229 & $35.3 \pm 11.9$ & \num[drop-exponent]{0.00370} & \num[drop-exponent]{0.00396} \\ 
\hline 
\multirow{4}{*}{\Lc\pip\Dstarp} & 0 & \num[tight-spacing=true]{0.0059} & 0.39 & 2.52 & 0.28 & \phz77 & $\phz1.9 \pm \phz1.5$ & 0.84 & 0.95 \\
  & 5 & \num[tight-spacing=true]{0.0045} & 0.22 & 2.61 & 0.77 & 161 & $\phz6.5 \pm \phz3.6$ & \num[drop-exponent]{0.00181} & \num[drop-exponent]{0.00204} \\
  & 10 & \num[tight-spacing=true]{0.0029} & 0.11 & 2.76 & 1.23 & 161 & $\phz9.2 \pm \phz4.4$ & \num[drop-exponent]{0.00231} & \num[drop-exponent]{0.00258} \\
  & 15 & \num[tight-spacing=true]{0.0015} & 0.05 & 2.96 & 1.64 & 165 & $12.2 \pm \phz5.1$ & \num[drop-exponent]{0.00280} & \num[drop-exponent]{0.00310} \\ 
\hline 
\multirow{4}{*}{\Lc\pim\Dz} & 0 & \num[tight-spacing=true]{0.00073} & 0.06 & 3.18 & 1.53 & 593 & $20.3 \pm \phz7.4$ & \num[drop-exponent]{0.00211} & \num[drop-exponent]{0.00225} \\
  & 5 & \num[tight-spacing=true]{0.00066} & 0.04 & 3.21 & 1.78 & 593 & $26.3 \pm \phz9.3$ & \num[drop-exponent]{0.00226} & \num[drop-exponent]{0.00238} \\
  & 10 & \num[tight-spacing=true]{0.00059} & 0.03 & 3.24 & 1.95 & 593 & $32.3 \pm 11.0$ & \num[drop-exponent]{0.00235} & \num[drop-exponent]{0.00247} \\
  & 15 & \num[tight-spacing=true]{0.00048} & 0.02 & 3.30 & 2.13 & 593 & $38.3 \pm 12.5$ & \num[drop-exponent]{0.00242} & \num[drop-exponent]{0.00254} \\ 
\hline 
\multirow{4}{*}{\Lc\pim\Dp} & 0 & \num[tight-spacing=true]{0.00012} & 0.01 & 3.67 & 2.32 & 153 & $21.4 \pm \phz6.9$ & \num[drop-exponent]{0.00299} & \num[drop-exponent]{0.00324} \\
  & 5 & \num[tight-spacing=true,round-mode=figures,round-precision=3]{2.07292e-05} &  \num[tight-spacing=true,round-mode=figures,round-precision=3]{0.00136265} &  \num[tight-spacing=true,round-mode=figures,round-precision=3]{4.09919786} & \num[tight-spacing=true,round-mode=figures,round-precision=3]{2.99713564} & 153 & $33.3 \pm \phz9.5$ & \num[drop-exponent]{0.00365} & \num[drop-exponent]{0.00388} \\
  & 10 & \num[tight-spacing=true,round-mode=figures,round-precision=3]{9.4421e-06} &  \num[tight-spacing=true,round-mode=figures,round-precision=3]{0.00049255} &  \num[tight-spacing=true,round-mode=figures,round-precision=3]{4.27769089} & \num[tight-spacing=true,round-mode=figures,round-precision=3]{3.29474974} & 153 & $43.1 \pm 11.6$ & \num[drop-exponent]{0.00406} & \num[drop-exponent]{0.00428} \\
  & 15 & \num[tight-spacing=true,round-mode=figures,round-precision=3]{5.96313e-06} &  \num[tight-spacing=true,round-mode=figures,round-precision=3]{0.00025403} &  \num[tight-spacing=true,round-mode=figures,round-precision=3]{4.37893152} & \num[tight-spacing=true,round-mode=figures,round-precision=3]{3.47647238} & 153 & $51.7 \pm 13.4$ & \num[drop-exponent]{0.00429} & \num[drop-exponent]{0.00448} \\ 
\hline 
\multirow{4}{*}{\Lc\pim\Dstarp} & 0 & \num[tight-spacing=true]{0.0023} & 0.17 & 2.84 & 0.97 & \phz73 & $\phz3.2 \pm \phz3.0$ & \num[drop-exponent]{0.00119} & \num[drop-exponent]{0.00135} \\
  & 5 & \num[tight-spacing=true]{0.00068} & 0.04 & 3.20 & 1.76 & \phz73 & $\phz5.7 \pm \phz3.3$ & \num[drop-exponent]{0.00171} & \num[drop-exponent]{0.00192} \\
  & 10 & \num[tight-spacing=true]{0.00085} & 0.04 & 3.14 & 1.79 & \phz73 & $\phz7.0 \pm \phz3.8$ & \num[drop-exponent]{0.00194} & \num[drop-exponent]{0.00219} \\
  & 15 & \num[tight-spacing=true]{0.0013} & 0.04 & 3.01 & 1.70 & \phz73 & $\phz7.6 \pm \phz4.2$ & \num[drop-exponent]{0.00210} & \num[drop-exponent]{0.00236} \\ 
\hline 
\multirow{4}{*}{\Sigmacstpp\Dz} & 0 & \num[tight-spacing=true]{0.012} & 0.88 & 2.27 & 0.00 & 113 & $\phz2.5 \pm \phz1.8$ & 0.63 & 0.72 \\
  & 5 & \num[tight-spacing=true]{0.0094} & 0.50 & 2.35 & 0.00 & 113 & $\phz3.2 \pm \phz2.2$ & 0.76 & 0.87 \\
  & 10 & \num[tight-spacing=true]{0.013} & 0.48 & 2.23 & 0.04 & 113 & $\phz3.6 \pm \phz2.4$ & 0.87 & \num[drop-exponent]{0.00100} \\
  & 15 & \num[tight-spacing=true]{0.018} & 0.52 & 2.11 & 0.00 & 113 & $\phz3.9 \pm \phz2.7$ & 0.95 & \num[drop-exponent]{0.00111} \\ 
\hline 
\multirow{4}{*}{\Sigmacstpp\Dp} & 0 & \num[tight-spacing=true]{0.006} & 0.48 & 2.51 & 0.05 & 133 & $\phz1.9 \pm \phz1.5$ & 0.70 & 0.81 \\
  & 5 & \num[tight-spacing=true]{0.0068} & 0.37 & 2.47 & 0.34 & 133 & $\phz2.7 \pm \phz1.9$ & 0.89 & \num[drop-exponent]{0.00102} \\
  & 10 & \num[tight-spacing=true]{0.0088} & 0.34 & 2.37 & 0.41 & 133 & $\phz3.0 \pm \phz2.1$ & 0.97 & \num[drop-exponent]{0.00111} \\
  & 15 & \num[tight-spacing=true]{0.012} & 0.35 & 2.27 & 0.38 & 133 & $\phz3.2 \pm \phz2.2$ & \num[drop-exponent]{0.00103} & \num[drop-exponent]{0.00118} \\ 
\hline 
\multirow{4}{*}{\Sigmacstz\Dz} & 0 & \num[tight-spacing=true,round-mode=figures,round-precision=3]{1.51544e-05} &  \num[tight-spacing=true,round-mode=figures,round-precision=3]{0.00167257} &  \num[tight-spacing=true,round-mode=figures,round-precision=3]{4.17113400} & \num[tight-spacing=true,round-mode=figures,round-precision=3]{2.93410134} & \phz89 & $\phz3.9 \pm \phz2.0$ & 0.68 & 0.76 \\
  & 5 & \num[tight-spacing=true,round-mode=figures,round-precision=3]{5.12846e-05} &  \num[tight-spacing=true,round-mode=figures,round-precision=3]{0.00369741} &  \num[tight-spacing=true,round-mode=figures,round-precision=3]{3.88443255} & \num[tight-spacing=true,round-mode=figures,round-precision=3]{2.67852020} & \phz89 & $\phz5.3 \pm \phz2.6$ & 0.90 & 0.99 \\
  & 10 & \num[tight-spacing=true,round-mode=figures,round-precision=3]{0.000105924} &  \num[tight-spacing=true,round-mode=figures,round-precision=3]{0.00547493} &  \num[tight-spacing=true,round-mode=figures,round-precision=3]{3.70445323} & \num[tight-spacing=true,round-mode=figures,round-precision=3]{2.54429507} & \phz89 & $\phz6.2 \pm \phz3.0$ & \num[drop-exponent]{0.00103} & \num[drop-exponent]{0.00114} \\
  & 15 & \num[tight-spacing=true,round-mode=figures,round-precision=3]{0.000166295} &  \num[tight-spacing=true,round-mode=figures,round-precision=3]{0.00668620} &  \num[tight-spacing=true,round-mode=figures,round-precision=3]{3.58849716} & \num[tight-spacing=true,round-mode=figures,round-precision=3]{2.47369432} & \phz89 & $\phz6.8 \pm \phz3.2$ & \num[drop-exponent]{0.00113} & \num[drop-exponent]{0.00125} \\ 
\hline 
\multirow{4}{*}{\Sigmacstz\Dp} & 0 & \num[tight-spacing=true]{0.018} & 1.20 & 2.10 & 0.00 & 325 & $\phz2.6 \pm \phz1.8$ & 0.88 & \num[drop-exponent]{0.00101} \\
  & 5 & \num[tight-spacing=true]{0.013} & 0.61 & 2.23 & 0.00 & \phz73 & $\phz3.5 \pm \phz2.4$ & \num[drop-exponent]{0.00112} & \num[drop-exponent]{0.00129} \\
  & 10 & \num[tight-spacing=true]{0.013} & 0.48 & 2.21 & 0.06 & \phz73 & $\phz3.8 \pm \phz3.2$ & \num[drop-exponent]{0.00129} & \num[drop-exponent]{0.00150} \\
  & 15 & \num[tight-spacing=true]{0.015} & 0.41 & 2.18 & 0.22 & \phz69 & $\phz4.7 \pm \phz3.7$ & \num[drop-exponent]{0.00151} & \num[drop-exponent]{0.00175} \\
\end{tabular}} 
\end{table}

%% file: Authorship_LHCb-PAPER-2023-018.tex
\centerline
{\large\bf LHCb collaboration}
\begin
{flushleft}
\small
R.~Aaij$^{33}$\lhcborcid{0000-0003-0533-1952},
A.S.W.~Abdelmotteleb$^{52}$\lhcborcid{0000-0001-7905-0542},
C.~Abellan~Beteta$^{46}$,
F.~Abudin{\'e}n$^{52}$\lhcborcid{0000-0002-6737-3528},
T.~Ackernley$^{56}$\lhcborcid{0000-0002-5951-3498},
B.~Adeva$^{42}$\lhcborcid{0000-0001-9756-3712},
M.~Adinolfi$^{50}$\lhcborcid{0000-0002-1326-1264},
P.~Adlarson$^{78}$\lhcborcid{0000-0001-6280-3851},
H.~Afsharnia$^{10}$,
C.~Agapopoulou$^{44}$\lhcborcid{0000-0002-2368-0147},
C.A.~Aidala$^{79}$\lhcborcid{0000-0001-9540-4988}
% ...

{\footnotesize \it

% ...
$^{10}$Universit{\'e} Clermont Auvergne, CNRS/IN2P3, LPC, Clermont-Ferrand, France\\
% ...
$^{33}$Nikhef National Institute for Subatomic Physics, Amsterdam, Netherlands\\
% ...
$^{42}$Instituto Galego de F{\'\i}sica de Altas Enerx{\'\i}as (IGFAE), Universidade de Santiago de Compostela, Santiago de Compostela, Spain\\
% ...
$^{44}$European Organization for Nuclear Research (CERN), Geneva, Switzerland\\
% ...
$^{46}$Physik-Institut, Universit{\"a}t Z{\"u}rich, Z{\"u}rich, Switzerland\\
% ...
$^{50}$H.H. Wills Physics Laboratory, University of Bristol, Bristol, United Kingdom\\
% ...
$^{52}$Department of Physics, University of Warwick, Coventry, United Kingdom\\
% ...
$^{55}$School of Physics and Astronomy, University of Glasgow, Glasgow, United Kingdom\\
$^{56}$Oliver Lodge Laboratory, University of Liverpool, Liverpool, United Kingdom\\
% ...
$^{64}$Syracuse University, Syracuse, NY, United States\\
% ...
$^{78}$Department of Physics and Astronomy, Uppsala University, Uppsala, Sweden, associated to $^{55}$\\
$^{79}$University of Michigan, Ann Arbor, MI, United States, associated to $^{64}$\\
% ...
\medskip
}
\end{flushleft}

%% file: main.bbl
\ifx\mcitethebibliography\mciteundefinedmacro
\PackageError{LHCb.bst}{mciteplus.sty has not been loaded}
{This bibstyle requires the use of the mciteplus package.}\fi
\providecommand{\href}[2]{#2}
\begin{mcitethebibliography}{10}
\mciteSetBstSublistMode{n}
\mciteSetBstMaxWidthForm{subitem}{\alph{mcitesubitemcount})}
\mciteSetBstSublistLabelBeginEnd{\mcitemaxwidthsubitemform\space}
{\relax}{\relax}

\bibitem{Gell-Mann:1964ewy}
M.~Gell-Mann, \ifthenelse{\boolean{articletitles}}{\emph{{A schematic model of baryons and mesons}}, }{}\href{https://doi.org/10.1016/S0031-9163(64)92001-3}{Phys.\ Lett.\  \textbf{8} (1964) 214}\relax
\mciteBstWouldAddEndPuncttrue
\mciteSetBstMidEndSepPunct{\mcitedefaultmidpunct}
{\mcitedefaultendpunct}{\mcitedefaultseppunct}\relax
\EndOfBibitem
\bibitem{Zweig:1964jf}
G.~Zweig, in {\em {An SU(3) model for strong interaction symmetry and its breaking. Version 2}}, D.~B. Lichtenberg and S.~P. Rosen, eds., pp.~22--101, 1964\relax
\mciteBstWouldAddEndPuncttrue
\mciteSetBstMidEndSepPunct{\mcitedefaultmidpunct}
{\mcitedefaultendpunct}{\mcitedefaultseppunct}\relax
\EndOfBibitem
\bibitem{Jaffe:1976yi}
R.~L. Jaffe, \ifthenelse{\boolean{articletitles}}{\emph{{Perhaps a stable dihyperon}}, }{}\href{https://doi.org/10.1103/PhysRevLett.38.195}{Phys.\ Rev.\ Lett.\  \textbf{38} (1977) 195}, Erratum \href{https://doi.org/10.1103/PhysRevLett.38.617}{ibid.\   \textbf{38} (1977) 617}\relax
\mciteBstWouldAddEndPuncttrue
\mciteSetBstMidEndSepPunct{\mcitedefaultmidpunct}
{\mcitedefaultendpunct}{\mcitedefaultseppunct}\relax
\EndOfBibitem
\bibitem{Horn:1977rq}
D.~Horn and J.~Mandula, \ifthenelse{\boolean{articletitles}}{\emph{{A model of mesons with constituent gluons}}, }{}\href{https://doi.org/10.1103/PhysRevD.17.898}{Phys.\ Rev.\  \textbf{D17} (1978) 898}\relax
\mciteBstWouldAddEndPuncttrue
\mciteSetBstMidEndSepPunct{\mcitedefaultmidpunct}
{\mcitedefaultendpunct}{\mcitedefaultseppunct}\relax
\EndOfBibitem
\bibitem{Fritzsch:1973pi}
H.~Fritzsch, M.~Gell-Mann, and H.~Leutwyler, \ifthenelse{\boolean{articletitles}}{\emph{{Advantages of the color octet gluon picture}}, }{}\href{https://doi.org/10.1016/0370-2693(73)90625-4}{Phys.\ Lett.\  \textbf{B47} (1973) 365}\relax
\mciteBstWouldAddEndPuncttrue
\mciteSetBstMidEndSepPunct{\mcitedefaultmidpunct}
{\mcitedefaultendpunct}{\mcitedefaultseppunct}\relax
\EndOfBibitem
\bibitem{BaBar:2003oey}
BaBar collaboration, B.~Aubert {\em et~al.}, \ifthenelse{\boolean{articletitles}}{\emph{{Observation of a narrow meson decaying to $D_s^+ \pi^0$ at a mass of 2.32~GeV/c$^2$}}, }{}\href{https://doi.org/10.1103/PhysRevLett.90.242001}{Phys.\ Rev.\ Lett.\  \textbf{90} (2003) 242001}, \href{http://arxiv.org/abs/hep-ex/0304021}{{\normalfont\ttfamily arXiv:hep-ex/0304021}}\relax
\mciteBstWouldAddEndPuncttrue
\mciteSetBstMidEndSepPunct{\mcitedefaultmidpunct}
{\mcitedefaultendpunct}{\mcitedefaultseppunct}\relax
\EndOfBibitem
\bibitem{Belle:2003nnu}
Belle collaboration, S.~K. Choi {\em et~al.}, \ifthenelse{\boolean{articletitles}}{\emph{{Observation of a narrow charmonium-like state in exclusive $B^\pm \to K^\pm \pi^+ \pi^- J/\psi$ decays}}, }{}\href{https://doi.org/10.1103/PhysRevLett.91.262001}{Phys.\ Rev.\ Lett.\  \textbf{91} (2003) 262001}, \href{http://arxiv.org/abs/hep-ex/0309032}{{\normalfont\ttfamily arXiv:hep-ex/0309032}}\relax
\mciteBstWouldAddEndPuncttrue
\mciteSetBstMidEndSepPunct{\mcitedefaultmidpunct}
{\mcitedefaultendpunct}{\mcitedefaultseppunct}\relax
\EndOfBibitem
\bibitem{Belle:2007hrb}
Belle collaboration, S.~K. Choi {\em et~al.}, \ifthenelse{\boolean{articletitles}}{\emph{{Observation of a resonance-like structure in the $\pion^\pm \psi^\prime$ mass distribution in exclusive $B \to K \pion^\pm \psi^\prime$ decays}}, }{}\href{https://doi.org/10.1103/PhysRevLett.100.142001}{Phys.\ Rev.\ Lett.\  \textbf{100} (2008) 142001}, \href{http://arxiv.org/abs/0708.1790}{{\normalfont\ttfamily arXiv:0708.1790}}\relax
\mciteBstWouldAddEndPuncttrue
\mciteSetBstMidEndSepPunct{\mcitedefaultmidpunct}
{\mcitedefaultendpunct}{\mcitedefaultseppunct}\relax
\EndOfBibitem
\bibitem{LHCb-PAPER-2015-029}
LHCb collaboration, R.~Aaij {\em et~al.}, \ifthenelse{\boolean{articletitles}}{\emph{{Observation of $\jpsi\proton$ resonances consistent with pentaquark states in \mbox{\decay{\Lb}{\jpsi\proton\Km}} decays}}, }{}\href{https://doi.org/10.1103/PhysRevLett.115.072001}{Phys.\ Rev.\ Lett.\  \textbf{115} (2015) 072001}, \href{http://arxiv.org/abs/1507.03414}{{\normalfont\ttfamily arXiv:1507.03414}}\relax
\mciteBstWouldAddEndPuncttrue
\mciteSetBstMidEndSepPunct{\mcitedefaultmidpunct}
{\mcitedefaultendpunct}{\mcitedefaultseppunct}\relax
\EndOfBibitem
\bibitem{LHCB-PAPER-2016-009}
LHCb collaboration, R.~Aaij {\em et~al.}, \ifthenelse{\boolean{articletitles}}{\emph{{Model-independent evidence for $\jpsi\proton$ contributions to \mbox{\decay{\Lb}{\jpsi\proton\Km}} decays}}, }{}\href{https://doi.org/10.1103/PhysRevLett.117.082002}{Phys.\ Rev.\ Lett.\  \textbf{117} (2016) 082002}, \href{http://arxiv.org/abs/1604.05708}{{\normalfont\ttfamily arXiv:1604.05708}}\relax
\mciteBstWouldAddEndPuncttrue
\mciteSetBstMidEndSepPunct{\mcitedefaultmidpunct}
{\mcitedefaultendpunct}{\mcitedefaultseppunct}\relax
\EndOfBibitem
\bibitem{LHCb-PAPER-2019-014}
LHCb collaboration, R.~Aaij {\em et~al.}, \ifthenelse{\boolean{articletitles}}{\emph{{Observation of a narrow pentaquark state, $P_c(4312)^+$, and of two-peak structure of the $P_c(4450)^+$}}, }{}\href{https://doi.org/10.1103/PhysRevLett.122.222001}{Phys.\ Rev.\ Lett.\  \textbf{122} (2019) 222001}, \href{http://arxiv.org/abs/1904.03947}{{\normalfont\ttfamily arXiv:1904.03947}}\relax
\mciteBstWouldAddEndPuncttrue
\mciteSetBstMidEndSepPunct{\mcitedefaultmidpunct}
{\mcitedefaultendpunct}{\mcitedefaultseppunct}\relax
\EndOfBibitem
\bibitem{LHCb-PAPER-2021-031}
LHCb collaboration, R.~Aaij {\em et~al.}, \ifthenelse{\boolean{articletitles}}{\emph{{Observation of an exotic narrow doubly charmed tetraquark}}, }{}\href{https://doi.org/10.1038/s41567-022-01614-y}{Nature Physics \textbf{18} (2022) 751}, \href{http://arxiv.org/abs/2109.01038}{{\normalfont\ttfamily arXiv:2109.01038}}\relax
\mciteBstWouldAddEndPuncttrue
\mciteSetBstMidEndSepPunct{\mcitedefaultmidpunct}
{\mcitedefaultendpunct}{\mcitedefaultseppunct}\relax
\EndOfBibitem
\bibitem{LHCb-PAPER-2021-032}
LHCb collaboration, R.~Aaij {\em et~al.}, \ifthenelse{\boolean{articletitles}}{\emph{{Study of the doubly charmed tetraquark $T^+_{cc}$}}, }{}\href{https://doi.org/10.1038/s41467-022-30206-w}{Nature Communications \textbf{13} (2022) 3351}, \href{http://arxiv.org/abs/2109.01056}{{\normalfont\ttfamily arXiv:2109.01056}}\relax
\mciteBstWouldAddEndPuncttrue
\mciteSetBstMidEndSepPunct{\mcitedefaultmidpunct}
{\mcitedefaultendpunct}{\mcitedefaultseppunct}\relax
\EndOfBibitem
\bibitem{Close:2003sg}
F.~E. Close and P.~R. Page, \ifthenelse{\boolean{articletitles}}{\emph{{The $D^{*0} \bar D^0$ threshold resonance}}, }{}\href{https://doi.org/10.1016/j.physletb.2003.10.032}{Phys.\ Lett.\  \textbf{B578} (2004) 119}, \href{http://arxiv.org/abs/hep-ph/0309253}{{\normalfont\ttfamily arXiv:hep-ph/0309253}}\relax
\mciteBstWouldAddEndPuncttrue
\mciteSetBstMidEndSepPunct{\mcitedefaultmidpunct}
{\mcitedefaultendpunct}{\mcitedefaultseppunct}\relax
\EndOfBibitem
\bibitem{Karliner:2015ina}
M.~Karliner and J.~L. Rosner, \ifthenelse{\boolean{articletitles}}{\emph{{New exotic meson and baryon resonances from doubly-heavy hadronic molecules}}, }{}\href{https://doi.org/10.1103/PhysRevLett.115.122001}{Phys.\ Rev.\ Lett.\  \textbf{115} (2015) 122001}, \href{http://arxiv.org/abs/1506.06386}{{\normalfont\ttfamily arXiv:1506.06386}}\relax
\mciteBstWouldAddEndPuncttrue
\mciteSetBstMidEndSepPunct{\mcitedefaultmidpunct}
{\mcitedefaultendpunct}{\mcitedefaultseppunct}\relax
\EndOfBibitem
\bibitem{Maiani:2004vq}
L.~Maiani, F.~Piccinini, A.~D. Polosa, and V.~Riquer, \ifthenelse{\boolean{articletitles}}{\emph{{Diquark-antidiquarks with hidden or open charm and the nature of $X(3872)$}}, }{}\href{https://doi.org/10.1103/PhysRevD.71.014028}{Phys.\ Rev.\  \textbf{D71} (2005) 014028}, \href{http://arxiv.org/abs/hep-ph/0412098}{{\normalfont\ttfamily arXiv:hep-ph/0412098}}\relax
\mciteBstWouldAddEndPuncttrue
\mciteSetBstMidEndSepPunct{\mcitedefaultmidpunct}
{\mcitedefaultendpunct}{\mcitedefaultseppunct}\relax
\EndOfBibitem
\bibitem{Maiani:2015vwa}
L.~Maiani, A.~D. Polosa, and V.~Riquer, \ifthenelse{\boolean{articletitles}}{\emph{{The new pentaquarks in the diquark model}}, }{}\href{https://doi.org/10.1016/j.physletb.2015.08.008}{Phys.\ Lett.\  \textbf{B749} (2015) 289}, \href{http://arxiv.org/abs/1507.04980}{{\normalfont\ttfamily arXiv:1507.04980}}\relax
\mciteBstWouldAddEndPuncttrue
\mciteSetBstMidEndSepPunct{\mcitedefaultmidpunct}
{\mcitedefaultendpunct}{\mcitedefaultseppunct}\relax
\EndOfBibitem
\bibitem{Dubynskiy:2008mq}
S.~Dubynskiy and M.~B. Voloshin, \ifthenelse{\boolean{articletitles}}{\emph{{Hadro-charmonium}}, }{}\href{https://doi.org/10.1016/j.physletb.2008.07.086}{Phys.\ Lett.\  \textbf{B666} (2008) 344}, \href{http://arxiv.org/abs/0803.2224}{{\normalfont\ttfamily arXiv:0803.2224}}\relax
\mciteBstWouldAddEndPuncttrue
\mciteSetBstMidEndSepPunct{\mcitedefaultmidpunct}
{\mcitedefaultendpunct}{\mcitedefaultseppunct}\relax
\EndOfBibitem
\bibitem{Bugg:2011jr}
D.~V. Bugg, \ifthenelse{\boolean{articletitles}}{\emph{{An explanation of Belle states $Z_b(10610)$ and $Z_b(10650)$}}, }{}\href{https://doi.org/10.1209/0295-5075/96/11002}{EPL \textbf{96} (2011) 11002}, \href{http://arxiv.org/abs/1105.5492}{{\normalfont\ttfamily arXiv:1105.5492}}\relax
\mciteBstWouldAddEndPuncttrue
\mciteSetBstMidEndSepPunct{\mcitedefaultmidpunct}
{\mcitedefaultendpunct}{\mcitedefaultseppunct}\relax
\EndOfBibitem
\bibitem{Pakhlov:2014qva}
P.~Pakhlov and T.~Uglov, \ifthenelse{\boolean{articletitles}}{\emph{{Charged charmonium-like Z$^+$\textup{(4430)} from rescattering in conventional $B$ decays}}, }{}\href{https://doi.org/10.1016/j.physletb.2015.06.074}{Phys.\ Lett.\  \textbf{B748} (2015) 183}, \href{http://arxiv.org/abs/1408.5295}{{\normalfont\ttfamily arXiv:1408.5295}}\relax
\mciteBstWouldAddEndPuncttrue
\mciteSetBstMidEndSepPunct{\mcitedefaultmidpunct}
{\mcitedefaultendpunct}{\mcitedefaultseppunct}\relax
\EndOfBibitem
\bibitem{Yang_2020}
G.~Yang, J.~Ping, and J.~Segovia, \ifthenelse{\boolean{articletitles}}{\emph{Doubly charmed pentaquarks}, }{}\href{https://doi.org/10.1103/physrevd.101.074030}{Phys.\ Rev.\  \textbf{D101} (2020) }\relax
\mciteBstWouldAddEndPuncttrue
\mciteSetBstMidEndSepPunct{\mcitedefaultmidpunct}
{\mcitedefaultendpunct}{\mcitedefaultseppunct}\relax
\EndOfBibitem
\bibitem{Padmanath:2013laa}
M.~Padmanath, R.~G. Edwards, N.~Mathur, and M.~Peardon, \ifthenelse{\boolean{articletitles}}{\emph{{Spectroscopy of doubly and triply-charmed baryons from lattice QCD}}, }{}\href{https://doi.org/10.22323/1.187.0247}{PoS \textbf{LATTICE2013} (2014) 247}, \href{http://arxiv.org/abs/1311.4354}{{\normalfont\ttfamily arXiv:1311.4354}}\relax
\mciteBstWouldAddEndPuncttrue
\mciteSetBstMidEndSepPunct{\mcitedefaultmidpunct}
{\mcitedefaultendpunct}{\mcitedefaultseppunct}\relax
\EndOfBibitem
\bibitem{LHCb-DP-2008-001}
LHCb collaboration, A.~A. Alves~Jr.\ {\em et~al.}, \ifthenelse{\boolean{articletitles}}{\emph{{The \lhcb detector at the LHC}}, }{}\href{https://doi.org/10.1088/1748-0221/3/08/S08005}{JINST \textbf{3} (2008) S08005}\relax
\mciteBstWouldAddEndPuncttrue
\mciteSetBstMidEndSepPunct{\mcitedefaultmidpunct}
{\mcitedefaultendpunct}{\mcitedefaultseppunct}\relax
\EndOfBibitem
\bibitem{LHCb-DP-2014-002}
LHCb collaboration, R.~Aaij {\em et~al.}, \ifthenelse{\boolean{articletitles}}{\emph{{LHCb detector performance}}, }{}\href{https://doi.org/10.1142/S0217751X15300227}{Int.\ J.\ Mod.\ Phys.\  \textbf{A30} (2015) 1530022}, \href{http://arxiv.org/abs/1412.6352}{{\normalfont\ttfamily arXiv:1412.6352}}\relax
\mciteBstWouldAddEndPuncttrue
\mciteSetBstMidEndSepPunct{\mcitedefaultmidpunct}
{\mcitedefaultendpunct}{\mcitedefaultseppunct}\relax
\EndOfBibitem
\bibitem{LHCb-DP-2019-001}
R.~Aaij {\em et~al.}, \ifthenelse{\boolean{articletitles}}{\emph{{Performance of the LHCb trigger and full real-time reconstruction in Run 2 of the LHC}}, }{}\href{https://doi.org/10.1088/1748-0221/14/04/P04013}{JINST \textbf{14} (2019) P04013}, \href{http://arxiv.org/abs/1812.10790}{{\normalfont\ttfamily arXiv:1812.10790}}\relax
\mciteBstWouldAddEndPuncttrue
\mciteSetBstMidEndSepPunct{\mcitedefaultmidpunct}
{\mcitedefaultendpunct}{\mcitedefaultseppunct}\relax
\EndOfBibitem
\bibitem{Borghi:2017hfp}
S.~Borghi, \ifthenelse{\boolean{articletitles}}{\emph{{Novel real-time alignment and calibration of the LHCb detector and its performance}}, }{}\href{https://doi.org/10.1016/j.nima.2016.06.050}{Nucl.\ Instrum.\ Meth.\  \textbf{A845} (2017) 560}\relax
\mciteBstWouldAddEndPuncttrue
\mciteSetBstMidEndSepPunct{\mcitedefaultmidpunct}
{\mcitedefaultendpunct}{\mcitedefaultseppunct}\relax
\EndOfBibitem
\bibitem{LHCb-DP-2012-004}
R.~Aaij {\em et~al.}, \ifthenelse{\boolean{articletitles}}{\emph{{The \lhcb trigger and its performance in 2011}}, }{}\href{https://doi.org/10.1088/1748-0221/8/04/P04022}{JINST \textbf{8} (2013) P04022}, \href{http://arxiv.org/abs/1211.3055}{{\normalfont\ttfamily arXiv:1211.3055}}\relax
\mciteBstWouldAddEndPuncttrue
\mciteSetBstMidEndSepPunct{\mcitedefaultmidpunct}
{\mcitedefaultendpunct}{\mcitedefaultseppunct}\relax
\EndOfBibitem
\bibitem{LHCb-DP-2016-001}
R.~Aaij {\em et~al.}, \ifthenelse{\boolean{articletitles}}{\emph{{Tesla: an application for real-time data analysis in high energy physics}}, }{}\href{https://doi.org/10.1016/j.cpc.2016.07.022}{Comput.\ Phys.\ Commun.\  \textbf{208} (2016) 35}, \href{http://arxiv.org/abs/1604.05596}{{\normalfont\ttfamily arXiv:1604.05596}}\relax
\mciteBstWouldAddEndPuncttrue
\mciteSetBstMidEndSepPunct{\mcitedefaultmidpunct}
{\mcitedefaultendpunct}{\mcitedefaultseppunct}\relax
\EndOfBibitem
\bibitem{Sjostrand:2006za}
T.~Sj\"{o}strand, S.~Mrenna, and P.~Skands, \ifthenelse{\boolean{articletitles}}{\emph{{PYTHIA 6.4 physics and manual}}, }{}\href{https://doi.org/10.1088/1126-6708/2006/05/026}{JHEP \textbf{05} (2006) 026}, \href{http://arxiv.org/abs/hep-ph/0603175}{{\normalfont\ttfamily arXiv:hep-ph/0603175}}\relax
\mciteBstWouldAddEndPuncttrue
\mciteSetBstMidEndSepPunct{\mcitedefaultmidpunct}
{\mcitedefaultendpunct}{\mcitedefaultseppunct}\relax
\EndOfBibitem
\bibitem{Sjostrand:2007gs}
T.~Sj\"{o}strand, S.~Mrenna, and P.~Skands, \ifthenelse{\boolean{articletitles}}{\emph{{A brief introduction to PYTHIA 8.1}}, }{}\href{https://doi.org/10.1016/j.cpc.2008.01.036}{Comput.\ Phys.\ Commun.\  \textbf{178} (2008) 852}, \href{http://arxiv.org/abs/0710.3820}{{\normalfont\ttfamily arXiv:0710.3820}}\relax
\mciteBstWouldAddEndPuncttrue
\mciteSetBstMidEndSepPunct{\mcitedefaultmidpunct}
{\mcitedefaultendpunct}{\mcitedefaultseppunct}\relax
\EndOfBibitem
\bibitem{LHCb-PROC-2010-056}
I.~Belyaev {\em et~al.}, \ifthenelse{\boolean{articletitles}}{\emph{{Handling of the generation of primary events in Gauss, the LHCb simulation framework}}, }{}\href{https://doi.org/10.1088/1742-6596/331/3/032047}{J.\ Phys.\ Conf.\ Ser.\  \textbf{331} (2011) 032047}\relax
\mciteBstWouldAddEndPuncttrue
\mciteSetBstMidEndSepPunct{\mcitedefaultmidpunct}
{\mcitedefaultendpunct}{\mcitedefaultseppunct}\relax
\EndOfBibitem
\bibitem{Lange:2001uf}
D.~J. Lange, \ifthenelse{\boolean{articletitles}}{\emph{{The EvtGen particle decay simulation package}}, }{}\href{https://doi.org/10.1016/S0168-9002(01)00089-4}{Nucl.\ Instrum.\ Meth.\  \textbf{A462} (2001) 152}\relax
\mciteBstWouldAddEndPuncttrue
\mciteSetBstMidEndSepPunct{\mcitedefaultmidpunct}
{\mcitedefaultendpunct}{\mcitedefaultseppunct}\relax
\EndOfBibitem
\bibitem{Golonka:2005pn}
P.~Golonka and Z.~Was, \ifthenelse{\boolean{articletitles}}{\emph{{PHOTOS Monte Carlo: A precision tool for QED corrections in $Z$ and $W$ decays}}, }{}\href{https://doi.org/10.1140/epjc/s2005-02396-4}{Eur.\ Phys.\ J.\  \textbf{C45} (2006) 97}, \href{http://arxiv.org/abs/hep-ph/0506026}{{\normalfont\ttfamily arXiv:hep-ph/0506026}}\relax
\mciteBstWouldAddEndPuncttrue
\mciteSetBstMidEndSepPunct{\mcitedefaultmidpunct}
{\mcitedefaultendpunct}{\mcitedefaultseppunct}\relax
\EndOfBibitem
\bibitem{Allison:2006ve}
Geant4 collaboration, J.~Allison {\em et~al.}, \ifthenelse{\boolean{articletitles}}{\emph{{Geant4 developments and applications}}, }{}\href{https://doi.org/10.1109/TNS.2006.869826}{IEEE Trans.\ Nucl.\ Sci.\  \textbf{53} (2006) 270}\relax
\mciteBstWouldAddEndPuncttrue
\mciteSetBstMidEndSepPunct{\mcitedefaultmidpunct}
{\mcitedefaultendpunct}{\mcitedefaultseppunct}\relax
\EndOfBibitem
\bibitem{Agostinelli:2002hh}
Geant4 collaboration, S.~Agostinelli {\em et~al.}, \ifthenelse{\boolean{articletitles}}{\emph{{Geant4: A simulation toolkit}}, }{}\href{https://doi.org/10.1016/S0168-9002(03)01368-8}{Nucl.\ Instrum.\ Meth.\  \textbf{A506} (2003) 250}\relax
\mciteBstWouldAddEndPuncttrue
\mciteSetBstMidEndSepPunct{\mcitedefaultmidpunct}
{\mcitedefaultendpunct}{\mcitedefaultseppunct}\relax
\EndOfBibitem
\bibitem{LHCb-PROC-2011-006}
M.~Clemencic {\em et~al.}, \ifthenelse{\boolean{articletitles}}{\emph{{The \lhcb simulation application, Gauss: Design, evolution and experience}}, }{}\href{https://doi.org/10.1088/1742-6596/331/3/032023}{J.\ Phys.\ Conf.\ Ser.\  \textbf{331} (2011) 032023}\relax
\mciteBstWouldAddEndPuncttrue
\mciteSetBstMidEndSepPunct{\mcitedefaultmidpunct}
{\mcitedefaultendpunct}{\mcitedefaultseppunct}\relax
\EndOfBibitem
\bibitem{LHCb-PAPER-2020-004}
LHCb collaboration, R.~Aaij {\em et~al.}, \ifthenelse{\boolean{articletitles}}{\emph{{Observation of new \Xicz baryons decaying to $\Lc\Km$}}, }{}\href{https://doi.org/10.1103/PhysRevLett.124.222001}{Phys.\ Rev.\ Lett.\  \textbf{124} (2020) 222001}, \href{http://arxiv.org/abs/2003.13649}{{\normalfont\ttfamily arXiv:2003.13649}}\relax
\mciteBstWouldAddEndPuncttrue
\mciteSetBstMidEndSepPunct{\mcitedefaultmidpunct}
{\mcitedefaultendpunct}{\mcitedefaultseppunct}\relax
\EndOfBibitem
\bibitem{TMVA4}
A.~Hoecker {\em et~al.}, \ifthenelse{\boolean{articletitles}}{\emph{{TMVA 4 --- Toolkit for Multivariate Data Analysis with ROOT. Users Guide.}}, }{}\href{http://arxiv.org/abs/physics/0703039}{{\normalfont\ttfamily arXiv:physics/0703039}}\relax
\mciteBstWouldAddEndPuncttrue
\mciteSetBstMidEndSepPunct{\mcitedefaultmidpunct}
{\mcitedefaultendpunct}{\mcitedefaultseppunct}\relax
\EndOfBibitem
\bibitem{Skwarnicki:1986xj}
T.~Skwarnicki, {\em {A study of the radiative cascade transitions between the Upsilon-prime and Upsilon resonances}}, PhD thesis, Institute of Nuclear Physics, Krakow, 1986, {\href{http://inspirehep.net/record/230779/}{DESY-F31-86-02}}\relax
\mciteBstWouldAddEndPuncttrue
\mciteSetBstMidEndSepPunct{\mcitedefaultmidpunct}
{\mcitedefaultendpunct}{\mcitedefaultseppunct}\relax
\EndOfBibitem
\bibitem{Gross:2010qma}
E.~Gross and O.~Vitells, \ifthenelse{\boolean{articletitles}}{\emph{{Trial factors for the look elsewhere effect in high energy physics}}, }{}\href{https://doi.org/10.1140/epjc/s10052-010-1470-8}{Eur.\ Phys.\ J.\  \textbf{C70} (2010) 525}, \href{http://arxiv.org/abs/1005.1891}{{\normalfont\ttfamily arXiv:1005.1891}}\relax
\mciteBstWouldAddEndPuncttrue
\mciteSetBstMidEndSepPunct{\mcitedefaultmidpunct}
{\mcitedefaultendpunct}{\mcitedefaultseppunct}\relax
\EndOfBibitem
\bibitem{PDG2022}
Particle Data Group, R.~L. Workman {\em et~al.}, \ifthenelse{\boolean{articletitles}}{\emph{{\href{http://pdg.lbl.gov/}{Review of particle physics}}}, }{}\href{https://doi.org/10.1093/ptep/ptac097}{Prog.\ Theor.\ Exp.\ Phys.\  \textbf{2022} (2022) 083C01}\relax
\mciteBstWouldAddEndPuncttrue
\mciteSetBstMidEndSepPunct{\mcitedefaultmidpunct}
{\mcitedefaultendpunct}{\mcitedefaultseppunct}\relax
\EndOfBibitem
\end{mcitethebibliography}
